\documentclass[reprint,amsfonts, amssymb, amsmath,  showkeys, nofootinbib,pra, superscriptaddress, twocolumn,longbibliography]{revtex4-2}

\usepackage{comment}
\usepackage{float}
\makeatletter
\let\newfloat\newfloat@ltx
\makeatother
\usepackage[english]{babel}
\usepackage[utf8]{inputenc}
\usepackage{graphics}
\usepackage{selinput}
\usepackage{enumerate}

\usepackage{braket}
\usepackage{amsthm}
\usepackage{mathtools}
\usepackage{physics}
\usepackage[dvipsnames]{xcolor}
\usepackage{graphicx}
\usepackage[left=16mm,right=16mm,top=35mm,columnsep=15pt]{geometry} 
\usepackage{adjustbox}
\usepackage{placeins}
\usepackage[T1]{fontenc}
\usepackage{lipsum}
\usepackage{csquotes}
\usepackage{mathbbol} 
\usepackage[linesnumbered,ruled,vlined]{algorithm2e}
\usepackage[svgnames]{xcolor}
\SetKwInput{kwInit}{Init}
\usepackage{dsfont}

\def\HC{\mathcal{H}}

\def\LC{\mathcal{L}}

\def\ad{^{\dagger}}

\def\a{\alpha}

\usepackage[makeroom]{cancel}
\usepackage[colorlinks=true,citecolor=blue,linkcolor=magenta]{hyperref}

\usepackage{tikz}
\tikzset{every picture/.style=remember picture}

\newcommand{\ketbraq}[1]{\ketbra*{#1}{#1}}
\newcommand{\bramatket}[3]{\langle #1 \hspace{1pt} | #2 | \hspace{1pt} #3 \rangle}
\newcommand{\bramatketq}[2]{\bramatket{#1}{#2}{#1}}

\newcommand{\ipa}[2]{\langle #1,#2\rangle}      

\newcommand{\poly}{\operatorname{poly}}

\newcommand{\rg}{G}

\newcommand{\Ebb}{\mathbb{E}}

\newcommand{\Rbb}{\mathbb{R}}
\newcommand{\Fbb}{\mathbb{F}}

\newcommand{\DC}{\mathcal{D}}

\newcommand{\MC}{\mathcal{M}}

\newcommand{\OC}{\mathcal{O}}

\newcommand{\SC}{\mathcal{S}}
\newcommand{\TC}{\mathcal{T}}
\newcommand{\UC}{\mathcal{U}}

\newcommand{\YC}{\mathcal{Y}}
\newcommand{\ZC}{\mathcal{Z}}

\renewcommand{\geq}{\geqslant}
\renewcommand{\leq}{\leqslant}

\newcommand{\iter}[1]{^{(#1)}}

\DeclareMathOperator*{\argmax}{arg\,max}
\DeclareMathOperator*{\argmin}{arg\,min}
\renewcommand{\vec}[1]{\boldsymbol{#1}}  
\newcommand{\ot}{\otimes}

\newcommand{\bs}{\textsf{BS}}

\newcommand{\ep}{\epsilon}
\newcommand{\vep}{\varepsilon}

\renewcommand{\th}{\theta } 

\newcommand{\Th}{\Theta }

\newcommand{\thv}{\vec{\theta}}

\newcommand{\Nora}{N_{\rm oracle}}
\newcommand{\D}{\Delta}

\newcommand{\alv}{\vec{\alpha} }

\def\be{\begin{equation}}
\def\ee{\end{equation}}
\def\bs{\begin{split}}
\def\e{\end{split}}
\def\ba{\begin{eqnarray}}
\def\bea{\begin{eqnarray}}
\def\nn{\nonumber\\}
\def\tea{\end{eqnarray}}
\def\ea{\end{eqnarray}}
\def\eea{\end{eqnarray}}

\def\d{\delta}

\def\k{^{(k)}}

\def\su{\mathfrak{s}\mathfrak{u}}

\def\a{\alpha}

\def\l{\lambda}
\def\d{\delta}
\def\a{\alpha}

\def\t{\tau}

\def\k{^{(k)}}

\def\su{\mathfrak{s}\mathfrak{u}}

\def\a{\alpha}

\newcommand{\toterr}{\Delta}
\newcommand{\ett}{\vep_{\rm tt}}
\newcommand{\eptt}{\ep_{\rm TT}}
\newcommand{\egc}{\vep_{\rm GC}}
\newcommand{\epgc}{\ep_{\rm GC}}
\newcommand{\ttt}{r_{\rm min}^{\rm tt}}
\newcommand{\tgc}{r_{\rm min}^{\rm GC}}
\newcommand{\NU}{N_{U_0}}

\newtheorem{theorem}{Theorem}
\newtheorem{lemma}{Lemma}
\newtheorem{corollary}{Corollary}

\newtheorem{proposition}{Proposition}
\newtheorem*{proposition*}{Proposition}

\newtheorem{assumption}{Assumption}
\newtheorem{definition}{Definition}

\newtheorem{remark}{Remark}

\usepackage{amssymb}
\usepackage{dsfont}

\usepackage[normalem]{ulem}
\usepackage{hyperref}

\usepackage{ulem}

\newcommand{\cdel}[1]{{\color{red}\sout{#1}}}        
\newcommand{\oadd}[1]{{\color{orange!90!black} #1}}   
\definecolor{ricardviolet}{HTML}{7A3EB1}

\usepackage{minitoc}
\usepackage{tocloft}

\makeatletter
\renewcommand{\part}[1]{%
  \refstepcounter{part}
  \addcontentsline{toc}{part}{#1}
  \mtcaddpart[part]{#1}
}
\makeatother

\newcommand{\deff}{d_{\rm eff}}

\makeatletter
\renewcommand\onecolumngrid{
\do@columngrid{one}{\@ne}
\def\set@footnotewidth{\onecolumngrid}
\def\footnoterule{\kern-6pt\hrule width 1.5in\kern6pt}
}
\renewcommand\twocolumngrid{
        \def\footnoterule{
        \dimen@\skip\footins\divide\dimen@\thr@@
        \kern-\dimen@\hrule width.5in\kern\dimen@}
        \do@columngrid{mlt}{\tw@}
}
\makeatother

\begin{document}

\doparttoc 
\faketableofcontents 
\part{} 

\title{ 
Autonomous quantum machine learning via Riemannian gradient-based optimization 
}

\author{Kasidit Srimahajariyapong}
\email{kasidit.quantum@gmail.com}
\affiliation{Chula Intelligent and Complex Systems Center of Excellence, Department of Physics, Faculty of Science, Chulalongkorn University, Bangkok, Thailand, 10330}
\affiliation{Siam Quantum Square, Faculty of Science, Chulalongkorn University, Bangkok, Thailand}
\affiliation{Nanyang Quantum Hub, School of Physical and Mathematical Sciences, Nanyang Technological University, 637371, Singapore}

\author{Thiparat Chotibut}
\email{thiparatc@gmail.com}
\affiliation{Chula Intelligent and Complex Systems Center of Excellence, Department of Physics, Faculty of Science, Chulalongkorn University, Bangkok, Thailand, 10330}
\affiliation{Siam Quantum Square, Faculty of Science, Chulalongkorn University, Bangkok, Thailand}
\affiliation{Nanyang Quantum Hub, School of Physical and Mathematical Sciences, Nanyang Technological University, 637371, Singapore}

\author{Suwannachad Suwannajitt}
\affiliation{Chula Intelligent and Complex Systems Center of Excellence, Department of Physics, Faculty of Science, Chulalongkorn University, Bangkok, Thailand, 10330}
\affiliation{Siam Quantum Square, Faculty of Science, Chulalongkorn University, Bangkok, Thailand}

\author{Tithivich~Paka}
\affiliation{Chula Intelligent and Complex Systems Center of Excellence, Department of Physics, Faculty of Science, Chulalongkorn University, Bangkok, Thailand, 10330}
\affiliation{Siam Quantum Square, Faculty of Science, Chulalongkorn University, Bangkok, Thailand}

\author{Ricard Puig}
\affiliation{Institute of Physics, Ecole Polytechnique Fédérale de Lausanne (EPFL),  Lausanne 1015, Switzerland}
\affiliation{Centre for Quantum Science and Engineering, Ecole Polytechnique Fédérale de Lausanne (EPFL),  Lausanne 1015, Switzerland}

\author{Yudai Suzuki}
\affiliation{Institute of Physics, Ecole Polytechnique Fédérale de Lausanne (EPFL),  Lausanne 1015, Switzerland}
\affiliation{Centre for Quantum Science and Engineering, Ecole Polytechnique Fédérale de Lausanne (EPFL),  Lausanne 1015, Switzerland}

\author{Marek Gluza}
\affiliation{Nanyang Quantum Hub, School of Physical and Mathematical Sciences, Nanyang Technological University, 637371, Singapore}

\author{Zoë Holmes}
\affiliation{Institute of Physics, Ecole Polytechnique Fédérale de Lausanne (EPFL),  Lausanne 1015, Switzerland}
\affiliation{Centre for Quantum Science and Engineering, Ecole Polytechnique Fédérale de Lausanne (EPFL),  Lausanne 1015, Switzerland}

\author{Supanut Thanasilp}
\email{supanut.thanasilp@gmail.com}
\affiliation{Chula Intelligent and Complex Systems Center of Excellence, Department of Physics, Faculty of Science, Chulalongkorn University, Bangkok, Thailand, 10330}
\affiliation{Siam Quantum Square, Faculty of Science, Chulalongkorn University, Bangkok, Thailand}

\date{\today}

\begin{abstract}
Quantum machine learning models typically rely on repeated quantum measurements and classical optimization during training, making the quantum-to-classical information-extraction process a fundamental bottleneck for scalability. We propose a framework for autonomous self-training quantum machine learning, based on Riemannian gradient optimization, in which the optimization proceeds entirely within the quantum domain without requiring quantum measurements or classical feedback. 
We derive the Riemannian gradient update rules for a broad class of learning tasks and identify sufficient conditions under which the resulting training dynamics can be realized fully autonomously. 
Within this regime, we establish an exponential convergence guarantee, in the training iterations, for sufficiently small step sizes, and further analyze the training cost in terms of the growth of the circuit depth. 
Importantly, we further show that the trained model can automatically discover symmetries hidden in the training data without \textit{a priori} specification.
For the canonical examples of quantum phase classification and unitary learning with a hidden symmetry, our numerical study indicates that learning can be achieved at low circuit depths with only a few distinct training examples. Our results thus establish autonomous quantum machine learning as a promising paradigm that removes the classical optimization loop while retaining strong theoretical guarantees and practical performances.
\end{abstract}

\maketitle

\begin{figure*}[!htbp]
    \centering
    \includegraphics[width=0.95\linewidth]{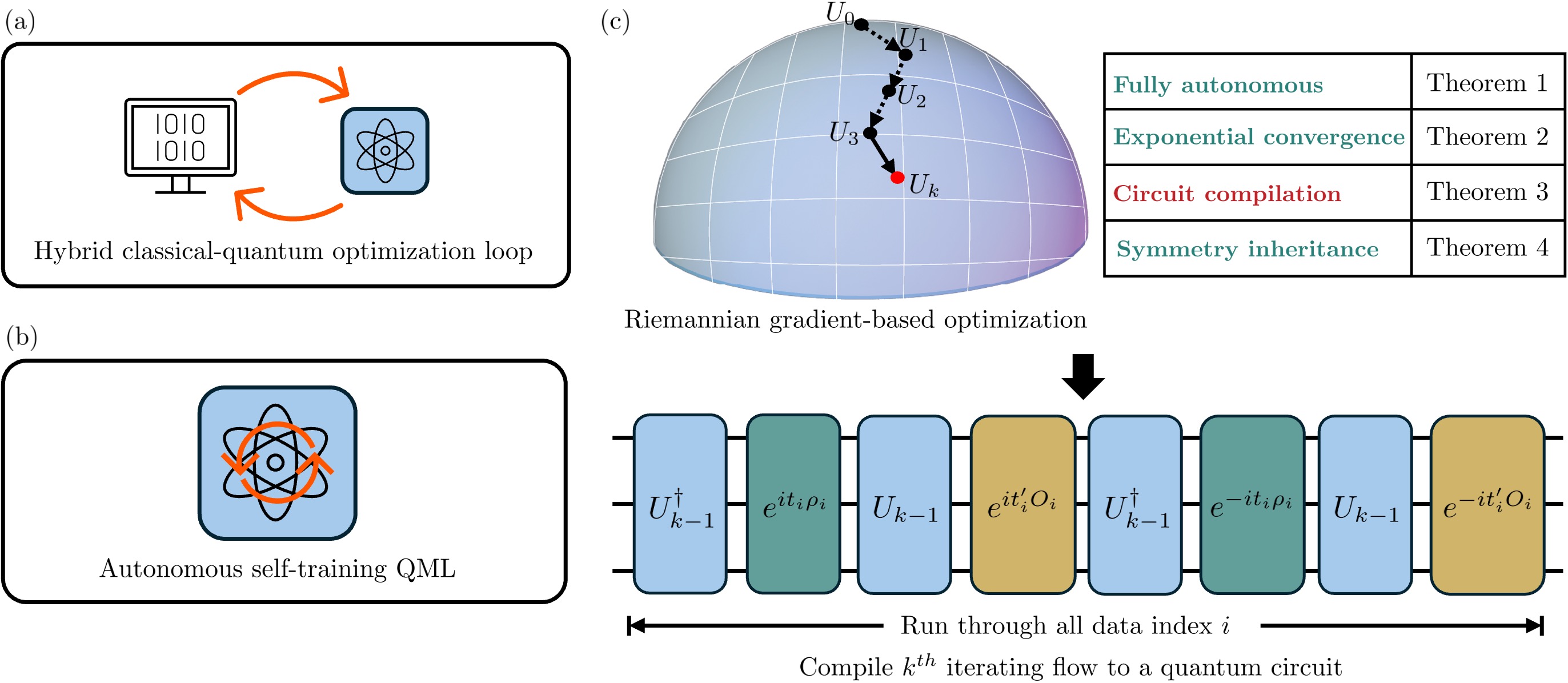}
    \caption{\textbf{Overview.} Schematic comparison of  the training processes for (a) conventional and (b) autonomous quantum machine learning. The former relies on a hybrid classical-quantum loop that exchanges information between a classical computer and a quantum device, while the latter operates entirely within the quantum system. (c) Schematic summary of Riemannian gradient-based optimization as an example of autonomous quantum machine learning under a specific condition (Theorem~\ref{thm:sufficient-cond-autonomous-qml}): the discretized Riemannian gradient flow on a unitary manifold is compiled into a quantum circuit, where each iteration is implemented by a sequence of data-dependent quantum gate blocks. The evolution times $t_i$ and $t_i'$ depend on the learning rate of the algorithm and the loss function. In addition to operating autonomously under the condition, this framework offers a loss decrease guarantee (Theorem~\ref{thm:auto_RG_learning_guarantee}) and symmetry inheritance shared between the data and the problem (Theorem~\ref{thm:sym-inherence}),  while its main challenge lies in circuit compilation (Theorem~\ref{thm:resource}).}
    \label{fig:summary}
\end{figure*}


\section{Introduction}

Training Quantum Machine Learning (QML) models typically requires extracting classical information through quantum measurements. Variational approaches~\cite{cerezo2020variationalreview,bharti2022noisy} such as parametrized quantum circuits~\cite{benedetti2019parameterized,peruzzo2014variational,farhi2014quantum} require repeated estimation of loss functions and gradients with hybrid quantum–classical feedback loops, whose statistical errors are fundamentally constrained by shot noise. On the other hand, although less apparent, non-variational quantum models face similar limitations. While their training procedures take place entirely on the classical side, quantum measurements are still required to extract information from quantum states before classical post-processing can be performed. In particular, quantum kernel methods require estimating data-independent quantities such as Gram matrices~\mbox{\cite{schuld2019quantum,havlivcek2019supervised,huang2021power}}, whereas quantum reservoir computing relies on expectation values of observables~\mbox{\cite{fujii2017harnessing,nakajima2019boosting,mujal2021opportunities,xiong2025role}}.

The core of the problem in all these cases is  arguably the use of a classical optimizer to solve a quantum problem. This means that information in the quantum model needs to be repeatedly converted to a classical cost function for classical optimization. The classical loss function values (or kernel values) are obtained statistically by repeatedly measuring the output of a quantum circuit. If loss function gradients are very small then the statistical noise resulting from this quantum-to-classical conversion process will wash out any meaningful information about how to train the quantum model~\cite{xiong2025role,mcclean2018barren,thanasilp2022exponential,cerezo2023does,srimahajariyapong2025connecting,aghaei2026pitfalls,larocca2024review}. 
However, if we could train on the exact loss function values, then variational quantum algorithms are guaranteed to converge in the over-parameterized regime~\cite{you2022convergence}. Thus it is the combination of the exponential concentration and the inevitability of shot noise that inhibits the scaling of these methods. This motivates the fundamental question: \textit{could we banish the classical optimizer from the optimization loop and optimize entirely within the quantum domain?} 

The concept of autonomous self-training quantum algorithms has begun to be explored. Approaches based on encoding optimization problem into dynamical or dissipative dynamics have been proposed in Ref.~\cite{motlagh2024ground,catli2025exponentially,leng2023quantum,wanjura2025quantum}. In addition, although not originally framed as self-training protocols, a series of works on circuit synthesis for Riemannian gradient–based optimization (known as double-bracket quantum algorithms) can be effectively viewed as operating under the same principles for imaginary time evolution~\cite{wiersema2023optimizing,gluza2024doublebracket,robbiati2024double,gluza2026double, wright2026double}. Together, these developments provide the first solid step into fully quantum, self-contained learning dynamics. However, for all intents and purposes, these methods have not yet been explored in the context of quantum machine learning.

In this work, we propose and analyze the performance of autonomous self-training quantum models based on Riemannian gradient optimization~\cite{Helmke1995Optimization,absil2008optimization,schulte2010gradient,wiersema2023optimizing}, summarized in Fig.~\ref{fig:summary}. First, we define autonomous quantum machine learning as a regime in which, after encoding the training data, no further interaction between quantum and classical resources is required. That is, the optimal model is obtained entirely through coherent quantum evolution without intermediate information extraction and hence the training process proceeds fully within the quantum domain. 

We then introduce a family of QML models based on Riemannian gradient optimization, for which the gradient expressions for general nonlinear loss functions are rigorously derived from first principles. This framework is flexible and can accommodate a wide range of learning tasks, including classification, regression, and generative modeling, through appropriate choices of loss functions. However, the resulting training dynamics are not fully autonomous in general, as estimating certain intermediate quantities may still require measurement and classical processing. 

We then proceed to identify sufficient conditions under which autonomous quantum machine learning naturally emerges within this Riemannian framework. We prove that the training loss  decreases exponentially in the number of iterations for sufficiently small step sizes. 
However, similarly to other ``double-bracket'' methods~\cite{gluza2024doublebracket,robbiati2024double,suzuki2025double,gluza2026double, wright2026double}, our circuits have a recursive structure which means that the circuit depth grows exponentially with the number of training iterations, with the base of the exponent increasing with the size of the training dataset. This could pose a potential limitation if good prediction accuracy cannot be achieved with a small number of training iterations and limited training data. 

Encouragingly, when we consider canonical examples of unitary learning with hidden symmetry~\cite{huang2022learning,caro2022outofdistribution,jerbi2023power,favre2026universal} and quantum phase classification~\cite{monaco2022quantum,caro2021generalization} (tasks that can be classically hard in general~\cite{schuster2025hardness} and that naturally fit within the autonomous learning paradigm), our results suggest that effective learning can still be achieved at low, constant circuit depths.
One reason for this is that if the training data and the underlying learning problem possess hidden symmetry, the trained model automatically becomes aware of this symmetry without any additional information being provided. This constitutes another key advantage of our approach, enabling it to exploit symmetries present in the training data even when they are not known \textit{a priori}.

\section{Framework}\label{sec:framework}
Consider supervised learning with (repeated) access to a training dataset of size $N$
\begin{align}\label{eq:training-data}
    \SC_{\rm train} = \{ \rho_i, y_i \}_{i=1}^N \;\;,
\end{align}
where $\rho_i$ are input $n$-qubit quantum states with $\rho_i \in \mathbb{S}(\HC_{\rm in})$~\footnote{The set $\mathbb{S}(\HC_{\rm in})$ contains all density operators on the $n$-qubit input Hilbert space $\HC_{\rm in}$.}, and $y_i \in \YC$ are the corresponding labels. 
Here, each $\rho_i$ may arise from a data-embedding procedure that maps classical data to quantum states, or from intrinsically a physical quantum process. The latter case is commonly referred to as quantum data~\cite{cerezo2022challenges}, and it includes the two canonical tasks considered in Section~\ref{sec:canonical-example}. 
Furthermore, the outputs $\{y_i\}_i$ can take different forms, including scalar-valued quantities (as in standard classification or regression tasks), vector-valued quantities, and even quantum states themselves.


These quantum states are then processed by quantum computers according to a chosen model to match the associated labels. The loss function is used to quantify the average discrepancy between model predictions and training labels. 
For concreteness, we consider the loss function of the form
\begin{align}\label{eq:qml-loss}
    \LC(U)
    =
    \frac{1}{N}\sum_{i=1}^N
    f\!\left(\ell_i(U),y_i\right) \;\;,
\end{align}
where the mismatch function $f(\cdot\,,\cdot)$ measures the discrepancy between the model output and the label of each sample. The  $\ell_i$ terms are expectation values of the form
\begin{align}
   \ell_i(U)
    = \Tr\!\left[U\rho_iU^\dagger O_i\right] \;\;,
\end{align}
where $U$ is a unitary operator and $O_i$ is an observable that may depend on the data index. This form of loss covers a wide range of learning tasks, including classification~\cite{caro2021generalization}, regression~\cite{huang2022learning,caro2022outofdistribution}, and, perhaps less obviously, generative modeling~\footnote{Generative modeling can also be formulated within this framework by omitting the labels $y_i$, where $\rho_i$ may represent computational basis states sampled from the target data distribution.}~\cite{rudolph2023trainability}.


The model is then trained by minimizing the loss to find an optimal unitary $U_*$. More particularly, a training procedure $\TC$ updates the parameters from an initial value $U_0$ to $U_*$ by minimizing the loss. That is, we have
\begin{equation}
\mathcal{T} \;:\; U_0,\, \SC_{\mathrm{train}}  \;\longrightarrow\; U_*
\quad {\rm s.t.} \quad
\LC(U_*) \le \LC(U_0) \;\; .
\end{equation}
We do not require $\TC$ to reach a global minimum. A successful training procedure only needs to reduce the loss significantly enough such that the trained model performs well, and $\TC$ itself may be stochastic.

The majority of the existing training proposals are based on extracting classical information from quantum states generated by the model through quantum measurements, which is a feature common to both variational~\cite{cerezo2020variationalreview,bharti2022noisy,benedetti2019parameterized,peruzzo2014variational,farhi2014quantum} and non-variational approaches~\cite{schuld2019quantum,havlivcek2019supervised,huang2021power,fujii2017harnessing,nakajima2019boosting,mujal2021opportunities,xiong2025role}. As an example of $\TC$, consider a common model family of parametrized quantum circuits, in which model parameters $\thv$ are embedded directly into the quantum evolution $U(\thv)$, and consider a regression task with the loss of the form $\LC(\thv) = \frac{1}{N}\sum_{i=1}^N \left( \Tr[U(\thv)\rho_iU^\dagger(\thv)O] - y_i \right)^2$ with some fixed observable $O$. Minimizing this loss via gradient-based methods requires estimating expectation values, which in turn necessitates extracting classical information from the quantum states $U(\thv)\rho_iU^\dagger(\thv)$. 
A similar mechanism also  arises in non-variational quantum models. Although the training procedure may be carried out classically, it still relies implicitly on estimating data-dependent quantities obtained from quantum measurements. For instance, in quantum kernel methods~\cite{schuld2019quantum,havlivcek2019supervised,huang2021power}, one must compute the Gram matrix by evaluating pairwise overlaps $k_{ij} = \Tr[\rho_i \rho_j]$ for all training data pairs. 

The information-extraction step is fundamentally inefficient as it attempts to access properties of objects in exponentially large Hilbert spaces through limited measurement data. This bottleneck is arguably one of the root causes of scalability challenges in QML. A prominent manifestation is the phenomenon of barren plateaus (or more broadly exponential concentration) where relevant variable-dependent quantities, such as loss values, become exponentially concentrated with system size. Avoiding such effects typically requires restricting the dynamics to an effective low-dimensional subspace, which in turn introduces a tension between trainability and classical surrogatability~\cite{schreiber2023classical}. 

One alternative is to bypass the inefficiency of information extraction during training altogether. This motivates the concept of autonomous quantum machine learning, in which models undergo self-training entirely within the quantum domain, without intermediate measurements or classical feedback.
\begin{definition}[Autonomous self-training of a QML model]\label{def-extended:autonomus-qml}
A QML framework is \emph{autonomously self-training} if its training process $\mathcal{T}$ involves \textbf{no quantum measurements} including POVM, projective, weak, or post-selective measurements. 
\end{definition}
In other words, the mapping $U_0 \rightarrow U_*$ is realized without any quantum-to-classical information extraction throughout the training protocol. 

This autonomous QML definition is intentionally narrower than the broad colloquial use of ``autonomous'' in classical machine learning, where a system may adapt its decisions using feedback~\cite{hutter2019automated}.
It is closer to coherent quantum optimization~\cite{motlagh2024ground,catli2025exponentially,leng2023quantum} where the information about the loss is encoded directly into an underlying Hamiltonian and the minimum is obtained through the coherent quantum evolution without any measurements.
Related idea also appears in engineered dissipative approaches to quantum optimization~\cite{wanjura2025quantum}.
Most relevantly, despite not originally framing as a coherent training, double-bracket flow algorithms for Riemannian-gradient circuit synthesis can be seen as a powerful framework for coherently implementing iterative dynamics of Riemannian-gradient for optimization problems~\cite{gluza2024doublebracket,robbiati2024double,suzuki2025double,gluza2026double, wright2026double}. Our approach builds on this perspective, extending to the QML paradigm. 


Lastly, the ultimate goal of QML is to achieve good generalization. That is, after training, the model should perform well on unseen data drawn from the same distribution as the training data. To make this precise, we regard the training loss in Eq.~\eqref{eq:qml-loss} as a proxy for the true risk $\LC_\DC(U)$, defined as the average loss over the data distribution $\DC$. The generalization error quantifies the difference between the two,
\begin{align}
\label{eq:gen-error}
{\rm gen}(U)= \LC(U)-\LC_\DC(U)\;\;.
\end{align}
In general, the generalization error decreases as the number of training examples increases, since the empirical average becomes a better estimate of the true risk~\cite{banchi2021generalization,huang2021information,caro2021generalization,caro2021encodingdependent,gibbs2022dynamical}. For parametrized quantum circuits with $m$ trainable parameters, good generalization can be achieved with $N\in\OC(m\log m)$ training examples~\cite{caro2021generalization}, and good generalization may remain possible even when the training and test distributions differ~\cite{caro2022outofdistribution,favre2026universal}. For the autonomous models considered here, no explicit parametrization is involved, and the relevant notion of model complexity is instead set by the subspace that the training dynamics explore. We return to this point in Section~\ref{sec:unitary-learning}, where symmetry inheritance leads to a generalization bound for unitary learning.

\section{Description of Autonomous Training Algorithm}

In the previous section, we motivate the general notion of autonomous quantum machine learning. We now develop a general training framework based on Riemannian gradient optimization. Although the framework is not autonomous in its most general form, we will identify a sufficient condition under which it is fully autonomous.

\subsection{Background on Riemannian Optimization}
\label{sec:bg-rg-opt}

Let us first introduce the Riemannian gradient for the loss function $\LC$ in Eq.~\eqref{eq:qml-loss} from an operational perspective. We regard a unitary $U$ as a point on the unitary manifold and seek the infinitesimal unitary transformation that decreases the loss most rapidly. Since the tangent space of the unitary group consists of anti-Hermitian operators, 
the unitary updated by a small displacement $s$ generated by an operator $A$ takes the form
\begin{align}
    U_A(s) = e^{sA} U \;\;,
    \qquad A^\dagger= -A \;\;.
\end{align}

The Riemannian gradient $G$ can therefore be understood as the tangent-space direction that locally decreases the loss most rapidly. More explicitly, the steepest-descent direction is determined by
\begin{align}
    \frac{G}{\|G\|_2}
    =
    -\argmin_{\substack{A^\dagger=-A\\\|A\|_2=1}}
    \left.\left(
    \frac{\partial \LC(e^{sA}U)}{\partial s}\right)
    \right|_{s=0},
    \label{eq derivative of L}
\end{align}
where $\|\cdot\|_2$ denotes the Schatten $2$-norm.

To identify $G$, we exploit the fact that the optimization is performed over the tangent space of the unitary group. This allows us to characterize the variation of the loss along every tangent direction generated by an arbitrary anti-Hermitian operator $A$. The Riemannian gradient $G$ is then defined through
\begin{align}
    \left.
    \left(\frac{\partial \LC(e^{sA}U)}{\partial s}
    \right)\right|_{s=0}
    =
    \langle A,G\rangle \;\;,
    \label{eq variational G}
\end{align}
where $\langle\cdot,\cdot\rangle$ denotes the Hilbert--Schmidt inner product.

By maximizing the Hilbert--Schmidt inner product in Eq.~\eqref{eq variational G}, the Riemannian gradient of the loss in Eq.~\eqref{eq:qml-loss} given some $U$ is of the form
\begin{align}\label{eq:RG-qml-explicit-form}
    G = -\frac{1}{N}\sum_{i=1}^N [U\rho_iU\ad,O_i] \,\left.\left(\frac{\partial f(\ell_i)}{\partial \ell_i} \right)\right|_{\ell_i=\ell_i(U)} \;\;.
\end{align}
The derivation is provided in the Appendix~\ref{appx:general-rg-derive} which follows  Refs.~\cite{Helmke1995Optimization,absil2008optimization,schulte2010gradient,wiersema2023optimizing}. 
Note that for $N=1$ and $f(\ell_1,y_1)=\ell_1$, the loss is a single expectation value $\Tr[U\rho_1U^\dagger O_1]$ and Eq.~\eqref{eq:RG-qml-explicit-form} reduces to $G=-[U\rho_1U^\dagger,O_1]$, which is the Riemannian gradient used for variational optimization over the unitary group in Ref.~\cite{wiersema2023optimizing}.

\subsection{Riemannian Optimization for Autonomous Quantum Machine Learning}
\label{sec:RG}

Having derived the Riemannian gradient, we now describe how it enables a training process $\TC$. The key idea is to iteratively update the model by following the gradient in Eq.~\eqref{eq:RG-qml-explicit-form}. This constructs the trained unitary through successive gradient steps.

More precisely, let $U_0$ denote an initial unitary, which may be obtained from prior knowledge of the problem, a classical optimization procedure (e.g., by training a parametrized quantum circuit), or simply chosen as the identity operator, $U_0=\mathbb{1}$. We first illustrate the training procedure explicitly for the initial iterations. At the first iteration, the Riemannian gradient at $U_0$ is given by
\begin{align}
    G_0 = -\frac{1}{N}\sum_{i=1}^N\left[\rho_i^{(0)},O_i\right]
    \left.
    \left(\frac{\partial f(\ell_i)}{\partial \ell_i}\right)
    \right|_{\ell_i=\ell_i(U_0)},
\end{align}
where
\begin{align}
    \rho_i^{(0)} = U_0\rho_iU_0^\dagger.
\end{align}
The unitary is then updated according to
\begin{align}
    U_1 = e^{-s_0G_0}U_0.
\end{align}
At the next iteration, the gradient at the updated unitary $U_1$ is given by
\begin{align}
    G_1 = -\frac{1}{N}\sum_{i=1}^N\left[\rho_i^{(1)},O_i\right]
    \left.
    \left(\frac{\partial f(\ell_i)}{\partial \ell_i}\right)
    \right|_{\ell_i=\ell_i(U_1)},
\end{align}
where $\rho_i^{(1)}=U_1\rho_iU_1^\dagger$, followed by the update
\begin{align}
    U_2=e^{-s_1G_1}U_1.
\end{align}

Continuing this procedure recursively, the trained unitary at iteration $k+1$ is obtained as
\begin{align}
    \label{eq:update_rule}
    U_{k+1}=e^{-s_kG_k}U_k,
\end{align}
where $s_k$ is the step size and $G_k$ is the Riemannian gradient at the current iterate,
\begin{align}
\label{eq:generic_RG_grad}
    G_k = -\frac{1}{N}\sum_{i=1}^N\left[\rho_i^{(k)},O_i\right]
    \left.
    \left(\frac{\partial f(\ell_i)}{\partial \ell_i}\right)
    \right|_{\ell_i=\ell_i(U_k)},
\end{align}
with the evolved states
\begin{align}
    \rho_i^{(k)}=U_k\rho_iU_k^\dagger\;\;.
\end{align}
where one can see that the gradient depends on the evolved training states $\{\rho_i^{(k)}\}_{i=1}^N$, which themselves depend recursively on all previous updates through $U_k$.

\hypertarget{sent:ii}{\paragraph*{Autonomous self-training condition.}}\label{sent:(ii)} 
Crucially, the Riemannian gradient $G_k$ and hence the trained unitary $U_{k+1}$ in Eq.~\eqref{eq:update_rule} depends on the coefficients $\{\left(\partial f/\partial \ell_i \right)_{\ell_i = \ell_i(U_k)}\}_{i=1}^N$. 
Evaluating these coefficients typically requires estimating the expectation values $\{\ell_i(U_k)\}_{i=1}^N$ through quantum measurements, thereby reintroducing a quantum-to-classical feedback loop at every training iteration. Eliminating this information-extraction step is therefore the central challenge in realizing fully autonomous quantum training.

We now identify a simple sufficient condition under which the entire training procedure becomes autonomous self-training. Specifically, if the mismatch function $f$ is linear in the expectation values $\{\ell_i(U)\}_{i=1}^N$, that is we have
\begin{align} \label{eq:linear-mismatch}
    f( \ell_i(U), y_i) = c_i \ell_i(U) + b_i \; \;,
\end{align}
where $c_i$ and $b_i$ are some real values independent of $ \ell_i(U)$ but may depend on $y_i$, then $\frac{\partial f}{\partial \ell_i}=c_i$ where $c_i$ are predetermined constants independent of the intermediate states. In this case, the Riemannian gradient no longer depends on quantities that must be estimated during training. Consequently, Eq.~\eqref{eq:update_rule} can be recursively applied to determine the complete quantum circuit corresponding to any desired training iteration, without performing intermediate quantum measurements. Training therefore reduces to executing the predetermined quantum circuit, i.e., \emph{simply running the circuit}. Prominent examples satisfying this condition include quantum phase classification~\cite{caro2021generalization} with an explicit form of the loss shown in Eq.~\eqref{eq:phase-loss-fn}, as well as unitary learning~\cite{huang2022learning,caro2022outofdistribution} with the loss shown in Eq.~\eqref{eq:unitary-loss-fn}. We will further investigate these two canonical examples later in details in Section~\ref{sec:canonical-example}.

\subsection{Circuit synthesis}
\label{sec:algo}

The recursive update rule in Eq.~\eqref{eq:update_rule} provides a constructive prescription for the optimized unitary after any desired number of training iterations. Realizing this update, however, requires implementing the unitary operators $e^{-s_kG_k}$, which are generally not available in native gate sets and therefore must be compiled into executable quantum circuits. 

We now describe how to compile the update rule in Eq.~\eqref{eq:update_rule} into a quantum circuit using oracle access to a set of unitaries~\footnote{We will later justify the construction of these oracles.}
\begin{align}\label{eq:oracle}
    \mathbb{U}_t = \{e^{-i t O_i },\,e^{-i t\rho_i} \}_{i=1}^N
\end{align}
for some evolution time $t$, together with the initial unitary $U_0$ and its inverse $U^\dagger_0$. 

Since $G_k$ in Eq.~\eqref{eq:generic_RG_grad} is a sum of commutators, the first step begins by approximating the update unitary using a product-formula (Lie--Trotter) decomposition,
\begin{equation}\label{eq:trotter-apprx}
    e^{\sum_{i=1}^N s^{(k)}_{i}A^{(k)}_{i}} = \left(\prod_{i=1}^Ne^{s^{(k)}_iA^{(k)}_i/r}\right)^r + 
    \OC(s_k^2/r)\;\;,
\end{equation}
where $s^{(k)}_i=\frac{s_k}{N}\left|c_i\k\right|$ and $A^{(k)}_i=\operatorname{sign}(c_i\k)[\rho^{(k)}_i,O_i]$ with $c_i\k = \left(\partial f/\partial \ell_i \right)_{\ell_i = \ell_i(U_k)}$. 
This approximation decomposes the evolution generated by the full Riemannian gradient into a sequence of evolutions generated by the individual commutators $[\rho^{(k)}_i,O_i]$, reducing the compilation problem to implementing each elementary factor $e^{s^{(k)}_iA^{(k)}_i/r}$.

To implement the remaining exponential of each commutator, we follow the framework of double-bracket quantum algorithms ~\cite{suzuki2025grover,gluza2024doublebracket,gluza2026double,wright2026double} and employ an order-$m$ group-commutator product formula
\begin{align}
\label{eq:group_commutator_apprx}
    e^{t^2A^{(k)}_i} ={}&e^{i\a_1t\operatorname{sign}(c_i\k)O_i}\,e^{i\a_2t\rho^{(k)}_i}\cdots\nonumber\\
    &\times e^{i\a_{N_m-1}t\operatorname{sign}(c_i\k)O_i}\,e^{i\a_{N_m}t\rho^{(k)}_i}+\OC(t^{m+1})\;\;,
\end{align}
where the coefficients $\{\a_j\}_{j=1}^{N_m}$ and the number of exponentials $N_m$ depend on the chosen product formula. For a negative $c_i\k$, the evolutions under $O_i$ are simply reversed. Hence, implementing the Riemannian update reduces to realizing only two elementary classes of unitaries: $e^{-itO_i}$ and $e^{-it\rho^{(k)}_i}$. 

The above construction implements a single Riemannian gradient update, namely the transition from the $k$-th to the $(k+1)$-th iteration, in terms of the evolved training states $\{\rho_i^{(k)}\}_{i=1}^N$. To express the resulting circuit entirely in terms of the elementary primitives $e^{-it\rho_i}$, $e^{-itO_i}$, and the initial unitary $U_0$, one recursively expands each evolved state $\rho_i^{(k)}=U_k\rho_iU_k^\dagger$ using the implementation of $U_k$. Repeating this procedure recursively until reaching the initial iteration $k=0$ eliminates all intermediate quantities.

\subsection{Summary of algorithm}

Here we summarize the step-by-step procedure of the autonomous QML algorithm for a given learning task with training dataset $\SC_{\rm train}=\{\rho_i,y_i\}_{i=1}^N$ in Eq.~\eqref{eq:training-data}.
\begin{enumerate}
    \item \textbf{Check the autonomous self-training condition.} Check whether the loss in Eq.~\eqref{eq:qml-loss} can be casted in a linear form with the mismatch $f$ as in Eq.~\eqref{eq:linear-mismatch}. If so, one can run the Riemannian gradient algorithm in a fully autonomous self-training manner.
    \item \textbf{Choose hyperparameters of the model.} Choose a prior unitary $U_0$ and the total number of training iterations $k_{\rm tot}$ and the  associated step sizes for each training iterations $\{s_k\}_{k=0}^{k_{\rm tot}-1}$. These step sizes should be sufficiently small such that the training can be guaranteed (see Theorem~\ref{thm:auto_RG_learning_guarantee}). 
    \item \textbf{Construct the Riemannian gradient update.} To generate a circuit corresponding to the training of $k_{\rm tot}$ iterations, one first considers the recursive update rule for the $k_{\rm tot}$ iteration in Eq.~\eqref{eq:update_rule} and write down $U_{k_{\rm tot}}$ as a function of $U_{k_{\rm tot} - 1}$ and $\{\rho_i^{(k_{\rm tot}-1)}=U_{k_{\rm tot}-1}\rho_iU_{k_{\rm tot}-1}^\dagger\}_i$ and $\{O_i\}_i$.
    \item \textbf{Compile the update into quantum operations.}
    The exponential of the Riemannian gradient 
    $e^{-s_{k_{\rm tot}-1}G_{k_{\rm tot}-1}}=e^{\frac{s_{k_{\rm tot}-1}}{N}\sum_{i=1}^N c_i[\rho_i^{(k_{\rm tot}-1)},O_i]}$
    is first decomposed into a product of exponentials of the individual commutators using a product-formula approximation in Eq.~\eqref{eq:trotter-apprx}. Each commutator exponential is then compiled using a group-commutator construction in Eq.~\eqref{eq:group_commutator_apprx}, reducing the implementation to elementary unitaries of the form $e^{-itO_i}$ and $e^{-it\rho_i^{(k_{\rm tot}-1)}}$.
    \item \textbf{Recursively unroll the circuit till the end.}
    Since $\rho_i^{(k_{\rm tot}-1)} = U_{k_{\rm tot}-1}\rho_iU_{k_{\rm tot}-1}^\dagger$, one recursively applies Eq.~\eqref{eq:update_rule}, further expressing $U_{k_{\rm tot}-1}$ in terms of $U_{k_{\rm tot}-2}$ and $\{\rho_i^{(k_{\rm tot}-2)},O_i\}_i$. This is then followed by repeating Step 4. above to have the circuit architecture $U_k$ in terms of $e^{-itO_i}$ and $e^{-it\rho_i^{(k_{\rm tot}-2)} }$. Furthermore, recursively unrolling this construction expresses the entire trained unitary in terms of the initial unitary $U_0$ and the elementary oracles $e^{-itO_i}$ and $e^{-it\rho_i}$.
    \item \textbf{Run the resulting quantum circuit.}
    Once the trained circuit structure is obtained, the resulting circuit can be executed without intermediate quantum measurements or classical feedback. The final unitary $U_k$ therefore realizes the trained QML model autonomously.
\end{enumerate}

We end this section by formalizing the result in the following theorem with the proof detailed in Appendix~\ref{appx:suff-for-autonomous}.
\begin{theorem}[Autonomous QML with the Riemannian gradient descent]\label{thm:sufficient-cond-autonomous-qml}
Consider the loss function in Eq.~\eqref{eq:qml-loss} together with the update rule in Eq.~\eqref{eq:update_rule} and the Riemannian gradient in Eq.~\eqref{eq:generic_RG_grad}. If, for every training iteration,
\begin{align}
    \left.\left(
    \frac{\partial f(\ell_i)}{\partial \ell_i}
    \right)\right|_{\ell_i=\ell_i(U_k)} = c_i \;\;,
\end{align}
where $c_i$ are predetermined constants independent of $U_k$, then given a repeated access to the oracle $\mathbb{U}_t$ in Eq.~\eqref{eq:oracle} and to $U_0$ and $U_0\ad$, there exists an autonomous self-training protocol (satisfying Definition~\ref{def-extended:autonomus-qml}) that follows the Riemannian gradient descent trajectory.
\end{theorem}

\section{Properties and Scaling of Algorithm}

Having introduced the algorithm in the previous section, we now discuss its fundamental properties, including guarantees on training-loss decrease, circuit complexity, and the ability of the trained model to recognize hidden symmetries in the learning task.

\subsection{Convergence guarantees}
\textit{General Riemannian model.} Since the proposed training method belongs to the broader family of gradient-based optimization methods, one naturally expects the loss to decrease for sufficiently small step sizes, similar to standard gradient descent over Euclidean parameter spaces. 
As derived in Appendix~\ref{appx:training_loss_guarantee} for a general Riemannian gradient algorithm, we establish a monotonic descent guarantee for the loss function under the iterative update with step sizes $s_0,s_1,\ldots,s_k$.
Specifically, at the $k$-th iteration, let $M_k = \sup_{\zeta \in [0, s_k]}|\partial_s^2 \LC(e^{-s \rg_k} U_k)|_{s=\zeta}$ denote the maximum curvature of the loss function along the Riemannian gradient direction over the interval $[0,s_k]$. If the step size satisfies $s_k \leq \frac{2\|\rg_k\|_2^2}{M_k}$ where $\|\cdot\|_2$ is the Schatten 2-norm, then the loss decreases monotonically 
\begin{equation}
    \LC(U_{k+1}) - \LC(U_k) \leq 0 \;\;.
\end{equation}

\textit{The autonomous self-training framework} admits a stronger training loss decrease guarantee than the general result established previously. While general Riemannian gradient descent only ensures monotonic descent, the autonomous setting enables us to establish an exponential convergence rate (in $k$). Provided that the step size $s_k$ is chosen sufficiently small, the training loss decreases exponentially fast with the number of training iterations under a single additional condition adopted from the classical machine learning literature.
In particular, we assume that the training run satisfies a Polyak--$\text{\L}$ojasiewicz (PL) inequality~\cite{karimi2020linearconvergencegradientproximalgradient,absil2005convergence} relative to its own limiting loss,
\begin{equation}
\label{eq:PL-informal}
    \tfrac12\,\|G_k\|_2^2\ \geq\ \mu\,\big(\LC(U_k)-\LC_{\infty}\big)\qquad\text{for all }k\geq0\;\;,
\end{equation}
for some constant $\mu>0$, where $\LC_{\infty}:=\lim_{k\rightarrow\infty} \LC(U_k)$ is the loss reached by the run itself. Intuitively, the optimization process continues to feel a proportional ``downward pull'' as long as the loss remains above its eventual limiting loss. This condition
keeps the gradient large enough that each step removes at
least a fixed positive fraction of the remaining loss, and guarantees a rapid convergence. 

Three remarks clarify the status of this assumption. First, Eq.~\eqref{eq:PL-informal} is a condition on the specific training trajectory $\{U_k\}_{k}$ rather than a global property of the loss landscape, and the constant $\mu$ may depend on the initial unitary $U_0$. Second, the monotonic descent established above is proven without it. The PL inequality therefore only enters in upgrading descent to an exponential rate, so the argument is not circular: the inequality is not used to show that the loss decreases but to show how fast it decreases. Third, the inequality can be established analytically in some cases. For example, for ground-state preparation with a gapped Hamiltonian, we prove in Appendix~\ref{subsec:PLvalid-pl} that it holds with $\mu=\Delta/2$ once the energy is within $\Delta/2$ of the ground-state energy, where $\Delta$ is the energy gap. Its validity for the two learning tasks studied in this work is verified numerically, as discussed below.
\begin{theorem}[Exponential convergence in autonomous QML, Informal]\label{thm:auto_RG_learning_guarantee}
Under the same assumption as in Theorem~\ref{thm:sufficient-cond-autonomous-qml} and further suppose that the autonomous training loss satisfies the PL condition in Eq.~\eqref{eq:PL-informal} with some $\mu$. If the step sizes for all iterations are below some threshold as $s_k \in \OC\left(1/\xi \right)$ with $\xi = \frac{1}{N}\sum_{i} |c_i|\, \|O_i\|_{\infty}$, then the loss decreases monotonically at every step, converging exponentially fast to $\LC_{\infty}$. Concretely, for any training iteration $k$, we have
\begin{equation}
\label{eq:informal-rate}
    \LC(U_k)\ \leq\ \LC_{\infty}+c\,q^{\,k}
    \;\;,
\end{equation}
where $q =  1 - \Theta(\mu\min_k(s_k)) \in(0,1)$ and $c =\LC(U_0) - \LC_{\infty}$. 
\end{theorem}
Theorem~\ref{thm:auto_RG_learning_guarantee} provides an explicit upper bound on the step size required to guarantee descent at every training iteration. It further establishes that, once the PL inequality holds, the loss decays exponentially toward its asymptotic floor $\LC_{\infty}$, with the decay rate determined by $\mu$ and the minimum step size.

A key subtle point is that the PL condition can always be satisfied by choosing $\mu$ arbitrarily small, but this does not necessarily yield a meaningful convergence rate. In particular, if $\mu$ is too small, the resulting decay factor $q$ approaches $1$, rendering the convergence guarantee uninformative. In Appendix~\ref{subsec:numerical-pl}, we numerically demonstrate that, for both learning examples considered in this work, $\mu$ does not vanish exponentially with system size and remains large as the number of training iterations increases (for a fixed system size).

Note that this guarantee ensures convergence toward a stationary point (or local minimum) of the loss landscape, but does not imply convergence to the global optimum.
One sufficient condition for reaching it is that the observables take the form $O_i=\alpha_i O+\beta_i\mathbb{1}$, where $\alpha_i$ and $\beta_i$ are index-dependent scalars and $O$ is some fixed observable, as shown in Appendix~\ref{subsec:global-QML}. In this case, the observables share a common structure across the training data. One example which satisfies this condition is a binary phase classification. With two labels $y_i\in\{0,1\}$ encoded in a single ancilla qubit, the class projectors are $P_0=\mathbb{1}\otimes\ketbraq{0}_a$ and $P_1=\mathbb{1}\otimes\ketbraq{1}_a=\mathbb{1}-P_0$, so every observable $O_i=P_{y_i}$ is an affine function of the single operator $O=P_0$. This is a special case of the phase classification task studied in Section~\ref{sec:canonical-example}, in which only two phases are to be distinguished. 

\subsection{Sample complexity}

This guarantee on exponential loss decrease in $k$ comes at the cost of resources exponential in $k$ to implement the circuit. Note that this is also the same for other double bracket flow papers. In particular, this recursive expansion comes at the cost of an increasing circuit depth. Since each update contains the full implementation of the previous iterate, the circuit depth grows multiplicatively with the number of training iterations. After fully unfolding the recursion, the depth scales as $\mathcal{O}(N^k)$ for a fixed number of Lie--Trotter steps $r$ and fixed order $m$. 

Theorem~\ref{thm:resource} formalizes the above intuition by quantifying the resource scaling of the compiled training circuit. The complete compilation protocol and the corresponding resource analysis are presented in Appendix~\ref{appx:resource}.
\begin{theorem}[Resource Scaling]
    \label{thm:resource}
    Consider the same setting as in Theorem~\ref{thm:sufficient-cond-autonomous-qml}, and let $s$ be an upper bound on the step sizes, $s_k\leq s$ for all $k$, with $\xi$ as defined in Theorem~\ref{thm:auto_RG_learning_guarantee}. Suppose the recursive unitary $U_k$ in Eq.~\eqref{eq:update_rule} is implemented by the Lie--Trotter formula in Eq.~\eqref{eq:trotter-apprx} and the order-$m$ commutator product formula in Eq.~\eqref{eq:group_commutator_apprx}. To implement $U_k$ with the total error $\ep$, the number of oracle calls $\Nora\k$ from $\mathbb{U}_t$ in Eq.~\eqref{eq:oracle} scales as
    \begin{equation}
        \Nora\k \leq \left[s\cdot B_m(\ep)\right]^k\;\;,
    \end{equation}
    where we have
    \begin{align}
        B_m(\ep) \in \OC\left(\max\left(\frac{N(4 {\xi} s+1)^k}{\ep},\left[\frac{(4 {\xi} s+1)^k}{\ep}\right]^\frac{2}{m-1}\right)\right)\,.
    \end{align}
\end{theorem}
We call the training protocol efficient if the number of oracle calls scales polynomially in the number of qubits $n$. Whether this holds depends on how the number of training iterations $k$ and the size of the training dataset $N$ scale with $n$. Intuitively, Theorem~\ref{thm:resource} contains two sources of exponential growth. The base $s B_m(\ep)$ is raised to the power $k$ because each iteration contains a full copy of the previous one, and $B_m(\ep)$ itself contains the factor $(4\xi s+1)^k$ because an error made at an early iteration is amplified by every subsequent one. For constant precision $\ep$, step size $s$, data constant $\xi$, and product-formula order $m\geq 3$ (and the per-step constant $C_{\max}^{(m)}$ of Appendix~\ref{appx:resource}), the number of oracle calls therefore scales as $[sN(4\xi s+1)^k/\ep]^{k}$ up to a factor $c^k$ with $c$ independent of $n$, $N$, $k$, and $\ep$, that is, exponentially in $k$ with an additional factor exponential in $k^2$. 

Consequently, if the number of training data remains constant, $N\in\OC(1)$, the number of training iterations can grow at most as $k\in\OC(\sqrt{\log n})$ while keeping the protocol efficient. If instead $N\in\OC(\poly(n))$, the number of training iterations must remain constant, $k\in\OC(1)$. The derivation of these scalings from Theorem~\ref{thm:resource} is given in Appendix~\ref{appx:resource}. How many training iterations are required for good training and generalization will in general be highly problem dependent and can likely only be determined  heuristically.

We now discuss how to implement the oracle unitaries $e^{-itO_i}$ and $e^{-it\rho_i}$. The former is typically straightforward, as the observables can be interpreted as Hamiltonians and simulated using standard Hamiltonian-simulation algorithms~\cite{lloyd1996universal,van2020circuit,childs2021theory} 
The latter requires implementing the exponential of a density operator. This can be achieved via a state-to-Hamiltonian transformation using density matrix exponentiation~\cite{lloyd2014quantum,kimmel2017hamiltonian,kjaergaard2022demonstration,wada2025state,marvian2016universal,son2025QDP,alghadeer2025double,Schwartzman2026ImaginaryTE,wei2024simulating,wilde1,wilde2,son2025quantum}, which is based on the identity
\begin{equation}
    e^{-is\rho_j}\rho_i e^{is\rho_j}=\Tr_B(e^{-is\Fbb_{AB}}(\rho_i\ot\rho_j)_{AB}e^{is\Fbb_{AB}}) + \OC(s^2)\;\;,
\end{equation}
where $\Fbb_{AB}$ denotes the SWAP operator acting on systems $A$ and $B$. Operationally, this realizes the oracle $e^{-it\rho_i}$ by coupling the quantum state in the data register to a copy of the training state prepared in an auxiliary register through a SWAP interaction.

\subsection{Symmetry inheritance}\label{sec:symmetry-main}
An appealing feature of the proposed framework is its ability to automatically inherit symmetries hidden in the training data and learning problem. Specifically, if the training data, observables, and initial model share a common symmetry, then the training dynamics preserve that symmetry throughout optimization. Consequently, the learned unitary is symmetry preserving without requiring any prior knowledge of the underlying symmetry or explicitly enforcing symmetry constraints during training. We state this explicitly in the theorem below together with the proof in Appendix~\ref{appx:autonomous_sym_inherence}.
\begin{theorem}[Symmetry Inheritance]
\label{thm:sym-inherence}
    Let $S$ be an operator, and suppose that the training data $\{\rho_i,O_i\}_{i=1}^N$ together with the initial unitary $U_0$ satisfy
    \begin{equation}
    \label{eq:sym-inherence-hypo}
        [S,\rho_i]=[S,O_i]=[S,U_0]=0,
    \end{equation}
    for all $i$. Then the iterates generated by the update rule in Eq.~\eqref{eq:update_rule}, under the conditions of Theorem~\ref{thm:sufficient-cond-autonomous-qml}, satisfy
    \begin{equation}
        [S,U_k]=0,
    \end{equation}
    for every training iteration $k$.
\end{theorem}

The theorem shows that the gradient flow remains within the symmetry sector throughout the optimization process. If the dimension of this sector is sufficiently small, the confinement translates into good generalization because the hypothesis class of the trained model is reduced from the full unitary group to a smaller effective subspace, requiring far fewer samples to narrow the gap between training and test performance. Later in Section~\ref{sec:unitary-learning}, we make this precise for the unitary learning, where symmetry inheritance plays a central role and further provide analytical results showing a generalization error bound.



\subsection{Potential refinements}\label{sec:potential-refinement}

Although the resource analysis indicates an exponential growth in circuit depth with the number of training iterations, this worst-case scaling should not be interpreted as a practical limitation of every implementation. Several techniques developed in both classical optimization and quantum compilation can substantially reduce the required resources while preserving good empirical performance.

One natural approach is to replace the full Riemannian gradient with a stochastic approximation. Instead of evaluating the entire sum over all training data at every iteration, one may estimate the gradient using only a randomly selected mini-batch of training examples, analogous to stochastic gradient descent in classical machine learning. In our framework, this corresponds to implementing only a subset of the commutator terms appearing in Eq.~\eqref{eq:generic_RG_grad} during each update. As a result, the number of elementary unitaries required for a single training iteration is reduced from scaling with the full dataset size $N$ to the mini-batch size $M_B\ll N$. Consequently, the exponential base governing the compiled circuit depth becomes $M_B$ rather than $N$, leading to a substantial reduction in the implementation cost. Moreover, under standard assumptions on the stochastic gradient estimator, convergence guarantees analogous to those of stochastic gradient methods can be established.

Beyond modifying the optimization procedure itself, significant reductions in circuit depth may also be achieved through quantum circuit simplification and recompilation. Modern compilation frameworks employ a variety of algebraic simplifications~\cite{duncan2020graph}, gate cancellations~\cite{nam2018automated}, commutation analysis~\cite{iten2022exact}, and hardware-aware optimizations to compress quantum circuits while preserving their functionality~\cite{maslov2017basic,nash2020quantum,tan2020optimal,schmid2024computational}. Since the recursive construction of our trained unitary naturally generates repeated circuit motifs, these compilation techniques are expected to eliminate a considerable amount of redundancy in practice, potentially yielding circuits that are substantially shallower than the worst-case theoretical construction.

\section{Canonical examples of learning tasks}
\label{sec:canonical-example}

With the general framework in place, we now demonstrate its utility through two concrete learning tasks. Moreover, although various techniques can be employed to reduce the circuit depth, it is encouraging to observe that effective training is possible even without relying on such heuristic approaches. To illustrate both the practical operation of our framework and its performance, we consider (i)~quantum phase classification and (ii)~learning an unknown quantum unitary.


\begin{figure*}[!ht]
    \centering
    \includegraphics[width=\linewidth]{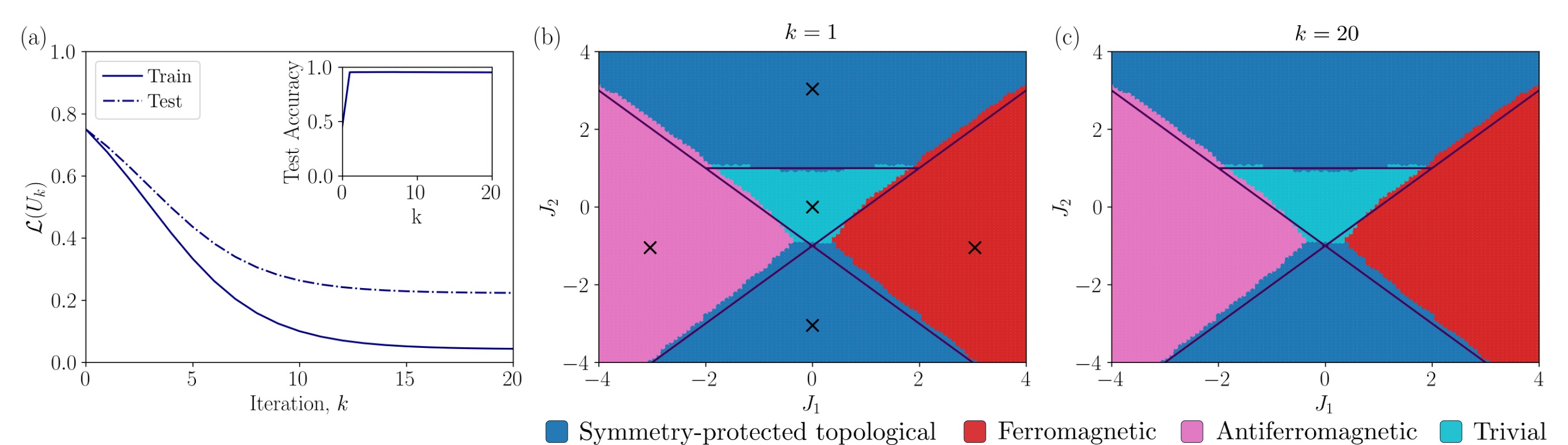}
    \vspace{-0.5 cm}
    \caption{\textbf{Performance of phase classification for the generalized cluster model} (a) The training (solid) and testing (dashed) loss functions are plotted against the number of training iterations. In the inset of (a), the test accuracy is plotted against the number of training iterations. (b-c) The predicted phases in $J_1-J_2$ plane after $1$ and $20$ iterations are shown for all testing points, respectively. Each color represents a different phase (blue, red, pink, and cyan are symmetry-protected topological, ferromagnetic, antiferromagnetic,    and trivial phases respectively) with black crosses $(\times)$ indicating the sparse $5$ training samples.} 
    \label{fig:p1-classif-result}
\end{figure*}

\subsection{Quantum Phase Classification}

\subsubsection{Set-up}
Consider a Hamiltonian $H(\alv)$ that exhibits $Q$ distinct phases depending on the parameter $\alv$. The training data consist of $N$ pairs $\{\rho_{{\rm s},i},y_i\}_{i=1}^N$, where $\rho_{{\rm s},i}$ is the ground state of $H(\alv_i)$ for a particular value $\alv_i$ and $y_i\in[Q]$ is its phase label. The task is to train a model to predict the phase of an unseen ground state. To cast this task in the form of Eq.~\eqref{eq:qml-loss}, we encode each training state as a composite system--ancilla state 
\begin{equation}
\rho_i=\rho_{\rm s,i}\ot\ketbraq{\phi}_a\;\;,
\end{equation}
where $|\phi\rangle_a$ is the initial ancilla state, chosen as an equal superposition of the $Q$ computational basis states with each individual corresponding to the phase label
\begin{equation}
|\phi\rangle_a = \frac{1}{\sqrt{Q}}\sum_{y=1}^Q|y\rangle_a\;\;.
\end{equation}
The choice of a uniform superposition is not arbitrary. If the ancilla were instead initialized in a label basis state $\ket{y}_a$, every encoded state $\rho_i$ would commute with every class projector $P_{y_i}$ defined below, so the Riemannian gradient in Eq.~\eqref{eq:generic_RG_grad} would vanish at $U_0=\mathbb{1}$ and training could not start. The uniform superposition gives the ancilla coherence across all labels and assigns equal prior weight to each class, so that the initial gradient is nonzero and no class is favored before training.
The loss function can then be chosen to be of the form
\begin{equation}\label{eq:phase-loss-fn}
    \LC(U) = 1-\frac{1}{N}\sum_{i=1}^N \Tr(P_{y_i}U\rho_iU\ad)\;\;,
\end{equation}
where 
\begin{equation}
    P_{y_i}=\mathbb{1}\ot\ketbraq{y_i}_a\;\;
\end{equation}
denotes the class-dependent projector onto the subspace associated with the label $y_i$.
The quantity $\Tr(P_{y_i}U\rho_iU\ad)$ is the probability that the model predicts the correct label $y_i$ for the input $\rho_i$. Thus, training the model amounts to learning a unitary $U$ that maps the initial ancilla state $|\phi\rangle_a$ to the basis state corresponding to the correct phase label. Since this loss is a linear-loss instance of Eq.~\eqref{eq:qml-loss}, the training process can be implemented autonomously according to Theorem~\ref{thm:sufficient-cond-autonomous-qml}.


After obtaining the optimal unitary $U_*$, we predict the phase of an unseen ground state $\rho_s$ by introducing the corresponding system–ancilla state $\rho=\rho_s\ot\ketbraq{\phi}_a$ and selecting the label $\hat{y}$ with the largest output probability, namely,
\begin{equation}
\label{eq:predict_rule}
    \hat{y}= \argmax_{y\in[Q]}  \Tr(P_{y}U_*\rho U_*\ad) 
\end{equation}
as a class prediction on $\rho$. 

\subsubsection{Results}
\label{sec:numerics_classif}
As a proof of concept, we apply the framework to two physical models: the 1D generalized cluster model, treated here, and the axial next-nearest-neighbor Ising model, treated in Appendix~\ref{appx:extended_numerics}. The generalized cluster model with the periodic boundary condition has the Hamiltonian
\begin{equation}
    H_{\rm GC} = \sum_{i=1}^{n'} (Z_i - J_1 X_i X_{i+1} - J_2 X_{i-1}Z_i X_{i+1})\;\;,
\end{equation}
which exhibits four phases of matter ($Q=4$) parameterized by $J_1$ and $J_2$~\cite{verresen2017one}. Each input quantum state corresponds to a many-body ground state at a specific point $(J_1,J_2)$ in the parameter space. 

In particular, we initialize the ancilla state in the uniform superposition state of $2$-bit strings, representing one of $4$ distinct classes, $ |\phi\rangle=\frac{1}{2}\sum_{q=1}^4\ket{q}$. This ensures unbiased initial state across the entire label subspace. We simulate an $n'=8$ qubit system (with the total of $n=10$ qubits to run the algorithm) using a step size of $s=1$. All results are obtained from exact state-vector simulations, so no shot noise is included. The algorithm is provided with an extremely sparse training dataset consisting of only five data points, indicated by the black crosses $(\times)$ in Fig.~\ref{fig:p1-classif-result}(b).

Fig.~\ref{fig:p1-classif-result}(a) shows monotonic convergence of the training loss together with a decrease in the testing loss, reaching final values of $0.04$ and $0.22$, respectively, after $k=20$ training iterations. Despite this severe data sparsity, the algorithm achieves a test accuracy of $0.95$ after only $k=1$ training iterations across a dense grid of $10,\!201$ unseen test samples, as shown in the inset of Fig.~\ref{fig:p1-classif-result}(a). The phase diagram is accurately reconstructed for both \(k=1\) and \(k=20\), yielding the same results as shown in Fig.~\ref{fig:p1-classif-result}(b,c). The high prediction accuracy despite the nonzero testing loss arises because correct classification requires only that the probability assigned to the correct phase be the largest among all classes. We note that the residual $5\%$ error of the prediction is concentrated around the phase boundaries. This comes from the fact that we take the phase boundaries from Ref.~\cite{verresen2017one} as the reference rather than constructing labels from finite-size order parameters, whose boundaries depend on the choice of order-parameter operators. Therefore, it is a limitation of the data rather than the optimizer.

Note that the underlying mechanism for distinguishing the phases can, in this case, be shown to rely solely on global comparisons of overlaps between different quantum states. From this perspective, our framework operates under the same underlying principle as global fidelity kernels, but in a fully autonomous manner. 

In addition, recognizing a phase of matter without access to training data can require quantum computational time exponential in the correlation range in the worst case, under standard cryptographic assumptions~\cite{schuster2025hardness}. On the other hand, with access to labeled training data, efficient recognition can exploit additional structure in the states. For example, when the phases can be distinguished using local information, a classical model trained on labeled classical-shadow data provably classifies them efficiently~\cite{huang2021provably}.

\subsection{Unitary Learning}\label{sec:unitary-learning}
\subsubsection{Set-up}
The goal of the unitary learning task is to learn an unknown unitary $V$~\cite{huang2022learning,caro2022outofdistribution,jerbi2023power,favre2026universal} through the $N$ available training data pairs of input and output states $\{\ket{\psi_i},V\ket{\psi_i}\}_{i=1}^N$ where $\ket{\psi_i}$ is drawn from some distribution such that after training $U_* |\psi\rangle \approx V|\psi\rangle$ for an unseen $|\psi\rangle$ drawn from the same distribution. The loss function in this case can be expressed as 
\begin{align}
\label{eq:unitary-loss-fn}
    \LC(U) =1 - \frac{1}{N}\sum_{i=1}^N 
    \Tr \left( U\rho_i U^\dagger \sigma_i\right)\;\;,
\end{align}
where $\rho_i=\ketbraq{\psi_i}$ and $\sigma_i=V\rho_iV\ad $.
 This loss function is linear in the expectation values of the observables $\sigma_i$. Therefore, the corresponding Riemannian gradient satisfies the autonomy condition in Theorem~\ref{thm:sufficient-cond-autonomous-qml}. 

As discussed in Section~\ref{sec:framework}, the training loss in Eq.~\eqref{eq:unitary-loss-fn} serves as a proxy for the true risk $\LC_\DC(U)$ over the input distribution $\DC$, and the generalization error ${\rm gen}(U)$ in Eq.~\eqref{eq:gen-error} quantifies the gap between the two.

\subsubsection{Results}

\begin{figure*}[!ht]
    \centering
    \includegraphics[width=\linewidth]{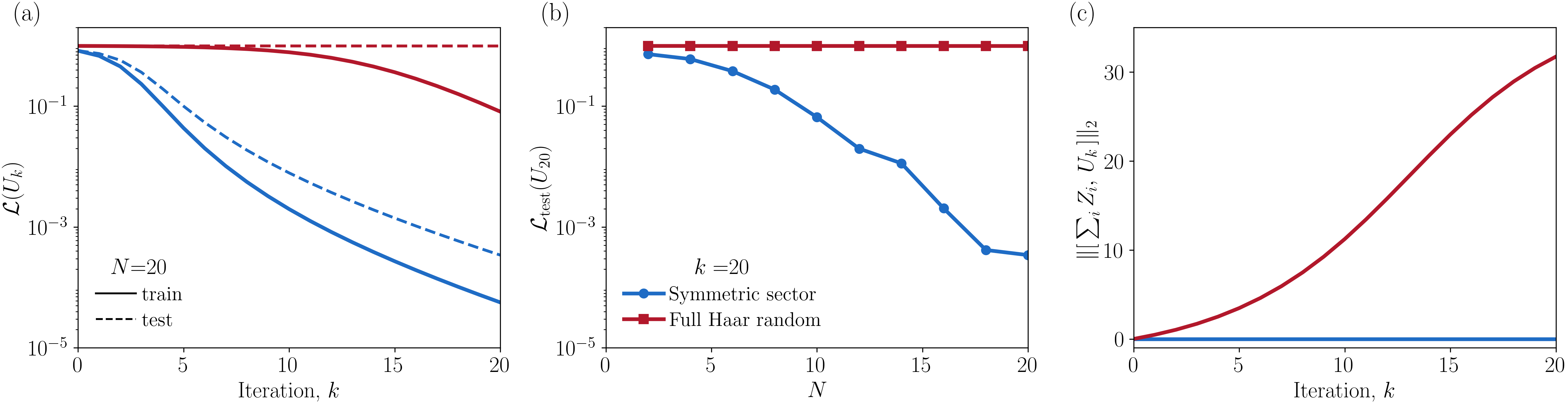}
    \vspace{-0.6cm}
    \caption{ \textbf{Performance of unitary learning} (a) The training (solid) and testing (dashed) loss functions using $N=20$ training samples are plotted against the number of training iterations. (b) The testing loss after $k=20$ training iterations is plotted against the number of training samples $N$. (c) The commutator norm $\|[\sum_iZ_i, U_k]\|_2$ between the symmetry generator and the trained unitary are plotted against the number of training iterations for verifying that the algorithm is aware of the problem's symmetry from the training data (Theorem~\ref{thm:sym-inherence}). In all panels, the label ``Symmetric sector'' and ``Full Haar random'' refer to the training dataset uniformly drawn from the invariant subspace of the excitation number equal to $1$ and the whole Hilbert space, respectively.}
    \label{fig:p2_unitary_learning_result}
\end{figure*}

We aim to leverage the model's inherent ability to detect hidden symmetries. Given only a training dataset, determining whether the underlying learning problem possesses symmetry is, in general, a difficult task. 
Even when a candidate symmetry group is specified, testing whether a state or a channel respects it is QSZK- and QMA-complete, respectively~\cite{laborde2021testing,rethinasamy2025quantum}. 
Symmetry-aware QML models therefore typically assume the symmetry group to be known in advance~\cite{larocca2022group,nguyen2022atheory,ragone2022representation,meyer2023exploiting}, while learning it from data is a task in its own right~\cite{lu2024learning,zhan2024learning}. In contrast, our approach automatically identifies and exploits such symmetries directly from the training data. 

As shown in Theorem~\ref{thm:sym-inherence}, the trained model remains well within the symmetry subspace. For the unitary learning task where training data states are drawn from such sector and target unitary preserve such symmetry, we can derive a generalization error bound, showing that the model complexity depends on the dimension of the effective symmetry subspace rather than that of the full Hilbert space. 
To formalize this relationship, we establish the following theorem 
\begin{theorem}[Generalization bound for symmetry unitary learning]
\label{thm:sample-complexity-u-learning}
    Consider the loss defined in Eq.~\eqref{eq:unitary-loss-fn} and the same setting of Theorem~\ref{thm:sym-inherence}, with $S=S^\dagger$ and $[S,V]=0$. Suppose the training states $\{\rho_i\}_{i=1}^N$ are drawn independently from a distribution $\DC$ supported on a single eigenspace of $S$ of dimension $\deff\geq 2$. Then, with probability at least $1-\delta$ over the draw of the training set, simultaneously for every iteration $k$, the generalization error in Eq.~\eqref{eq:gen-error} satisfies
    \begin{gather}
        \lvert{\rm gen}(U_k)\rvert \in \OC\left(\sqrt{\frac{\deff^2 \log(\deff N) +\log(1/\d)}{N}}\right)\;\;.
    \end{gather}

    In addition, if $\DC$ is a state $2$-design on that eigenspace, the true risk of every iterate $U_k$, and of any unitary $U$ with $[S,U]=0$, is given by
    \begin{equation}
        \LC_\DC(U)=1-\frac{|\Tr(\Pi_{\deff} U\ad V)|^2+\deff}{\deff(\deff+1)}\;\;,
    \end{equation}
    where $\Pi_{\deff}$ is the projector onto that eigenspace. 
\end{theorem}
Consequently, Theorem~\ref{thm:sample-complexity-u-learning} implies that the sample complexity required for this setting to generalize to any fixed accuracy is $N \in\OC(\poly(n))$, which is in agreement with our findings. We note that the bound holds for every iterate $U_k$ and does not require the training loss to vanish. That is, no perfect training is needed for the generalization guarantee.

Two features distinguish this bound from generic alternatives. First, the same argument without the symmetry assumption gives the bound with $\deff$ replaced by the full dimension $2^n$, which is nontrivial only when $N$ grows exponentially with $n$. Second, the bound does not depend on the iteration $k$, although the circuit depth grows exponentially with $k$ (Theorem~\ref{thm:resource}). By contrast, generalization bounds for parameterized circuits typically scale with the number of trainable gates~\cite{caro2021generalization}. 

The second part of the theorem represents the average-infidelity over Haar distribution restricted to that sector. In other words, the mismatching in that subspace $|\Tr(\Pi_{\deff} U\ad V)|$ and the dimension of the subspace directly controls the expected test loss. If the unitary faithfully resemble the target unitary in such subspace i.e., $\Pi_{\deff} U\ad V=\mathbb{1}_{\deff}$, then the expected loss vanishes. The proof of these results can be found in Appendix~\ref{appx:autonomous_sym_inherence}.

To further illustrate this, we consider unitary learning with some symmetry. Specifically, we consider the quantum dynamics, $V(t)=e^{-itH}$ generated by the Hamiltonian
\begin{equation}
    H=\sum_{i=1}^{n-1}(X_iX_{i+1}+Y_iY_{i+1})\;\;.
\end{equation}
This Hamiltonian evolution conserves the total excitation number
$(n-\sum_i Z_i)/2$ counting the number of qubits in the $-1$ eigenspace of $Z_i$, since $[\sum_i Z_i,H]=0$. Consequently, the Hilbert space can be decomposed into invariant subspaces labeled by the number of excitations. We further consider a symmetry-tailored dataset with the input state of the form
\begin{equation}
    \ket{\psi_a} = \frac{1}{\ZC}\sum_x {(a_x+ib_x)}\ket{x}\;\;,
\end{equation}
where the coefficients $a_x$ and $b_x$ are drawn from the normal distribution, i.e, $a_x,b_x\in\mathcal{N}(0,1)$, $\ket{x}$ represents a computational basis state with Hamming weight $1$, and $\ZC$ is the normalization factor. Essentially, these input states are drawn from a Haar random ensemble of the $n$-dimensional single-excitation subspace. We note that using $U_0=\mathbb{1}$, this setting satisfies all assumptions in Theorem~\ref{thm:sym-inherence}. 

To make a comparison and show the power of symmetry aware, we compare this symmetrized training dataset to a random dataset, where the training dataset is drawn from the Haar ensemble of the full $2^n$-dimensional Hilbert space. 
As shown in Fig.~\ref{fig:p2_unitary_learning_result}(a), with $N_{\rm train}=20$ at $k=20$ iterations and step size of $s=4$, the model achieves a testing loss of $6.85\times 10^{-4}$ and demonstrates robust generalization, i.e., the testing loss decreases alongside the training loss. Fig.~\ref{fig:p2_unitary_learning_result}(b) further shows that the testing loss after $k=20$ iterations decreases steadily with the number of training samples, so that a symmetry-tailored dataset successfully recovers the unitary action on the invariant subspace using $\OC({\rm poly}(n))$ samples. Conversely, $N_{\rm train}=20$ samples from the random training dataset are insufficient to achieve generalization, as the testing loss does not change its order of magnitude despite a decrease in the training loss. This generalization gap is a direct manifestation of Theorem~\ref{thm:sym-inherence}. As illustrated in Fig.~\ref{fig:p2_unitary_learning_result}(c), because the symmetry-tailored dataset satisfies all the assumptions of the theorem, the algorithm discovers the conserved $U(1)$ symmetry from the structure of the training data without requiring a manually engineered equivariant ansatz. Panel (c) makes the mechanism behind this gap visible. The commutator norm $\|[\sum_i Z_i,U_k]\|_2$ vanishes identically for the symmetry-tailored dataset, confirming that every iterate commutes with the symmetry generator as Theorem~\ref{thm:sym-inherence} predicts. On the other hand, the norm grows with $k$ for the Haar-random dataset, whose training states break the symmetry. The trained model therefore stays within the single-excitation sector only because the training data respect that sector. This is precisely the situation in which the generalization bound of Theorem~\ref{thm:sample-complexity-u-learning} applies, with the model complexity set by $\deff=n$ rather than by $2^n$.


A closely related approach is the universal quantum emulator~\cite{marvian2016universal}, which is likewise given copies of input--output pairs $\{\ket{\psi_i}, V\ket{\psi_i}\}_{i=1}^N$ and emulates the action of the unknown $V$ on the span of the inputs, $\mathrm{Span}(\{\ket{\psi_i}\}_{i=1}^N)$. Since the emulator acts only within this span, it also exploits the symmetric sector of Theorem~\ref{thm:sample-complexity-u-learning} automatically. Its construction requires resources that scale polynomially with the dimension of the input subspace. Moreover, because it implements reflections about the fixed training samples rather than the evolved states $U_k\rho_iU_k^\dagger$, it avoids the recursive circuit depth arising in Theorem~\ref{thm:resource}. This, however, comes at the cost of additional ancilla qubits and controlled density-matrix exponentiation of both the input and output states. In contrast, our framework applies to a broader class of learning tasks defined by losses that are linear in observable expectation values, including classification, and provides guarantees on loss decrease and symmetry inheritance. In addition, the two approaches pursue different goals. The emulator reproduces $V$ on the span of the inputs with a worst-case guarantee, whereas our framework minimizes a training loss and is analyzed in terms of its generalization to unseen data drawn from the training distribution.

Finally, although not implemented in an autonomous self-training manner, Riemannian optimization has also been applied to quantum circuit compression~\cite{kotil2022riemannian,le2025riemannian,le2026hessian}, a task closely related to unitary learning. In this setting, the target is a known time-evolution unitary, and the goal is to find a shallow-depth circuit that approximates its action. This has been studied both in the conventional setting, where no training data are involved~\cite{kotil2022riemannian,le2025riemannian}, and in a QML setting~\cite{le2026hessian}. A key distinction from our approach is that these works combine Riemannian optimization with tensor-network methods, allowing the gradients to be computed through analytical contractions and the optimization to be performed entirely using classical resources.

\medskip

\section{Discussion}

We have proposed an autonomous self-training framework for quantum machine learning based on Riemannian gradient optimization, in which the training process takes place entirely in the quantum domain. 
Under suitable conditions, our framework guarantees a monotonic exponential decrease in the training loss and enables the trained model to recognize symmetries encoded in the data. These guarantees, however, come with an important resource trade-off: the depth of the resulting quantum circuit can grow exponentially with the number of training iterations. This places a fundamental constraint on the practical utility of the approach, as the benefits of autonomous training must be weighed against the increasing coherent-resource requirements. Encouragingly, our numerical results for the two canonical examples of quantum phase classification and unitary learning demonstrate that good predictive accuracy can be achieved with relatively small training datasets and a modest number of training iterations.

These promising results should be viewed as a starting point for exploring autonomous approaches to QML, rather than a complete solution. Several aspects of the framework remain open for further investigation. One immediate direction is to develop heuristic and systematic strategies for reducing its resource requirements. These could include combining the framework with quantum circuit compression techniques, as discussed in Section~\ref{sec:potential-refinement}, and exploring adaptive or heuristic step-size schedules. The latter may offer a practical route to faster convergence, potentially reducing the number of iterations and hence the circuit depth required for training. Another possibility is to mitigate compilation errors through Richardson extrapolation~\cite{richardson1927deferred,sidi2003practical}. Since this technique combines measured expectation values through classical post-processing, it can be applied at the inference stage without compromising the autonomy of the training process. On a related topic, the resource requirements of deployment after training also warrant consideration. In the current framework, each prediction requires executing the deep trained circuit. One possible way to avoid this cost is to extract a classical surrogate of the trained model, in the spirit of shadow models~\cite{jerbi2023shadows}, which could then be used to process new queries without repeatedly implementing the quantum circuit.

Another important direction is to understand the role of inductive bias, particularly the choice of initial unitary \(U_0\). An initialization closer to the desired solution may provide a form of warm start, potentially improving both the convergence rate and the overall resource requirements. This is closely related to warm-start strategies explored in conventional variational quantum algorithms~\cite{puig2024variational,mhiri2025unifying,lerch2026iqp,puig2026warm}, and raises the question of how such strategies can be adapted to autonomous self-training. One possible approach is to obtain such an initialization using classical methods, for example, by exploiting prior information from tensor-network methods before carrying out autonomous training. A particularly interesting theoretical question is how the choice of \(U_0\) influences the PL parameter \(\mu\) in Eq.~\eqref{eq:PL-informal}, which governs the convergence rate under the PL condition. While one would expect an initialization closer to the solution to facilitate faster convergence, establishing a quantitative relationship between the initialization and the PL parameter remains an open problem. Inductive bias could also be incorporated through a fixed unitary applied after the trained unitary, providing an additional degree of freedom immediately before measurement. Together, these choices offer several possible ways to control the inductive bias of the model. Their systematic study, potentially from a resource-theoretic perspective, could provide further insight into the fundamental capabilities and limitations of the framework, as well as identify promising regimes for practical applications.


We also emphasize that the framework presented here is one of several possible approaches to autonomous QML. Alternative strategies based on encoding optimization problems into quantum dynamical or dissipative processes have recently been proposed~\cite{motlagh2024ground,catli2025exponentially,leng2023quantum,wanjura2025quantum}. Although these approaches have primarily been formulated in the context of optimization, their underlying ideas may also be extended to learning tasks. More broadly, while coherent self-training offers an appealing direction, it remains an open question which approach, if any, provides the most favorable balance between learning performance and resource requirements. Continued exploration of different paradigms will be essential to understanding the potential of autonomous QML.

Finally, assessing the practical value of autonomous QML requires a careful comparison with alternative approaches, including variational, kernel-based, dissipative, and geometric QML methods. Such comparisons should account for both measurement complexity and coherent-resource requirements, rather than considering either in isolation. Otherwise, the term ``autonomous'' may obscure rather than clarify the computational trade-offs involved. Ultimately, from a practitioner's perspective, the central question is which approach achieves a given level of learning performance with the lowest overall resource cost.

\section*{Acknowledgments}
K.S. and M.G. were  supported by the startup grant of the Nanyang Assistant Professorship by the Nanyang Technological University. T.C. acknowledges the Start-Up Grant (SUG) from Nanyang Technological University.
R.P. acknowledges the support of the SNF Quantum Flagship Replacement Scheme (Grant No. 215933). 
 Z.H. acknowledges support from the Sandoz Family Foundation-Monique de Meuron program for Academic Promotion. 
 S.T. acknowledges Exchange Faculty Travel Grant: TG168033 from Chulalongkorn University, as well as funding from National Research Council of Thailand (NRCT) [grant number N42A680126]. S.T. further acknowledges Thailand Science research and Innovation Fund Chulalongkorn University (IND\_FF\_69\_258\_2300\_062).

\section*{Artificial Intelligence Disclosure}
The authors acknowledge the use of Claude and ChatGPT for assistance with manuscript writing and revision, analytical calculations, and the improvement of existing code. All scientific content, analyses, and conclusions were independently reviewed and verified by the authors.

\bibliography{quantum.bib,quantum2.bib}
\onecolumngrid
\newpage

\appendix
{\huge \noindent \textbf{Appendix}}

\bigskip

\part{}
\parttoc 

\bigskip

\section{Derivation of the Riemannian gradient}
\label{appx:A}

We begin by deriving the Riemannian gradient of the QML loss in Eq.~\eqref{eq:qml-loss} from first principles in Appendices~\ref{appx:prelim} and ~\ref{appx:general-rg-derive}. Then we prove the two results that the main text relies on, i.e., the monotonic-descent guarantee for a general loss in Appendix~\ref{appx:training_loss_guarantee} and the condition for autonomous QML of our protocol in Appendix~\ref{appx:suff-for-autonomous}. Finally, in Appendix~\ref{appx:specific-exams}, we give an explicit expression of the Riemannian gradients for two canonical tasks: quantum phase classification and unitary learning.

\subsection{Preliminaries}
\label{appx:prelim}

First, we recall the definition of the Riemannian gradient. For a more comprehensive discussion on the Riemannian gradient, we refer the reader to Ref.~\cite{wiersema2023optimizing}.

\begin{definition}[Riemannian Gradient]
\label{def:riemannian_grad}
    Let $\MC$ be a manifold and $h:\MC\to\Rbb$ be differentiable. The Riemannian gradient ${\rm grad}(h)_x$ is the unique tangent vector at x such that for all $A \in T_x\MC$:
    \begin{align*}
        \langle {\rm grad}(h)_x, A\rangle_x = \grad_A h(x)\;\;,
    \end{align*}
    where $\grad_A h(x)$ is the directional derivative of $h$ at $x$ along $A$ and $\ipa{\cdot}{\cdot}_x$ is an inner product on $T_x\MC$.
\end{definition}

In particular, our manifold of interest is the unitary group $\mathcal U(d)=\{M\in\mathbb C^{d\times d}:M^{-1}=M^\dagger\}$. 
Since the curve $e^{sA}U$ passes through $U$ with velocity $AU$, we identify $T_U\mathcal U(d)=\{AU:A\in\mathfrak u(d)\}$ with the Lie algebra $\mathfrak u(d)=\{A\in\mathbb C^{d\times d}:A^\dagger=-A\}$. 
We equip $\mathcal U(d)$ with the Hilbert-Schmidt inner product, which is invariant under right multiplication by unitaries,
\begin{equation}
\langle XU,YU\rangle_{\rm HS}=\Tr \big((XU)^\dagger YU\big)=\Tr\big(X^\dagger Y\big)=\langle X,Y\rangle_{\rm HS} ,
\end{equation}
so the inner product on $T_U\mathcal U(d)$ does not depend on $U$. The Riemannian gradient of a differentiable function $h:\mathcal U(d)\to\Rbb$ at $U$ is the anti-Hermitian matrix $\grad(h)_U\in\mathfrak u(d)$ satisfying
\begin{equation}
\langle\grad(h)_U,A\rangle_{\rm HS}=\nabla_Ah(U)\quad\text{for all }A\in\mathfrak u(d)\;\;,
\end{equation}
where $\nabla_Ah(U)$ is the directional derivative along $A$ such that
\begin{align}
\nabla_Ah(U)=\lim_{s\to0}\frac{h(e^{sA}U)-h(U)}{s}\;\;.
\end{align}

\subsection{General QML loss gradient}
\label{appx:general-rg-derive}

To compute the Riemannian gradient of the QML loss in Eq.~\eqref{eq:qml-loss}, we first rewrite it in a compact form. Let  $\ell_i(U)=\Tr(U\rho_iU\ad O_i)$ denote the expectation value for $i^{\rm th}$ sample index. Since the label $y_i$ is fixed by the data, we absorb it into the sample index and write the mismatch function $f(x,y_i)=f_i(x)$. This lets us regard $f_i:\Rbb\to\Rbb$ as a function of the expectation value alone. The loss then becomes
\begin{equation}
\label{eq:appx-generic-qml-loss}
    \LC(U)=\frac 1 N \sum_{i=1}^Nf_i(\ell_i(U))\;\;.
\end{equation}
Assuming each $f_i$ is differentiable, we write 
\begin{equation}
\label{eq:fprime}
    f'_i(U):=\left.\left(\frac {\partial f_i(\ell_i)}{\partial\ell_i}\right)\right|_{\ell_i=\ell_i(U)}\;\;,
\end{equation}
which is a real number since $f$ is real-valued. 

Let $A\in \mathfrak{u}(d)$. The anti-Hermicity of $A$ gives $(e^{sA})\ad=e^{-sA}$, so that $\ell_i(e^{sA}U)=\Tr(e^{sA}U\rho_iU\ad e^{-sA}O_i)$. The directional derivative of the loss along $A$ is 
\begin{align}
    \nabla_A\LC(U) &=\lim_{s\to0}\frac{1}{N}\sum_{i=1}^N\frac{f_i(\ell_i(e^{sA}U))-f_i(\ell_i(U))}{s}\label{eq:generic_diff_1}\\
    &=\frac{1}{N}\sum_{i=1}^N \left[\frac{\partial}{\partial s}f_i(\ell_i(e^{sA}U))\right]\\
    &=\frac{1}{N}\sum_{i=1}^N f_i'(U)\,\partial_s\Tr\big(e^{sA}U\rho_iU\ad e^{-sA}O_i\big)\Big|_{s=0}\label{eq:generic_diff_2}\\
    &=\frac{1}{N}\sum_{i=1}^N f_i'(U)\Big(\Tr\big(AU\rho_iU\ad O_i\big)-\Tr\big(U\rho_iU\ad AO_i\big)\Big)\label{eq:generic_diff_3}\\
    &=\frac{1}{N}\sum_{i=1}^N f_i'(U) \Tr\big([U\rho_iU\ad,O_i]\,A\big)\label{eq:generic_diff_4}\\
    &=-\Big\langle\frac{1}{N}\sum_{i=1}^N f_i'(U)[U\rho_iU\ad,O_i],\;A\Big\rangle_{\rm HS}\;\;.\label{eq:generic_diff_5}
\end{align}
Eq.~\eqref{eq:generic_diff_2} is due to the chain rule with $f'_i(U)$ defined in Eq.~\eqref{eq:fprime}. Then, Eq.~\eqref{eq:generic_diff_3} follows from differentiating the expectation value at $s=0$ with $\partial_se^{sA}=Ae^{sA}$, and Eq.~\eqref{eq:generic_diff_4} from the cyclic property of the trace. Finally, Eq.~\eqref{eq:generic_diff_5} uses the fact that $f'_i(U)$ is real and $([X,Y])\ad = -[X,Y]$ for Hermitian $X$ and $Y$.  

Since $\frac{1}{N}\sum_{i=1}^N f_i'(U)[U\rho_i U\ad,O_i]\in\mathfrak{u}(d)$ because it is anti-Hermitian. Comparing Eq.~\eqref{eq:generic_diff_5} with the defining property $\langle{\rm grad}(\LC)_U,A\rangle_{\rm HS}=\nabla_A\LC(U)$ for all $A\in\mathfrak u(d)$ gives
\begin{equation}
\label{eq:appx-generic-rg-form}
    \rg(U):={\rm grad}(\LC)_U =-\frac{1}{N}\sum_{i=1}^N f_i'(U)\,[U\rho_i U \ad,O_i]\;\;.
\end{equation}
This ensures that the matrix exponential $e^{-sG}$ is unitary for real $s$, so that the update rule 
\begin{equation}
\label{eq:appx-unitary-update}
    U_{k+1} = e^{-s_kG_k}U_k\;\;,
\end{equation}
where $\rg_k:=G(U_k)$, stays on $\UC(d)$ at every iteration. Moreover, $A=-G_k$ is the direction of locally steepest descent among all $A\in\mathfrak{u}(d)$ with $\|A\|_2=\|G_k\|_2$, where $\|\cdot\|_2$ is the Schatten 2-norm. This is given by the Cauchy-Schwarz inequality: $\nabla_A \LC(U)=\ipa{G(U)}{A}_{\rm HS}\geq -\|G(U)\|_2\|A\|_2$.

\subsection{Training loss guarantee}
\label{appx:training_loss_guarantee}

With the optimization procedure established in Appendix~\ref{appx:general-rg-derive}, we now use the curvature of the loss along the update direction to establish that loss does not increase.

\begin{theorem}[Monotonic descent of the training loss]
\label{thm:monotonic_descent_RG_QML}
Let $\LC$ be a loss of the form of Eq.~\eqref{eq:appx-generic-qml-loss} with each $f_i$ and $\partial_s f_i$ is differentiable, let $\rg_k$ be the Riemannian gradient of Eq.~\eqref{eq:appx-generic-rg-form} at the $k$-th iterate $U_k$, and let $h_k(s):=\LC(e^{-sG_k}U_k)$. Suppose that $\rg_k\neq 0$ and that $M_k > 0$ satisfies $\sup_{\zeta \in [0, s_k]} |h''_k(\zeta)| \leq M_k$. If $0 < s_k \leq  2\|G_k\|_2^2/M_k$, then the update $U_{k+1}=e^{-s_kG_k}U_k$ of Eq.~\eqref{eq:appx-unitary-update} satisfies
\begin{equation}
    \LC(U_{k+1}) - \LC(U_k) \leq -s_k\|G_k\|_2^2+\frac {s_k^2} 2M_k \leq 0 \;\;.
\end{equation}
\end{theorem}

\begin{proof}
Since each $f_i$ is twice continuously differentiable and $s\mapsto\ell_i(e^{-s\rg_k}U_k)$ is smooth, $h_k$ is twice continuously differentiable, and Taylor's theorem with the Lagrange remainder gives, for some $\zeta\in(0,s_k)$,
\begin{equation}
    \LC(U_{k+1})=h_k(s_k)=h_k(0)+s_kh_k'(0)+\frac{s_k^2}{2}h_k''(\zeta)\;\;.
\end{equation}
The first derivative is the directional derivative of $\LC$ at $U_k$ along $W=-\rg_k$. By the defining property of the Riemannian gradient and $\rg_k\ad=-\rg_k$,
\begin{align}
    h_k'(0)&=\nabla_{-\rg_k}\LC(U_k)=\langle\rg_k,-\rg_k\rangle_{\rm HS} \\
    &=-\Tr(\rg_k\ad\rg_k)=-\|\rg_k\|_2^2\;\;.
\end{align}
Explicitly, Eq.~\eqref{eq:generic_diff_4} with $U=U_k$ and $W=-\rg_k$ gives $h_k'(0)=-\frac{1}{N}\sum_{i=1}^Nf_i'(U_k)\Tr([\rho_i\iter{k},O_i]\rg_k)=\Tr(\rg_k^2)=-\|\rg_k\|_2^2$, where the second equality uses Eq.~\eqref{eq:appx-generic-rg-form}. Using $h_k(0)=\LC(U_k)$ and $|h_k''(\zeta)|\leq M_k$, we obtain
\begin{align}
    \LC(U_{k+1})-\LC(U_k)&\leq-s_k\|\rg_k\|_2^2+\frac{s_k^2}{2}M_k \label{eq: dL with second order}\\
    &=-s_k\|\rg_k\|_2^2\left(1-\frac{s_kM_k}{2\|\rg_k\|_2^2}\right)\;\;,
\end{align}
and the right-hand side is nonpositive if and only if $s_k\leq2\|\rg_k\|_2^2/M_k$.
\end{proof}

\begin{remark}
For linear mismatch functions, Lemma~\ref{lem:curvature-fp} in Appendix~\ref{appx:convergence_guarantee} gives the $s_k$-independent bound $M_k=4\xi\|\rg_k\|_\infty^2$, which turns the implicit step-size condition of Theorem~\ref{thm:monotonic_descent_RG_QML} into the explicit window of Proposition~\ref{prop:mono_descent}.
\end{remark}

\subsection{Sufficient condition for autonomous QML}
\label{appx:suff-for-autonomous}

In Appendix~\ref{appx:general-rg-derive} we derived the Riemannian gradient without any constraint on the mismatch function. In general, the coefficients $f_i'(U_k)$ in Eq.~\eqref{eq:appx-generic-rg-form} depend on $\ell_i(U_k)$ and must be estimated by measurement at every iteration, which violates Definition~\ref{def-extended:autonomus-qml}. A simple sufficient condition that removes this dependence is a linear mismatch function in Eq.~\eqref{eq:linear-mismatch} where the coefficients then reduce to constants fixed by the task, and no measurement feedback is needed during training. We first record this reduction as a lemma and then assemble the proof of Theorem~\ref{thm:sufficient-cond-autonomous-qml}.

\begin{lemma}[Constant gradient coefficients for linear mismatch functions]
\label{lem:linear-rg-appx}
Suppose that the mismatch function in Eq.~\eqref{eq:appx-generic-qml-loss} is linear, as in Eq.~\eqref{eq:linear-mismatch},
\begin{align}
    f_i(\ell)=f(\ell,y_i)=c_i\ell+b_i\;\;,\label{eq:generic-linear-loss}
\end{align}
with real constants $b_i$ and $c_i$ that may depend on $y_i$ but not on $U$. Then
\begin{align}
    f_i'(U)=c_i
\end{align}
for every unitary $U$, and the Riemannian gradient of Eq.~\eqref{eq:appx-generic-rg-form} reduces to
\begin{align}
    \rg_k=-\frac{1}{N}\sum_{i=1}^Nc_i\,[\rho_i\iter{k},O_i]\;\;.\label{eq:linear-rg-appx}
\end{align}
In particular, $\rg_k$ is determined without requiring the expectation value of the evolved state.
\end{lemma}
\begin{proof}
    The function $f_i$ is linear in $\ell$ with slope $c_i$, so $f_i'(U)=c_i$. Substituting $f_i'(U_k)=c_i$ into Eq.~\eqref{eq:appx-generic-rg-form} at $U=U_k$ gives Eq.~\eqref{eq:linear-rg-appx}.
\end{proof}

\subsubsection{Proof of Theorem~\ref{thm:sufficient-cond-autonomous-qml}: Autonomous QML with the Riemannian gradient descent}

We are now ready to state the proof of Theorem~\ref{thm:sufficient-cond-autonomous-qml} as follows.

\begin{proof}
By Lemma~\ref{lem:linear-rg-appx}, every linear mismatch function has $f_i'(U_k)=c_i$ for every iteration $k$ and $\rg_k=-\frac{1}{N}\sum_{i=1}^Nc_i[\rho_i\iter{k},O_i]$. That is, the Riemannian gradient is a linear combination of the commutators $[\rho_i\iter{k},O_i]$ whose coefficients $c_i$ are fixed before training. We construct the protocol by induction on $k$ and show that the compiled iterate $\tilde U_k$ of Appendix~\ref{appx:resource}, with $\tilde U_0=U_0$, is a product of $U_0$, of its inverse $U_0\ad$ (which Appendix~\ref{appx:oracle-count} counts as an oracle call), and of oracle calls from $\mathbb{U}_t$ in Eq.~\eqref{eq:oracle}, with evolution times of both signs as in Eq.~\eqref{eq:group_commutator_apprx}, none of which involves a measurement.

Suppose that $\tilde U_k$ has been compiled in this way; for $k=0$ there is nothing to show. Let $\tilde\rho_i\iter{k}:=\tilde U_k\rho_i\tilde U_k\ad$ and $\tilde\rg_k:=-\frac{1}{N}\sum_{i=1}^Nc_i[\tilde\rho_i\iter{k},O_i]$, as in Theorem~\ref{thm:appx-accum-err}. First, the update $e^{-s_k\tilde\rg_k}=\exp\big(\frac{s_k}{N}\sum_{i=1}^Nc_i[\tilde\rho_i\iter{k},O_i]\big)$ is the exponential of a sum of commutators, which the Lie--Trotter formula in Eq.~\eqref{eq:trotter-apprx} of Section~\ref{sec:algo} approximates by a product of the elementary factors $e^{s_i^{(k)}A_i^{(k)}/r}$ with $s_i^{(k)}=s_k|c_i|/N$ and $A_i^{(k)}={\rm sign}(c_i)[\tilde\rho_i\iter{k},O_i]$. Second, each elementary factor is approximated by the group-commutator formula in Eq.~\eqref{eq:group_commutator_apprx} with $t^2=s_i^{(k)}/r$, which is a product of unitaries of the form $e^{\pm i\alpha tO_i}$ and $e^{\pm i\alpha t\tilde\rho_i\iter{k}}$. Third, the factors $e^{\pm i\alpha tO_i}$ are oracle calls, and the factors generated by the evolved states are obtained from the oracle $e^{-it\rho_i}$ by conjugation,
\begin{equation}
    e^{-it\tilde\rho_i\iter{k}}=e^{-it\tilde U_k\rho_i\tilde U_k\ad}=\tilde U_k\,e^{-it\rho_i}\,\tilde U_k\ad\;\;,\label{eq:appx-conjugation-identity}
\end{equation}
which holds because $(\tilde U_k\rho_i\tilde U_k\ad)^p=\tilde U_k\rho_i^p\tilde U_k\ad$ for every power $p$, so the exponential series conjugates term by term. Here $\tilde U_k$ is the circuit compiled in the previous iterations, and $\tilde U_k\ad$ is the same circuit run in reverse order with every factor replaced by its inverse, that is, $e^{\pm itO_i}\mapsto e^{\mp itO_i}$, $e^{\pm it\rho_i}\mapsto e^{\mp it\rho_i}$, and $U_0\mapsto U_0\ad$. Hence $\tilde U_{k+1}:=W_k\tilde U_k$, where $W_k$ is the product of oracle calls produced by the two product formulas, is again a product of $U_0$, $U_0\ad$, and oracle calls, which completes the induction; this is the recursive unrolling described in Section~\ref{sec:algo}. The only data entering the construction are the classical constants $\{c_i\}_{i=1}^N$, the step sizes $\{s_k\}_k$, and the coefficients of the two product formulas, so the circuit for any number of iterations is fixed before the run and no step requires a quantum measurement. The protocol therefore satisfies Definition~\ref{def-extended:autonomus-qml}.

Finally, the two product formulas are approximations, so the compiled circuit $\tilde U_k$ differs from the exact iterate $U_k$. Appendix~\ref{appx:resource} bounds this difference (Theorem~\ref{thm:appx-accum-err}) and shows that it can be made smaller than any prescribed $\ep>0$ at the oracle cost stated in Theorem~\ref{thm:resource}. The compiled protocol therefore follows the Riemannian gradient descent trajectory to any desired accuracy, which proves the theorem.
\end{proof}

\subsection{Specific examples}
\label{appx:specific-exams}
We now apply Lemma~\ref{lem:linear-rg-appx} to the two learning tasks of Section~\ref{sec:canonical-example}, including quantum phase classification and unitary learning. Each task is specified by the input states $\rho_i$, the observables $O_i$, and the constants $b_i$ and $c_i$ in Eq.~\eqref{eq:linear-mismatch}. Table~\ref{tab:rg-collection} lists them together with the resulting Riemannian gradient.

\begin{table*}[!b]
\centering
\caption{\protect\textbf{Summary of task-specific parameters for the two canonical examples.} For each task of Section~\ref{sec:canonical-example}, the table lists the input states $\rho_i$, the observables $O_i$, and the constants $b_i$ and $c_i$ of the affine mismatch function in Eq.~\eqref{eq:linear-mismatch}, together with the Riemannian gradient $\rg_k$ at the $k$-th iterate, where $\rho_i\iter{k}=U_k\rho_iU_k\ad$.}
\label{tab:rg-collection}
\begin{tabular}{lccccc}
\hline\hline
Task & $\rho_i$ & $O_i$ & $b_i$ & $c_i$ & $G_k$\\
\hline
Quantum phase classification
  & $\rho_{{\rm s},i}\ot\ketbraq{\phi}_a$
  & $P_{y_i}=\mathbb{1}\ot\ketbraq{y_i}_a$
  & $1$ & $-1$ & $\frac{1}{N}\sum_{i=1}^N[\rho_i\iter{k},P_{y_i}]$\\
Unitary learning
  & $\ketbraq{\psi_i}$
  & $\sigma_i=V\ketbraq{\psi_i}V\ad$
  & $1$ & $-1$ & $\frac{1}{N}\sum_{i=1}^N[\rho_i\iter{k},\sigma_i]$\\
\hline\hline
\end{tabular}
\end{table*}

\subsubsection{\texorpdfstring{Quantum phase classification}{Quantum phase classification}}
The loss in Eq.~\eqref{eq:phase-loss-fn} is one minus the average probability of predicting the correct label. The input states are the encoded system--ancilla states $\rho_i=\rho_{{\rm s},i}\ot\ketbraq{\phi}_a$ and the observables are the label projectors $O_i=P_{y_i}=\mathbb{1}\ot\ketbraq{y_i}_a$ of Section~\ref{sec:canonical-example}, so
\begin{align}
    \LC(U)=1-\frac{1}{N}\sum_{i=1}^N\Tr(U\rho_iU\ad P_{y_i})\;\;,
\end{align}
which is Eq.~\eqref{eq:appx-generic-qml-loss} with the linear mismatch function $f_i(\ell)=1-\ell$, that is, $c_i=-1$ and $b_i=1$. Lemma~\ref{lem:linear-rg-appx} gives
\begin{align}
    \rg_k=\frac{1}{N}\sum_{i=1}^N[\rho_i\iter{k},P_{y_i}]\;\;.
\end{align}

\subsubsection{Unitary learning}
The loss in Eq.~\eqref{eq:unitary-loss-fn} is one minus the average fidelity $F(U\ket{\psi_i},V\ket{\psi_i})=|\bra{\psi_i}V\ad U\ket{\psi_i}|^2$ between the model output $U\ket{\psi_i}$ and the target $V\ket{\psi_i}$. With $\rho_i=\ketbraq{\psi_i}$ and $\sigma_i:=V\ketbraq{\psi_i}V\ad$ it reads
\begin{align}
    \LC(U)=1-\frac{1}{N}\sum_{i=1}^N\Tr(U\rho_iU\ad\sigma_i)\;\;,
\end{align}
which is Eq.~\eqref{eq:appx-generic-qml-loss} with $O_i=\sigma_i$ and the affine mismatch function $f_i(\ell)=1-\ell$, that is, $c_i=-1$ and $b_i=1$. Lemma~\ref{lem:linear-rg-appx} gives
\begin{equation}
    \rg_k = \frac{1}{N}\sum_{i=1}^N[\rho_i\iter{k},\sigma_i]\;\;.
\end{equation}

\section{\texorpdfstring{Training dynamics of the autonomous training update}{Training dynamics of the autonomous training update}}
\label{appx:convergence_guarantee}

Although the autonomous training update is motivated by double-bracket
flows and imaginary-time evolution, its dynamics generally differ from both, so
their convergence guarantees do not transfer automatically. In this section, we therefore address four questions about the exact update of Eq.~\eqref{eq:update_rule} (not about the compiled circuits). (i) Does each training step lower the loss ? (ii) Under what conditions does the loss approach its limiting value exponentially fast ? (iii) What evidence supports these conditions for specific learning tasks ? (iv) Under what conditions does the limiting loss equal the global minimum ?

To address these questions, we consider losses of the form covered
by the sufficient condition for autonomous QML in
Theorem~\ref{thm:sufficient-cond-autonomous-qml}:
\begin{equation}
\label{eq:auto-loss-fp}
    \LC(U)=\frac1N\sum_{i=1}^N\big[c_i\ell_i(U)+b_i\big]\;\;,
    \qquad
    \ell_i(U)=\Tr\big(U\rho_iU^\dagger O_i\big)\;\;.
\end{equation}
Here $N$ is the number of training samples. The states $\rho_i$,
Hermitian observables $O_i$, and real coefficients $b_i,c_i$ are fixed
by the task; the variable being optimized is $U\in{\rm U}(d)$, with
$d=\dim(\HC)$. We use $\|\cdot\|_2$ for the Schatten 2-norm (Frobenius norm), $\|A\|_2=\sqrt{\Tr(A^\dagger A)}$, and $\|\cdot\|_\infty$ for the operator norm. Writing $\rho_i\iter{k}:=U_k\rho_iU_k^\dagger$ as in the main text, the Riemannian gradient
generator and exact training dynamics are
\begin{equation}
\label{eq:notation-fp}
\begin{split}
    &G(U):=-\frac1N\sum_ic_i\big[U\rho_iU^\dagger,O_i\big]\ \in\ \su(d)\;\;,\\
    &U_{k+1}=e^{-s_kG_k}U_k\;\;,
\end{split}
\end{equation}
with $G_k:=G(U_k)$ and step size $s_k>0$. This is Eq.~\eqref{eq:generic_RG_grad} with $\partial f/\partial\ell_i=c_i$, so that $G_k=-\frac1N\sum_ic_i[\rho_i\iter{k},O_i]$. This $G(U)$ coincides with the Riemannian gradient $\rg(U)$ of Eq.~\eqref{eq:appx-generic-rg-form} (Eq.~\eqref{eq:RG-qml-explicit-form} in the main text) for the loss of Eq.~\eqref{eq:qml-loss} with the affine mismatch function, that is, with Eq.~\eqref{eq:linear-rg-appx} at $U=U_k$. The negative generator
$-G_k$ determines the descent direction. Note that, even if the negative
gradient points downhill locally, a large step size can leave the loss higher than it was before the update.

To choose a step size that keeps the loss from increasing, we bound
how much its slope can change during an update step. This bound uses the loss
coefficients and observables through the constant
\begin{equation}
    \xi:=\frac1N\sum_i|c_i|\,\|O_i\|_\infty\,.
\end{equation}

We impose the following step-size condition.
\begin{enumerate}
    \renewcommand{\theenumi}{A\arabic{enumi}}
    \renewcommand{\labelenumi}{(\theenumi)}

    \item \label{assump:A1} \emph{Step-size window.}
    Let $\xi>0$. For fixed $\theta\in(0,1)$ and $s_{\min}>0$,
    \begin{equation*}
        0<s_{\min}\leq s_k\leq\frac{\theta}{2\xi}
        \qquad\text{for all }k.
    \end{equation*}
\end{enumerate}

The following subsections address these questions in order. Under Assumption~\ref{assump:A1}, Appendix~\ref{subsec:mono_descent} proves monotone descent of the training loss. Appendix~\ref{subsec:baseline-fp} introduces the Polyak--$\text{\L}$ojasiewicz (PL) inequality, a standard technique in optimization and dynamical systems, which keeps the gradient large enough that each step removes a fixed positive fraction of the difference between the current and limiting loss, and thereby gives an exponential convergence rate. Appendix~\ref{subsec:PLvalid-pl} establishes a PL condition for pure-state ground-state preparation, and Appendix~\ref{subsec:numerical-pl} examines the PL condition numerically for phase classification and unitary learning. Finally, Appendix~\ref{subsec:global-QML} discusses global convergence for losses with a common-observable structure, including binary classification.

\subsection{Monotone descent}
\label{subsec:mono_descent}

We bound the loss change along the training dynamics in
Eq.~\eqref{eq:notation-fp}.

\begin{lemma}[Uniform curvature bound]
\label{lem:curvature-fp}
Hold $G_k$ fixed and define $h_k(t):=\LC(e^{-tG_k}U_k)$, the training
loss along the $k$-th update. Here $t=0$ and $t=s_k$ correspond to
the current and next iterates, respectively. Then $h_k$ is smooth and
\begin{equation}
\label{eq:curvature-fp}
    |h_k''(t)|\leq 4\xi\|G_k\|_\infty^2
    \qquad\text{for all }t\in\Rbb.
\end{equation}
\end{lemma}

\begin{proof}
Fix the iteration $k$ and hold $G_k$ constant as $t$ varies.
Since $G_k^\dagger=-G_k$, the matrix $e^{-tG_k}$ is unitary for
every real $t$. Thus
$\rho_i\iter{k}(t):=e^{-tG_k}\rho_i\iter{k}e^{tG_k}$ remains a positive
semidefinite operator with unit trace. Differentiating gives
\[
    \dot\rho_i\iter{k}(t)=-G_k\rho_i\iter{k}(t)+\rho_i\iter{k}(t)G_k
    =[\rho_i\iter{k}(t),G_k].
\]
Applying the cyclicity
of the trace to
$\Tr([\rho_i\iter{k}(t),G_k]X)=\Tr(\rho_i\iter{k}(t)[G_k,X])$, first with $X=O_i$ and then with
$X=[G_k,O_i]$, both independent of $t$, yields
\begin{align}
\label{eq:hprime-fp}
    h_k'(t)
    &=\frac1N\sum_i c_i\,\Tr\big(\rho_i\iter{k}(t)[G_k,O_i]\big),\\
\label{eq:hsecond-fp}
    h_k''(t)
    &=\frac1N\sum_i c_i\,\Tr\big(\rho_i\iter{k}(t)[G_k,[G_k,O_i]]\big).
\end{align}
Because $\rho_i\iter{k}(t)$ is positive semidefinite with unit trace,
H\"older's inequality gives
$|\Tr(\rho_i\iter{k}(t)X)|\leq\|X\|_\infty$.
The triangle inequality and submultiplicativity of the operator norm
give $\|[A,B]\|_\infty\leq2\|A\|_\infty\|B\|_\infty$.
Using H\"older's inequality in Eq.~\eqref{eq:hsecond-fp} and applying this commutator bound twice,
we obtain
\[
\begin{aligned}
    |h_k''(t)|
    &\leq\frac1N\sum_i|c_i|\,
        \|[G_k,[G_k,O_i]]\|_\infty\\
    &\leq\frac{2\|G_k\|_\infty}{N}
        \sum_i|c_i|\,\|[G_k,O_i]\|_\infty\\
    &\leq\frac{4\|G_k\|_\infty^2}{N}
        \sum_i|c_i|\,\|O_i\|_\infty
     =4\xi\|G_k\|_\infty^2.
\end{aligned}
\]
The final equality uses the definition of $\xi$. Since the bound
does not depend on $t$, Eq.~\eqref{eq:curvature-fp} holds uniformly
for all real $t$.
\end{proof}

\begin{proposition}[Monotone descent of the training loss]
\label{prop:mono_descent}
For $\xi>0$, suppose the step sizes satisfy Assumption~\ref{assump:A1},
so $0<s_{\min}\leq s_k\leq\theta/(2\xi)$ for all $k$, with $0<\theta<1$.
Then
\begin{equation}
\label{eq:descent-fp}
    \LC(U_{k+1})-\LC(U_k)
    \leq -(1-\theta)s_k\|G_k\|_2^2\leq0
    \qquad\text{for all }k.
\end{equation}
The decrease is strict when $G_k\neq0$. If $G_k=0$, the update fixes
$U_k$. For $\xi=0$, the loss is constant and every update fixes the
iterate.
\end{proposition}

\begin{proof}
If $\xi=0$, then every $|c_i|\|O_i\|_\infty$ is zero, so $c_iO_i=0$
for every $i$. Hence $\LC$ is constant and $G_k=0$, so every
update fixes the iterate. Now suppose $\xi>0$. 
The Taylor estimate that forms the first inequality of Eq.~\eqref{eq: dL with second order} in Appendix~\ref{appx:training_loss_guarantee} holds for every $s_k>0$ and every bound $M_k$ on $|h_k''|$ over $[0,s_k]$. With $M_k=4\xi\|G_k\|_\infty^2$ from Lemma~\ref{lem:curvature-fp}, it gives
\begin{align}
   \LC(U_{k+1})-\LC(U_k)
    \leq-s_k\|G_k\|_2^2+2\xi s_k^2\|G_k\|_\infty^2\;\;,
\end{align}
which holds trivially when $G_k=0$.
The norm comparison
$\|G_k\|_\infty\leq\|G_k\|_2$ and the step-size condition
$2\xi s_k\leq\theta$ give
\[
    2\xi s_k^2\|G_k\|_\infty^2
    \leq(2\xi s_k)s_k\|G_k\|_2^2
    \leq\theta s_k\|G_k\|_2^2.
\]
Substituting this estimate yields
\[
    \LC(U_{k+1})-\LC(U_k)
    \leq-(1-\theta)s_k\|G_k\|_2^2\leq0,
\]
which is Eq.~\eqref{eq:descent-fp}. Since $s_k>0$ and $\theta<1$,
the decrease is strict when $G_k\neq0$. If $G_k=0$, then
$e^{-s_kG_k}=\mathbb{1}$, so $U_{k+1}=U_k$.
\end{proof}

\subsection{Exponential convergence of the training loss under PL condition}
\label{subsec:baseline-fp}

For the convergence-rate analysis,
we additionally assume that the iterates approach a limiting unitary
$U_\infty$, which may depend on initialization:
\begin{enumerate}
    \renewcommand{\theenumi}{A\arabic{enumi}}
    \renewcommand{\labelenumi}{(\theenumi)}
    \setcounter{enumi}{1}

    \item \label{assump:A2} \emph{Convergence.}
    The iterates $U_k$ started from $U_0$ converge to a single unitary $U_\infty$.
\end{enumerate}
We expect Assumption~\ref{assump:A2} to follow from Assumption~\ref{assump:A1} by verifying that our update rule satisfies the conditions of the convergence results established in Refs.~\cite{schneider2015convergenceresultsprojectedlinesearch,absil2005convergence}. For simplicity, however, we retain Assumption~\ref{assump:A2} and leave this verification for future work.%

Now, denote
\begin{equation}
    \LC_\infty:=\LC(U_\infty)\,.
\end{equation}

We first show that the loss converges and the gradient norm vanishes along the run, and that the assumed limiting unitary $U_\infty$ is a critical point.

\begin{proposition}[Loss limit and vanishing gradient]
\label{prop:limit-fp}
Under Assumptions~\ref{assump:A1} and~\ref{assump:A2},
\begin{equation}
    \LC(U_k)\to\LC_\infty=\inf_k\LC(U_k),
    \qquad
    \|G_k\|_2\to0\,.
\end{equation}
In particular, the limiting unitary $U_\infty$ is a critical point
of the training loss:
\begin{equation}
    G(U_\infty)=0\,.
\end{equation}
\end{proposition}

\begin{proof}
By Eqs.~\eqref{eq:auto-loss-fp} and~\eqref{eq:notation-fp}, $\LC$ and $G$ are polynomial in the real and imaginary parts of the entries of $U$, and hence continuous. Compactness of ${\rm U}(d)$ therefore implies that $\LC$ is bounded below. By Eq.~\eqref{eq:descent-fp}, $\LC(U_k)$ is nonincreasing and thus converges to its infimum. Assumption~\ref{assump:A2} and continuity of $\LC$ identify this limit as $\LC(U_\infty)=\LC_\infty$. Hence,
\begin{equation}
    \LC_\infty=\inf_k\LC(U_k)\,.
\end{equation}

Summing Eq.~\eqref{eq:descent-fp} over $k=0,\ldots,K$ and using $s_k\geq s_{\min}$ gives
\begin{align}
    (1-\theta)s_{\min}\sum_{k=0}^K\|G_k\|_2^2
    &\leq\LC(U_0)-\LC(U_{K+1})
    \nonumber\\
    &\leq\LC(U_0)-\LC_\infty\,.
\end{align}
The total loss decrease is finite, so the sum of the squared gradient
norms is bounded. Letting $K\to\infty$ gives
$\sum_k\|G_k\|_2^2<\infty$. If the gradient norm exceeded a fixed
positive threshold infinitely often, this sum would diverge.
Therefore, $\|G_k\|_2\to0$. Finally, Assumption~\ref{assump:A2}
and continuity of $G$ give
\begin{equation}
    G(U_\infty)=\lim_{k\to\infty}G(U_k)=0\,.
\end{equation}
\end{proof}

Proposition~\ref{prop:limit-fp} shows that the training loss
converges to $\LC_\infty$, but does not specify how quickly.
We now seek conditions under which the training loss approaches
$\LC_\infty$ \emph{exponentially fast}. The Polyak--$\text{\L}$ojasiewicz (PL)
inequality provides a standard way to establish exponential
convergence: it relates the squared gradient norm to the remaining
loss, allowing us to use the
descent estimate in Proposition~\ref{prop:mono_descent}. Originating
in Polyak's analysis of gradient methods~\cite{polyak1963gradient},
its usual global form requires that, for an objective $\varphi$ with minimum
value $\varphi^\star$, there exists a constant $\mu>0$ such that
\begin{gather*}
\tfrac12\|\nabla\varphi(x)\|^2\geq\mu\big(\varphi(x)-\varphi^\star\big)
    \qquad\text{for all }x.
\end{gather*}

PL and related $\text{\L}$ojasiewicz gradient inequalities are established
tools for convergence analysis in optimization, machine learning, and
dynamical systems~\cite{karimi2020linearconvergencegradientproximalgradient,absil2005convergence}.
For an objective with a Lipschitz-continuous gradient, the PL condition
guarantees exponential decay of $\varphi(x_k)-\varphi^\star$ under gradient descent
with a suitable fixed step size, without requiring
convexity~\cite{karimi2020linearconvergencegradientproximalgradient}.

Combining the descent bound in Proposition~\ref{prop:mono_descent} with a PL inequality relative to $\LC_\infty$ shows that each step removes a fixed fraction of the remaining loss; this leads to the exponential bound in Theorem~\ref{thm:transversal-fp}. The limiting value $\LC_\infty$ need not be the global minimum.

We now make this condition precise.

\begin{definition}[PL inequality along the run]
\label{def:PL_inequality}
Under Assumption~\ref{assump:A2}, the autonomous training run satisfies a PL inequality if there exists a constant $\mu>0$ such that
\begin{equation}
\label{eq:PL-fp}
    \tfrac12\|G_k\|_2^2
    \geq\mu\big(\LC(U_k)-\LC_\infty\big)
    \qquad\text{for all }k\geq 0\,.
\end{equation}
For a fixed step-size schedule, the constant $\mu$
may depend on the initial unitary $U_0$, but is independent of the iteration $k$.
\end{definition}

\begin{theorem}[Exponential convergence of the remaining loss]
\label{thm:transversal-fp}
Let $\xi>0$ and consider the update in Eq.~\eqref{eq:notation-fp}
under Assumptions~\ref{assump:A1} and~\ref{assump:A2}. Suppose the
PL inequality in Definition~\ref{def:PL_inequality} holds. We may
choose its coefficient to satisfy $0<\mu\leq2\xi$, since decreasing
a valid PL coefficient preserves the PL condition. Define the remaining loss by
$\widehat{\LC}_k:=\LC(U_k)-\LC_\infty$, and set
\[
    q:=1-2(1-\theta)s_{\min}\mu\in[1/2,1),
\]
where $s_{\min}$ and $\theta$ are the constants in Assumption~\ref{assump:A1}. Then
\begin{equation}
\label{eq:main-rate-fp}
    0\leq\widehat{\LC}_k\leq q^k\widehat{\LC}_0
    \qquad(k\geq0).
\end{equation}
Each iteration reduces the remaining loss by at least
the fixed fraction $1-q$.
\end{theorem}

\begin{proof}
Proposition~\ref{prop:limit-fp} gives $\widehat{\LC}_k\geq0$.
For every $k\geq0$, combining the descent
bound~\eqref{eq:descent-fp}, the PL inequality~\eqref{eq:PL-fp},
and $s_k\geq s_{\min}$ gives
\[
    \widehat{\LC}_{k+1}
    \leq\widehat{\LC}_k-(1-\theta)s_k\|G_k\|_2^2
    \leq q\,\widehat{\LC}_k.
\]
Since $s_{\min}\leq\theta/(2\xi)$ and $\mu\leq2\xi$, we have $2(1-\theta)s_{\min}\mu\leq2\theta(1-\theta)\leq1/2$, and $2(1-\theta)s_{\min}\mu>0$ because $s_{\min}$, $\mu$, and $1-\theta$ are positive; hence $q\in[1/2,1)$.
Each iteration therefore reduces the remaining loss by at least the
same fraction $1-q$. Repeatedly applying this bound from the initial
iterate proves Eq.~\eqref{eq:main-rate-fp}. Since $q<1$, the
factor $q^k$ decays exponentially with $k$. This is the standard
argument combining a descent estimate with a PL inequality to obtain
an exponential convergence  rate~\cite{karimi2020linearconvergencegradientproximalgradient}.
\end{proof}

We note that Theorem~\ref{thm:auto_RG_learning_guarantee} in the main text is the informal version of Theorem~\ref{thm:transversal-fp} here. It is obtained by regarding $\theta$ in Assumption~\ref{assump:A1} as an absolute constant, so that the window $s_k\leq\theta/(2\xi)$ reads $s_k\in\OC(1/\xi)$ and the factor $2(1-\theta)$ is absorbed in the $\Theta$ notation of $q=1-\Theta(\mu\min_ks_k)$; there $\min_ks_k$ stands for the uniform lower bound $s_{\min}$, and the cap $\mu\leq2\xi$ is implicit in ``with some $\mu$''.

\subsection{\texorpdfstring{PL inequality for ground-state preparation}{PL inequality for ground-state preparation}}
\label{subsec:PLvalid-pl}

For pure-state ground-state preparation, double-bracket flow is equivalent to normalized imaginary-time evolution (imaginary-time evolution renormalized at every instant, $\partial_t\ket{\psi}=-(H-\langle H\rangle_\psi)\ket{\psi}$);
with a positive ground-state gap and nonzero initial ground-state overlap,
this evolution is guaranteed to converge to the ground-state subspace~\cite{gluza2026double}.
Here we connect this convergence behavior to the PL framework of
Theorem~\ref{thm:transversal-fp}. For the discrete update under Assumption~\ref{assump:A1}, we prove that, once the iterates approach some energy window near the ground-state energy $E_0$, the energy converges to $E_0$ exponentially fast. Proposition~\ref{prop:energy-gap-pl} establishes a PL inequality relative to $E_0$ in that window, and Corollary~\ref{cor:energy-rate} implies exponential convergence of the energy to $E_0$ and identifies $\LC_\infty=E_0$.

\begin{proposition}[PL bound from an energy gap]
\label{prop:energy-gap-pl}
Let $H=H^\dagger$ have ground-state energy $E_0$. Let $\Delta>0$ be the energy gap from $E_0$ to
the next excited-state eigenvalue $E_1>E_0$. For a pure input state $\rho_0$, set
\begin{gather*}
    \rho(U):=U\rho_0U^\dagger,\qquad
    \LC(U):=\Tr\bigl(\rho(U)H\bigr),\\
    e(U):=\LC(U)-E_0.
\end{gather*}
This is the loss of Eq.~\eqref{eq:auto-loss-fp} with $N=1$, $c_1=1$, $b_1=0$, and $O_1=H$, so $G(U)=[H,\rho(U)]$ by Eq.~\eqref{eq:notation-fp} and $\xi=\|H\|_\infty$.
 One has
\begin{equation}
\label{eq:energy-gap-pl}
    \frac12\|G(U)\|_2^2
    \geq\frac{\Delta}{2}\,e(U)
    \qquad\text{whenever }0\leq e(U)\leq\frac{\Delta}{2}.
\end{equation}
Thus $\mu=\Delta/2$ is a valid PL coefficient in this energy region.
\end{proposition}

\begin{proof}
For brevity, write $\rho=\rho(U)=|\psi\rangle\langle\psi|$
and $E=\Tr(\rho H)$. Since $\rho$ is a pure state,
\[
    \rho^2=\rho,\qquad
    \rho H\rho
    =|\psi\rangle\langle\psi|H|\psi\rangle\langle\psi|
    =E\rho.
\]
The commutator satisfies $[H,\rho]^\dagger=-[H,\rho]$.
Expanding its squared Frobenius norm and using cyclicity of the trace gives
\[
\begin{aligned}
    \|[H,\rho]\|_2^2
    &=-\Tr\bigl([H,\rho]^2\bigr)\\
    &=2\Tr(\rho^2H^2)-2\Tr(\rho H\rho H)\\
    &=2\Tr(\rho H^2)-2E^2.
\end{aligned}
\]
The last equality uses the pure-state identities above. Dividing by two gives
\begin{equation}
\label{eq:energy-variance-identity}
    \frac12\|[H,\rho(U)]\|_2^2
    =\Tr\bigl(\rho(U)H^2\bigr)
      -\Tr\bigl(\rho(U)H\bigr)^2
    =\operatorname{Var}_{\rho(U)}(H).
\end{equation}
Every eigenvalue $\lambda$ of $H-E_0\mathbb{1}$ is either zero or at least $\Delta$,
and therefore $\lambda^2-\Delta\lambda\geq0$. Since $H-E_0\mathbb{1}$ and $(H-E_0\mathbb{1})^2$ have
the same eigenvectors, this gives
\[
    (H-E_0\mathbb{1})^2-\Delta(H-E_0\mathbb{1})\succeq0,
\]
where $\succeq0$ means positive semidefinite. Taking the expectation
in the state $\rho$ preserves this inequality, yielding
\[
    \Tr\bigl(\rho(H-E_0\mathbb{1})^2\bigr)
    \geq\Delta\Tr\bigl(\rho(H-E_0\mathbb{1})\bigr)=\Delta e(U).
\]
Now consider the energy variance
\[
\begin{aligned}
    \operatorname{Var}_{\rho(U)}(H)
    &=\Tr\bigl(\rho(H-E_0\mathbb{1})^2\bigr)-e(U)^2\\
    &\geq\Delta e(U)-e(U)^2
    =e(U)\bigl(\Delta-e(U)\bigr).
\end{aligned}
\]
Finally, $0\leq e(U)\leq\Delta/2$ implies
$\Delta-e(U)\geq\Delta/2$, so
\[
    \operatorname{Var}_{\rho(U)}(H)\geq\frac{\Delta}{2}e(U).
\]
Combining this inequality with Eq.~\eqref{eq:energy-variance-identity} proves
Eq.~\eqref{eq:energy-gap-pl}.
\end{proof}

Additionally, the proof implies slightly more: $\tfrac12\|G(U)\|_2^2\geq\big(\Delta-e(U)\big)\,e(U)$ for every $U$.

\begin{corollary}[Exponential convergence of the energy]
\label{cor:energy-rate}
In the setting of Proposition~\ref{prop:energy-gap-pl}, let the step sizes satisfy Assumption~\ref{assump:A1} with $\xi=\|H\|_\infty$, write $\widehat e_k:=e(U_k)=\LC(U_k)-E_0$, and suppose $\widehat e_{k_0}\leq\Delta/2$ for some $k_0\geq0$. Then, for all $k\geq k_0$,
\begin{equation}
\label{eq:energy-rate}
    0\leq\widehat e_k\leq\big(1-(1-\theta)s_{\min}\Delta\big)^{k-k_0}\,\widehat e_{k_0},
\end{equation}
with $0<(1-\theta)s_{\min}\Delta\leq\theta(1-\theta)<1$. In particular, $\lim_{k\to\infty}\LC(U_k)=E_0$, so $\LC_\infty=E_0$ with $\LC_\infty:=\lim_{k\to\infty}\LC(U_k)$ as in Eq.~\eqref{eq:PL-informal} (under Assumption~\ref{assump:A2} this is $\LC(U_\infty)$), and Eq.~\eqref{eq:PL-fp} holds with $\mu=\Delta/2$ for all $k\geq k_0$.
\end{corollary}

\begin{proof}
The quantity $\widehat e_k$ is nonnegative because $E_0$ is the smallest eigenvalue of $H$. Suppose $\widehat e_k\leq\Delta/2$ for some $k\geq k_0$. Proposition~\ref{prop:energy-gap-pl} gives $\|G_k\|_2^2\geq\Delta\,\widehat e_k$, and the descent bound in Eq.~\eqref{eq:descent-fp} with $s_k\geq s_{\min}$ gives
$\widehat e_k-\widehat e_{k+1}\geq(1-\theta)s_k\|G_k\|_2^2\geq(1-\theta)s_{\min}\Delta\,\widehat e_k$,
so $\widehat e_{k+1}\leq\big(1-(1-\theta)s_{\min}\Delta\big)\widehat e_k\leq\widehat e_k\leq\Delta/2$.
By induction from $k=k_0$, the iterates never leave the window, and Eq.~\eqref{eq:energy-rate} follows.
Since $\Delta\leq E_{\max}-E_0\leq2\|H\|_\infty=2\xi$ and $s_{\min}\leq\theta/(2\xi)$, we have $(1-\theta)s_{\min}\Delta\leq\theta(1-\theta)<1$, and $(1-\theta)s_{\min}\Delta>0$ because $s_{\min}$, $\Delta$, and $1-\theta$ are positive.
Hence $\widehat e_k\to0$, that is, $\lim_k\LC(U_k)=E_0$, and Eq.~\eqref{eq:energy-gap-pl} evaluated along the run for $k\geq k_0$ is Eq.~\eqref{eq:PL-fp} with $\mu=\Delta/2$.
\end{proof}

Corollary~\ref{cor:energy-rate} is the precise form of the statement in the main text that the PL inequality holds with $\mu=\Delta/2$ once the energy is within $\Delta/2$ of the ground-state energy.

\subsection{Numerical evidence for the PL condition in some QML tasks}
\label{subsec:numerical-pl}

Unlike pure-state ground-state preparation, the quantum phase
classification and unitary-learning losses are not covered by the
energy-gap argument in Proposition~\ref{prop:energy-gap-pl}.
Establishing a PL inequality for these tasks remains challenging and
calls for further analysis of the loss-landscape geometry in general
Riemannian gradient descent. Here, we instead examine
the PL condition numerically, by estimating PL coefficients over finite
iteration windows. The reported finite-run diagnostics
provide evidence consistent with PL behavior, supporting the mechanism
for exponential decay of the remaining loss described in
Theorem~\ref{thm:transversal-fp}.

Following Appendix~\ref{subsec:baseline-fp}, we write
$\widehat{\LC}_k^{(r)}:=\LC(U_k^{(r)})-\LC_\infty^{(r)}$
for the remaining loss in realization $r$. In numerical evaluations,
$\LC_\infty^{(r)}$ is replaced by its numerical estimate by the final iterate loss at $k=k_{\rm max}$. The run terminates when the Schatten 2-norm of the gradient is less than floating-point error. We define the
numerical PL coefficient by
\begin{equation}
\label{eq:empirical-pl-ratio}
    \mu_r
    :=\min_{k\in I_r}
    \frac{\|G_k^{(r)}\|_2^2}
    {2\widehat{\LC}_k^{(r)}}.
\end{equation}
where $I_r:=\{k:0\leq k< k_{\rm max}\}$. Note that the final iterate is excluded because its estimated remaining loss vanishes by construction. The plotted coefficient is the mean of five such estimates,
$\bar\mu_{\mathrm{emp}}=\frac15\sum_{r=1}^5\mu_r$. Each realization uses independent training data. Specifically, we sample ground states at uniformly random $(J_1,J_2)$ for quantum phase classification and random input states for unitary learning problem, as  described in Section~\ref{sec:canonical-example}. Lastly, we set $U_0=\mathbb{1}$ and $s=1$.

\begin{figure}[!ht]
    \centering
    \begin{minipage}[t]{0.49\textwidth}
        \raggedright(a)\par
        \includegraphics[width=\linewidth]{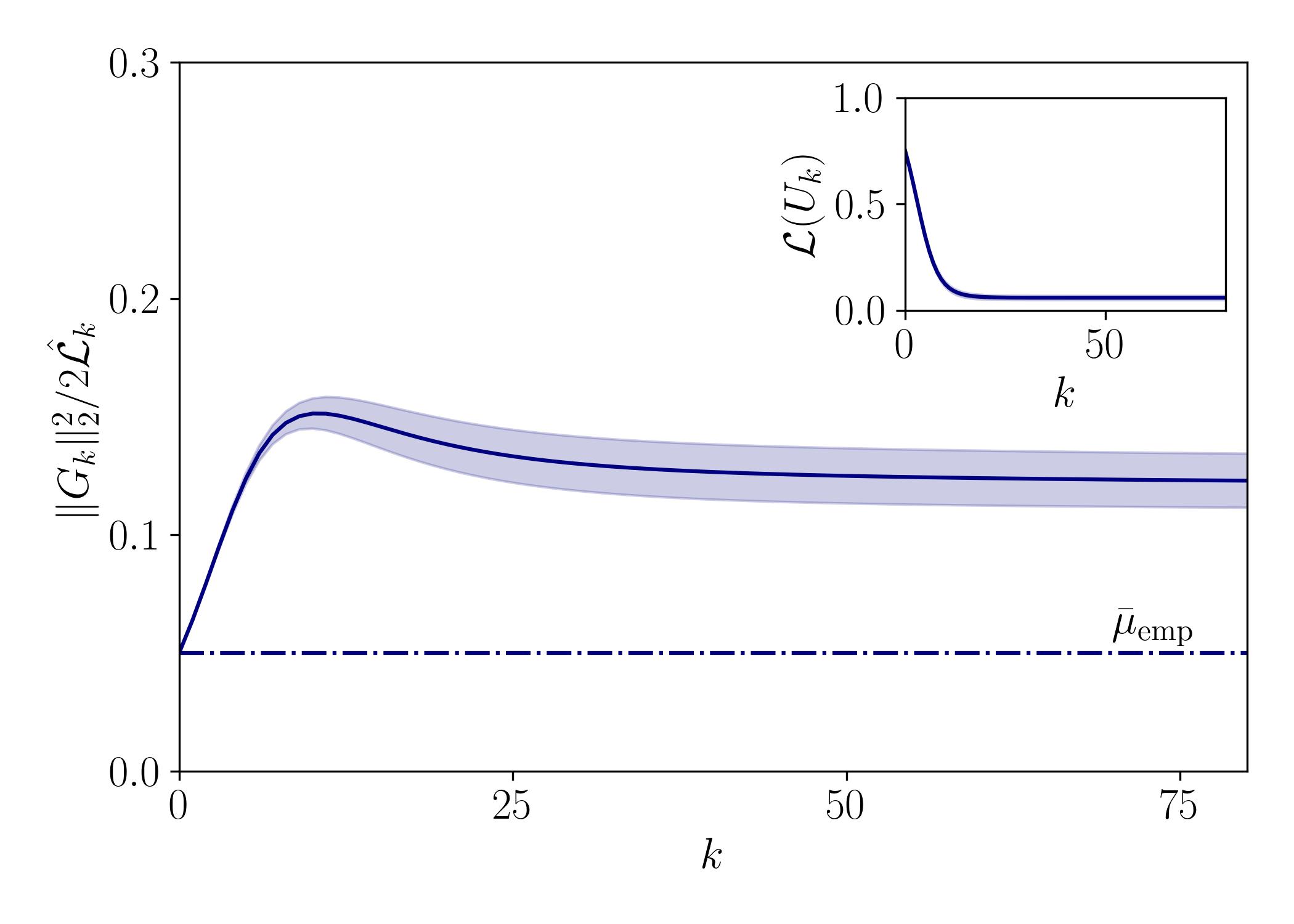}
    \end{minipage}\hfill
    \begin{minipage}[t]{0.49\textwidth}
        \raggedright(b)\par
        \includegraphics[width=\linewidth]{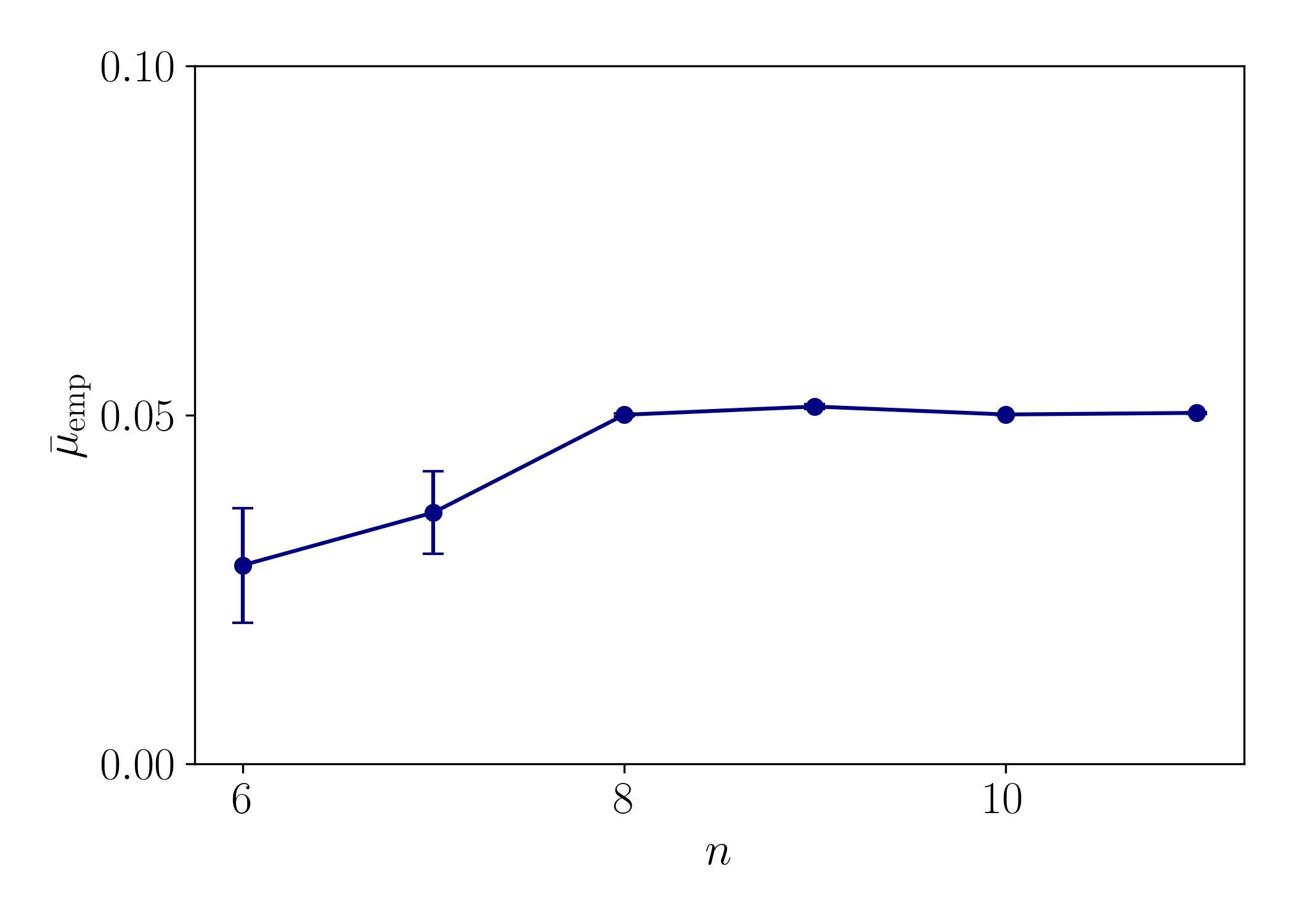}
    \end{minipage}
    \caption{\textbf{Empirical PL coefficients for quantum phase classification.} (a) The average of the ratio $\|G_k\|^2_2/(2\hat{\LC_k})$ against the  iteration $k$ at $n=10$. The shaded region indicates the sample standard deviation and the dash-dotted line displays the mean estimated PL coefficient $\bar{\mu}_{\rm emp}$. Inset represents the average training loss $\LC(U_k)$ against $k$. (b) $\bar{\mu}_{\rm emp}$ against the number of qubits $n$ for $n=4$ to $9$. The error bars indicate the sample standard deviation. The two plots use five realizations of $N=5$ training samples and step size $s=1$. }
    \label{fig:empirical-pl-p1}
\end{figure}

\begin{figure}[!h]
    \centering
    \begin{minipage}[t]{0.49\textwidth}
        \raggedright(a)\par
        \includegraphics[width=\linewidth]{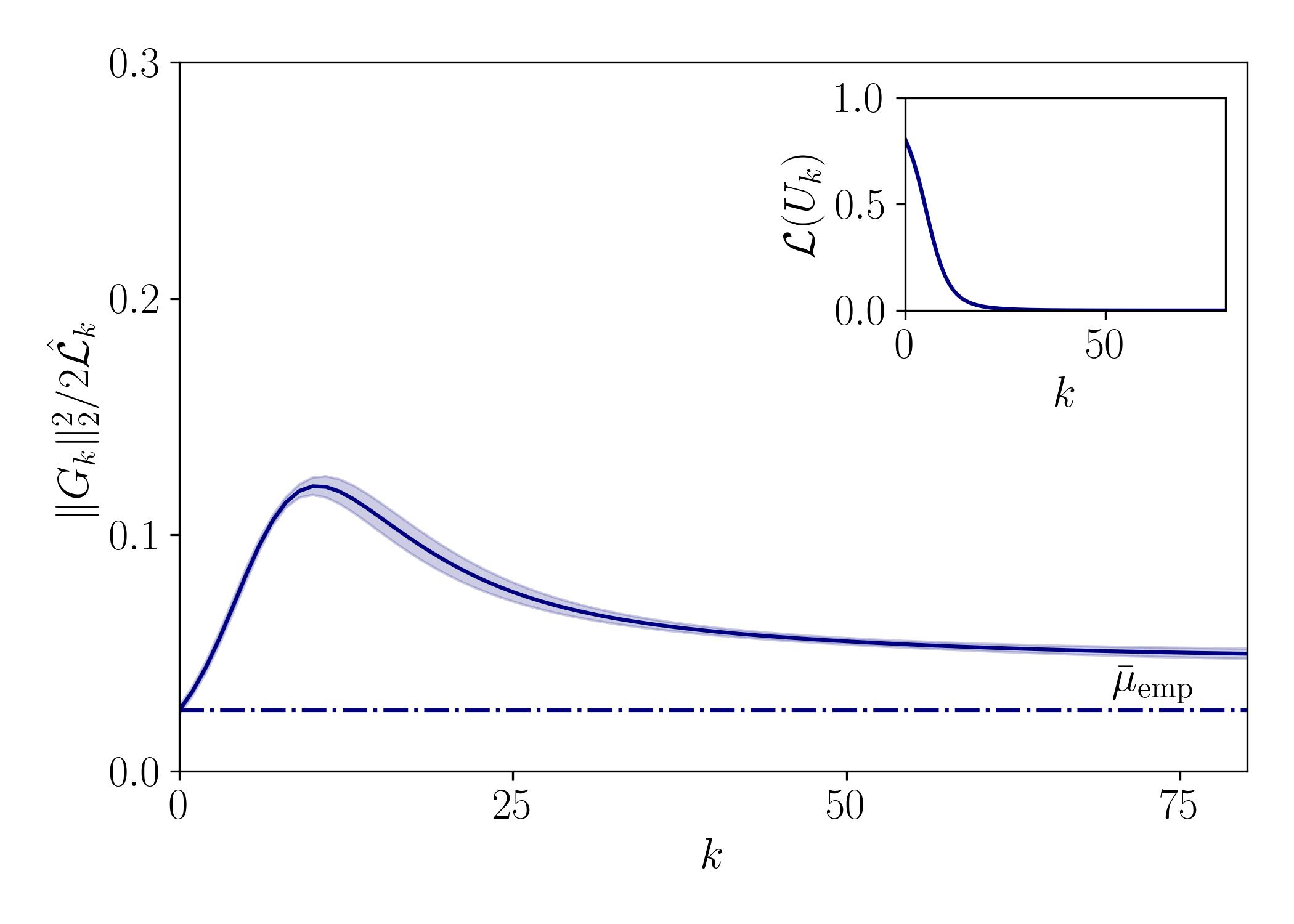}
    \end{minipage}\hfill
    \begin{minipage}[t]{0.49\textwidth}
        \raggedright(b)\par
        \includegraphics[width=\linewidth]{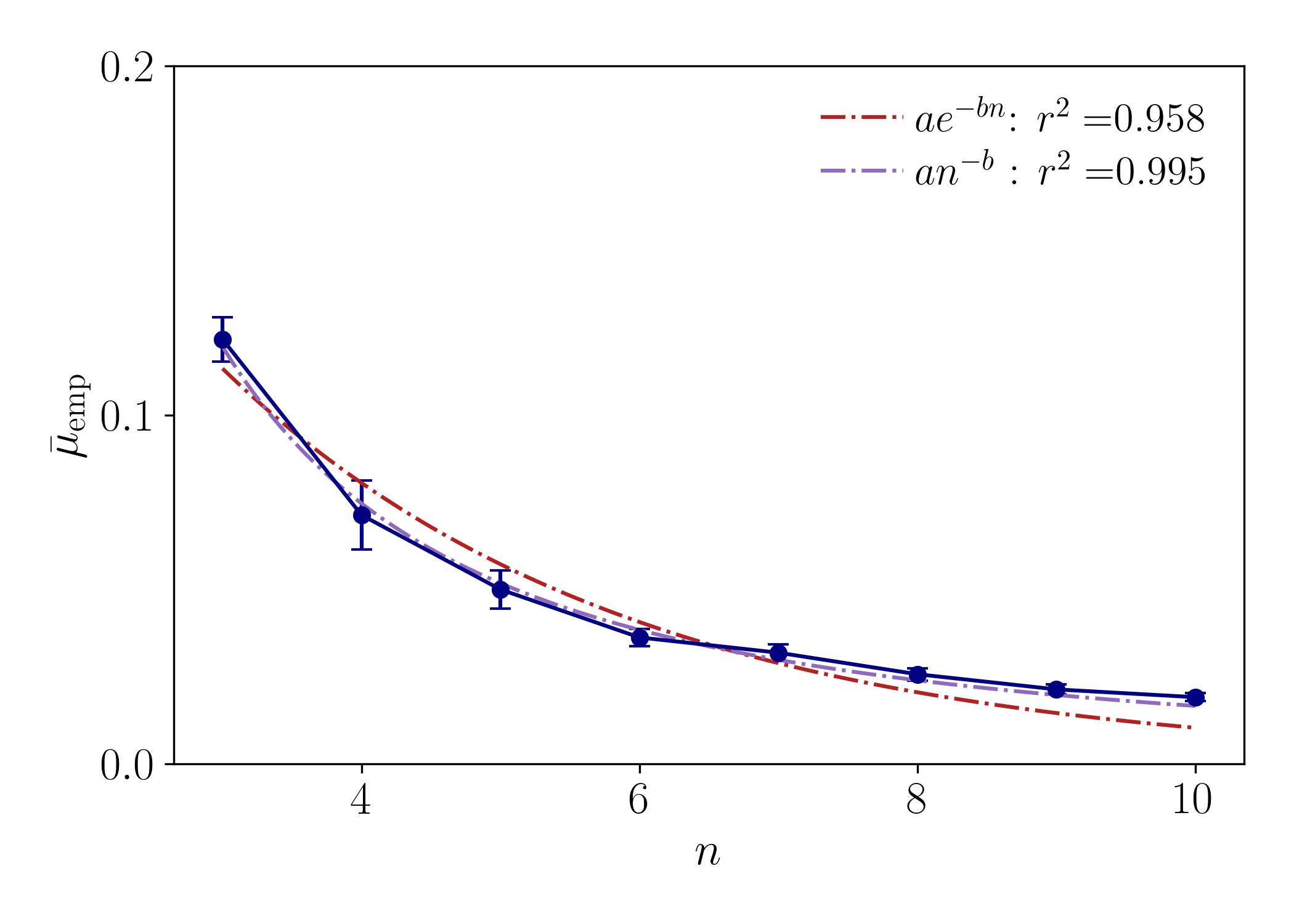}
    \end{minipage}
    \caption{\textbf{Empirical PL coefficients for unitary learning.} (a) The average of the ratio $\|G_k\|^2_2/(2\hat{\LC_k})$ against the the iteration $k$ at $n=8$. The shaded region indicates the sample standard deviation and the dash-dotted line displays the mean estimated PL coefficient $\bar{\mu}_{\rm emp}$. Inset represents the average training loss $\LC(U_k)$ against $k$. (b) $\bar{\mu}_{\rm emp}$ against the number of qubits $n$ for $n=3$ to $10$. The error bars indicate the sample standard deviation. The red dash-dotted curve is the exponential fit
    $ae^{-bn}$ ($r^2=0.958$), and the purple dash-dotted curve is the power-law fit $an^{-b}$ ($r^2=0.995$). The two plots use five realizations of $N=2n$ training samples and step size $s=1$. }
    \label{fig:empirical-pl-p2}
\end{figure}

First, we investigate the dependence of the ratio $\|G_k\|^2_2/(2\hat{\LC_k})$ on the iteration at fixed system size $n=10$ for quantum phase classification (Fig.~\ref{fig:empirical-pl-p1}(a)) and $n=8$ for unitary learning (Fig.~\ref{fig:empirical-pl-p2}(a)). In both tasks, the mean ratio stays away from zero throughout training and attains its minimum at the first iteration ($k=0$), which sets $\bar{\mu}_{\rm emp}$. 
Note that the choice $s=1$ lies outside the step size window of Assumption~\ref{assump:A1} for $c_i=-1$ (see Appendix~\ref{appx:specific-exams}). Nonetheless, the inset of Fig.~\ref{fig:empirical-pl-p1}(a) and Fig.~\ref{fig:empirical-pl-p2}(a) show that the training loss decreases monotonically, suggesting that this window could be relaxed.

Second, we study the dependence of $\bar{\mu}_{\rm emp}$ on the system size. For all $n$, the minimum ratio is again attained at $k=0$. So each point in Fig.~\ref{fig:empirical-pl-p1}(b) and Fig.~\ref{fig:empirical-pl-p2}(b) is determined by the initialization. These figures show the positive mean estimate for all system sizes investigated. For classification, $\bar{\mu}_{\rm emp}$
does not vanish with $n$. For unitary learning, it decreases overall with the sample count $N=2n$ increased along the system size to keep the ratio between the number of data and the effective dimension of the problem fixed (see Section~\ref{sec:unitary-learning}). Although exponential and polynomial fits are likely statistically indistinguishable over the range of system sizes considered, the data are better described by a polynomial decay ($r^2=0.995$) than by an exponential decay ($r^2=0.958$). Taken together, these results suggest that the optimization does not appear to suffer from the curse of dimensionality. 

\subsection{Convergence to a global minimum in specific QML tasks}
\label{subsec:global-QML}

The preceding subsections address loss descent and convergence
rates. We now ask whether the limiting loss is a global minimum.
For QML tasks whose observables share a common structure, such as a binary classification task, this
question can be studied by relating the training dynamics to
double-bracket flows,  whose global convergence properties have been studied in some cases~\cite{gluza2024doublebracket}.
This connection provides the basis for the following sufficient
condition for global convergence, which we justify below.
The required condition is that all observables are obtained from a single Hermitian observable $O$ by rescaling and adding a multiple of the identity:
\begin{equation}
    O_i=\alpha_i O+\beta_i\mathbb{1}\,,
    \qquad \alpha_i,\beta_i\in\mathbb{R}\,.
    \label{eq:common-observable}
\end{equation}

\medskip
\noindent\textbf{Sufficient condition for global convergence.}
For the autonomous loss in Eq.~\eqref{eq:auto-loss-fp}, suppose
the observables satisfy the common-observable structure in Eq.~\eqref{eq:common-observable}
and
the Riemannian gradient descent updates use a sufficiently small constant step size. If Assumption~\ref{assump:A2} holds almost surely for an initial unitary $U_0$ sampled from the Haar measure, then the iterates converge almost surely to a global minimum.
\medskip

To argue the global convergence guarantee in some non-trivial QML tasks, we seek to show that the gradient flow of the autonomous loss in Eq.~\eqref{eq:auto-loss-fp} reduces to a single double-bracket flow, with the Riemannian gradient descent update considered in
Appendix~\ref{appx:general-rg-derive} as its discrete counterpart. This allows us to draw on the convergence properties of double-bracket flows. 

Under the common-observable condition in Eq.~\eqref{eq:common-observable}, we can collect all training states into a
single effective operator. Substituting
Eq.~\eqref{eq:common-observable} into the loss gives
\begin{align}
    \LC(U)
    &=\frac1N\sum_i
      \left[b_i+c_i\Tr\!\left(
      U\rho_iU^\dagger(\alpha_iO+\beta_i\mathbb{1})
      \right)\right] \nonumber\\
    &=\frac1N\sum_i (c_i\beta_i+b_i)
      +\Tr\!\left[
      O\,U\left(\frac1N\sum_i c_i\alpha_i\rho_i\right)
      U^\dagger\right] \nonumber\\
    &=C+\Tr(OX)\,,
    \label{eq:single-observable-loss}
\end{align}
where we used $\Tr(\rho_i)=1$ and defined
$C:=N^{-1}\sum_{i=1}^N (c_i\beta_i+b_i)$,
$\sigma:=N^{-1}\sum_{i=1}^N c_i\alpha_i\rho_i$, and
$X:=U\sigma U^\dagger$.
Likewise, since the identity commutes with every operator,
the gradient becomes
\begin{align}\label{eq:grad_DBF}
    G(U)
    &=-\frac1N\sum_i c_i
      [U\rho_iU^\dagger,\alpha_iO+\beta_i\mathbb{1}]
      \nonumber\\
    &=\left[O,U\left(\frac1N\sum_i
      c_i\alpha_i\rho_i\right)U^\dagger\right]
      =[O,X]\,.
\end{align}
Since $G^\dagger=-G$, we have $\dot U^\dagger=U^\dagger G$ along the gradient flow $\dot U=-G(U)U$, and the continuous-time gradient flow 
therefore induces
\begin{equation}
    \dot X
    =\dot U\sigma U^\dagger+U\sigma\dot U^\dagger
    =-G(U)X+XG(U)
    =[[X,O],X]\,,
    \label{eq:single-DBF}
\end{equation}
which is a single double-bracket flow. The finite-step update correspondingly takes the form
\begin{equation}\label{eq:DBI}
    X_{k+1}
    =e^{-s[O,X_k]}X_ke^{s[O,X_k]}\,.
\end{equation}

The resulting update is a double-bracket iteration, as
defined in Eqs.~(1)--(2) of Ref.~\cite{gluza2024doublebracket} and analyzed, for a constant step size, in Ref.~\cite{doi:10.1137/S0036141092229732}.
When $\sigma$ is a pure-state projector, the continuous flow in
Eq.~\eqref{eq:single-DBF} coincides with normalized imaginary-time
evolution generated by $O$; it approaches the ground eigenspace
of $O$ whenever the initial state has nonzero overlap with that
subspace.
In general, $\sigma$ need not be pure or even
positive semidefinite.

Although the trajectory may converge to different stationary
points, the double-bracket convergence analysis of Ref.~\cite{Helmke1995Optimization} for the flow and its critical-point structure and of Ref.~\cite{doi:10.1137/S0036141092229732} for the discrete iteration, reviewed in Section~1.3 of Ref.~\cite{gluza2024doublebracket}, implies that nonglobal
stationary points are unstable, while global minima are the only stable stationary points.
Together with Assumption~\ref{assump:A2} and the saddle-avoidance
theorem of Ref.~\cite{lee2019firstorderstrictsaddles}, this gives
convergence, almost surely, to a global minimum for sufficiently small constant
step sizes when $U_0$ is drawn initially from a Haar distribution. Initializations whose trajectories converge to nonglobal stationary points form a set of Haar measure zero.

\noindent\textbf{Global convergence in binary classification.} Binary classification is the phase-classification task of Eq.~\eqref{eq:phase-loss-fn} with $Q=2$, as discussed in the main text. With two labels $y_i\in\{0,1\}$ encoded in a single ancilla qubit, the class projectors are $P_0=\mathbb{1}\ot\ketbraq{0}_a$ and $P_1=\mathbb{1}\ot\ketbraq{1}_a=\mathbb{1}-P_0$, and the sample weights are uniform. Writing $R_y:=\frac1N\sum_{i:y_i=y}\rho_i$ for $y\in\{0,1\}$, the loss is the average error probability,
\begin{align}
    \LC(U)
    &=1-\frac1N\sum_{i=1}^N\Tr(U\rho_iU^\dagger P_{y_i})
    \nonumber\\
    &=\Tr R_0+\Tr\!\left[P_0U(R_1-R_0)U^\dagger\right]\,,
\end{align}
where we used $P_1=\mathbb{1}-P_0$ and $\Tr R_0+\Tr R_1=1$. This is Eq.~\eqref{eq:single-observable-loss} with $O=P_0$ and $\sigma=R_1-R_0$, which is Hermitian but in general neither pure nor positive semidefinite, so the sufficient condition above applies.
The preceding argument therefore
gives almost-sure convergence to the global minimum for an initialization $U_0$ drawn from Haar measure.

\section{Implementation of the algorithm}
\label{appx:resource}

In this section we analyze the resource cost required for the implementation of the Riemannian gradient descent approach to QML training from Section~\ref{sec:algo} under the condition in Theorem~\ref{thm:sufficient-cond-autonomous-qml}. In particular, we need to implement the unitary at the $k$-th iteration step which is defined recursively
\begin{equation}
\label{eq:appx-update-rule}
    U_{k} = e^{-s_{k-1}\rg_{k-1}}U_{k-1}\;\;,
\end{equation}
where $\rg_k=-\frac{1}{N}\sum_{i=1}^Nc_i[U_k\rho_iU_k\ad ,O_i]$.
We will do so through queries to three types of unitaries,
\begin{enumerate}
    \item $e^{-it\rho_i}$ which is responsible for taking into account the training data set,
    \item $e^{-itO_i}$ which guides the protocol towards minimizing the cost function,
    \item the initial unitary $U_0$ and its inverse $U_0\ad$ which implements the initial knowledge about the problem.
\end{enumerate} 
As the resource cost, we study the query complexity, i.e., the total number of appearances of any of the unitaries $e^{-it\rho_i}$, $e^{-itO_i}$, $U_0$ or $U_0\ad$ in the compiled circuit for the unitary update in Eq.~\eqref{eq:update_rule}. We denote the total number of $e^{-it\rho_i}$ and $e^{-itO_i}$ by $\Nora\k$ and the number of calls to $U_0$ and $U_0\ad$ together by $\NU$, assuming that each has an equivalently efficient implementation. One could argue that if $O_i = Z_i$ then for near-term platforms this is an overcounting because the implementation of $e^{-it\rho_i}$ is more costly.
However, the result of our analysis is an exponential scaling and such insights do not change it to polynomial in the number of Riemannian gradient descent steps.

A resource cost analysis requires controlling two types of errors, both measured in operator norm, which we denote by $\|\cdot\|=\|\cdot\|_\infty$ throughout this appendix.  
\begin{enumerate}[(i)]
    \item The error of approximating the update in the direction of the Riemannian gradient $\rg_k$ by the implementation proposed in Section~\ref{sec:algo}. We define it as
    \begin{align}
        \vep_k:=\|e^{-s_k\rg_k}-W_k\|\;\;.    
    \end{align}
    The implementation unitary $W_k$ is defined in Appendix~\ref{appx:per-step-err} and the error stems from two product formula approximations, namely the Lie-trotter product formula and the group commutator product formula. 
    \item The error accumulated throughout the iteration by taking the approximation $\vep_k$ in each preceding step which we define as 
    \begin{equation}
    \toterr_k := \|U_k-\tilde{U}_k\|\;\;,
    \end{equation} 
    where we compare the ideal target unitary $U_k$ of the Riemannian gradient descent approach to QML training with the unitary $\tilde{U}_k$ which follows the double-bracket compilation protocol in Section~\ref{sec:algo} and is defined recursively through the use of the approximation to the Riemannian gradient descent steps
    \begin{equation}
    \tilde{U}_{k} = W_{k-1}\tilde{U}_{k-1}\;\;.
    \end{equation} 
\end{enumerate}
Then, we count the total oracle cost in Appendix~\ref{appx:oracle-count}. We note that the entire analysis is stated in $\infty$-norm (i.e. $\|\cdot\|=\|\cdot\|_{\infty}$). Furthermore, we recall a data-dependent constant of Theorem~\ref{thm:auto_RG_learning_guarantee},
\begin{equation}
\label{eq:data-const}
    \xi:=\frac{1}{N}\sum_{i=1}^N|c_i|\|O_i\|\;\;.
\end{equation}

\subsection{Local Error $\vep_k$}
\label{appx:per-step-err}

In this section, we bound the incurred error from compiling the exponentiation of the Riemannian gradient, $\rg=-\frac{1}{N}\sum_ic_i[\rho_i,O_i]$, given generic pairs $\{\rho_i,O_i\}$ of a density matrix and a Hermitian operator. Throughout this section, we write 
\begin{equation}
    \t_i = \frac{|c_i|s}{N}\;,\quad\th_i=\frac{\t_i}{r}\;,\quad \tilde{O}_i={\rm sign}(c_i)O_i\;,\quad A_i:=[\rho_i,\tilde{O}_i]\;\;,
\end{equation}
so that $e^{-sG}=e^{\sum_{i=1}^N\t_iA_i}$. Product of operators are ordered as $\Pi_{i=1}^NX_i:=X_N\cdots X_1$. This compilation relies on two procedures: the Lie-trotter product formula to approximate $e^{\sum_{i=1}^N\t_iA_i}$ by $(\prod_{i=1}^Ne^{\th_iA_i})^r$ with $r$ Lie-Trotter steps and an order-$m$ group commutator product formula to implement each $e^{\th_i A_i}$ via 
\begin{equation} \label{eq:group_approx_gates}
    S_i^{(m)}(\th_i):=e^{i\a_1\sqrt{\th_i}\tilde{O}_i}e^{i\a_2\sqrt{\th_i}\rho_i}\cdots e^{i\a_{N_m-1}\sqrt{\th_i}\tilde{O}_i}e^{i\a_{N_m}\sqrt{\th_i}\rho_i}\;\;.
\end{equation}
The number of exponentials $N_m$ depends on the chosen formula and parameter counting suggests $N_m\in\Omega(2^m/m)$~\cite{chen2022efficient, casas2025approximating}. Combining the two product formulas, the compiled circuit for one Riemannian gradient descent update is 
\begin{equation}
\label{eq:appx-W-def} W:=\left(\prod_{i=1}S_i^{(m)}(\th_i)\right)^r\;\;,
\end{equation}
whose error $\|e^{-s\rg}-W\|$ is bounded in Theorem~\ref{thm:appx-per-step-err} below.

We begin by introducing Lemma~\ref{lemma:tele}, a core mathematical tool used extensively in our bound. We then separate the per-step error into two contributions: the Lie-Trotter error in Lemma~\ref{lemma:trott_err} and the order-$m$ group commutator error in Lemma~\ref{lemma:GC_err}. Finally, we combine these results to establish the per-step error in Theorem~\ref{thm:appx-per-step-err}. Note that in Lemmas~\ref{lemma:trott_err}-~\ref{lemma:GC_err} and Theorem~\ref{thm:appx-per-step-err}, $\tilde{O}_i$ is interchangeable with $O_i$, because $\|O_i\|=\|\tilde{O}_i\|$.

\begin{lemma}[Telescoping bound of product of unitaries]
\label{lemma:tele}
For unitary matrices $U_1,U_2,\cdots,U_r,V_1,V_2,\cdots,V_r$ and unitary invariant norm $\|\cdot\|$, we have
    \begin{equation}
    \left\|\prod_{i=1}^rU_i-\prod_{i=1}^rV_i\right\|\leq \sum_{i=1}^r\left\|U_i-V_i\right\|\;\;.
    \end{equation}
\end{lemma}
\begin{proof}
    Consider the telescoping expansion
    \begin{align}
        \prod_{i=1}^rU_i-\prod_{i=1}^rV_i&=\prod_{i=1}^r U_i-\left(\prod_{i=2}^{r}U_i\right)V_1+\left(\prod_{i=2}^{r}U_i\right)V_1-\dots -U_r\prod_{i=1}^{r-1}V_i+U_r\prod_{i=1}^{r-1}V_i-\prod_{i=1}^rV_i\;\;,\nn
        &=\sum_{k=1}^{r}\left(\left(\prod_{i=k+1}^{r}U_i\right)\big(U_k-V_k\big)\left(\prod_{i=1}^{k-1}V_i\right)\right)\;\;.
    \end{align}
    Using the triangle inequality, together with the fact that the norm is unitary invariant, we arrive at 
    \begin{equation}
        \norm{\prod_{i=1}^rU_i-\prod_{i=1}^rV_i} \leq \sum_{k=1}^{r}\norm{\left(\prod_{i=k+1}^{r}U_i\right)\big(U_k-V_k\big)\left(\prod_{i=1}^{k-1}V_i\right)} = \sum_{k=1}^{r}\|U_k-V_k\|\;\;.
    \end{equation}
\end{proof}

\begin{lemma}[Lie-Trotter Error]
\label{lemma:trott_err}
Let $r$ be the number of the Lie-Trotter steps, $\th_i=\t_i/r\geq0$, $A_i=[\rho_i,O_i]$ be an anti-Hermitian matrix, and $\rho_i$ be a density matrix. We have
    \begin{equation}
    \norm{e^{\sum_{i=1}^N \t_i A_i}-\left(\prod_{i=1}^Ne^{\th_iA_i}\right)^r}\leq \frac{4}{r}\sum_{i<j}\t_i \t_j  \|O_i\|\|O_j\|\;\;.
    \end{equation}
\end{lemma}
\begin{proof}
    Consider
    \begin{align}
        \norm{|e^{\sum_{i=1}^N\th_iA_i}-\prod_i^N e^{\th_iA_i}}&=\norm{e^{\sum_{i=1}^NB_i}-\prod_i^Ne^{B_i}}\;\;,\nn 
        &\leq \frac{1}{2}\sum_{i<j}\|[B_j,B_i]\|\;\;,\nn
        &=\frac{1}{2}\sum_{i<j}\th_i\th_j\|[A_i,A_j]\|\;\;,\nn 
        &\leq \sum_{i<j}\th_i\th_j\|A_i\|\|A_j\| \;\;,\nn
        &\leq 4\sum_{i<j}\th_i \th_j \|\rho_i\|\|\rho_j\| \|O_i\|\|O_j\|\;\;,
    \end{align}
    First, we use $B_i:=\th_i A_i$. Then, we use the first-order Lie-Trotter formula~\cite{childs2021theory} to arrive at the first inequality. 
    We pull out the factors $\th_i$ from $B_i$ to get the third line. After that, we use the sub-multiplicative property of the norm with the triangle inequality to obtain the bound of the commutator, $\|[A,B]\| \leq 2\|A\|\|B\|$, for the last two lines. 
    
    We note that $e^{\sum_{i=1}^N\t_iA_i}=(e^{\sum_{i=1}^N\th_iA_i})^r$. Applying Lemma~\ref{lemma:tele} with $U_l=e^{\sum_{i=1}^N\th_iA_i}$ and $V_l=\prod_{i=1}^Ne^{\th_iA_i}$ for all $l=1,\dots,r$ together with the above inequality for $\th=\t_i/r$ and $\|\rho\|\leq1$, the claim follows.
\end{proof}

\begin{lemma}[Group Commutator Errors]
\label{lemma:GC_err}
    For $S_i$ defined in Eq.~\eqref{eq:group_approx_gates} and anti-Hermitian matrices $A_i=[\rho_i,O_i]$, there is a constant $C_i^{(m)}$ depending on $m$ and on the chosen product formula such that 
    \begin{equation}
        \left\|\left(\prod_i^Ne^{\theta_i A_i}\right)^r-\left(\prod_i^N S_i^{(m)}(\th_i)\right)^r\right\|  \leq r\sum_{i=1}^NC^{(m)}_{i}{\th_i}^{\frac{m+1}{2}}\;\;,
    \end{equation}
Specifically, for $m=2$,
\begin{equation}
C_i^{(2)}=\|[\rho_i,[\rho_i,O_i]]\|+\|[O_i,[O_i,\rho_i]]\|\;\;.
\end{equation}
\end{lemma}
\begin{proof}
We start by applying Lemma~\ref{lemma:tele} 
\begin{align}
    \left\|\left(\prod_i^Ne^{\theta_iA_i}\right)^r-\left(\prod_i^NS_i^{(m)}(\th_i)\right)^r\right\| &\leq r\sum_{i=1}^N\norm{e^{{\th_iA_i}}-S_i^{(m)}(\th_i)}\leq r\sum_{i=1}^N C_i {\th_i}^{(m+1)/2}\;\;.
\end{align}
The second inequality holds for all $\th_i\geq0$. For $\th_i\leq 1$, it follows by bounding the remainder of the Taylor expansion of $e^{\th_iA_i}-S_i^{(m)}(\th_i)$, which starts at order $\th_i^{(m+1)/2}$. For $\th_i\geq 1$, the unitarity of both terms gives $\|e^{\th_iA_i}-S_i^{(m)}(\th_i)\|\leq 2\leq 2\th^{(m+1)/2}$ from the triangle inequality, so that the constant $2$ suffices for this regime. Note that the relevant regime in practice is $\th\leq1$, where $C_i^{(m)}$ depends on the chosen product formula. Explicit forms of such constants are provided in Ref.~\cite{childs2013product,chen2022efficient}.

For the case $m=2$, Lemma~9 of Ref.~\cite{gluza2024doublebracket}, applies with anti-Hermitian $iO_i$ and $i\rho_i$, gives that 
\begin{equation}
\left\|\left(\prod_i^Ne^{\theta_iA_i}\right)^r-\left(\prod_i^NS_i^{(2)}(\th_i)\right)^r\right\|\leq r\sum_{i=1}^N{\th_i}^{\frac{3}{2}}\big(\|[\rho_i,[\rho_i,O_i]]\|+\|[O_i,[O_i,\rho_i]]\|\big)\;\;,
\end{equation}
where $S_i^{(2)}(\th_i) = e^{i\sqrt{\th_i}O_i}e^{i\sqrt{\th_i}\rho_i}e^{-i\sqrt{\th_i}O_i}e^{-i\sqrt{\th_i}\rho_i}$.
\end{proof}

\begin{theorem}[Per-step Error Bound]
\label{thm:appx-per-step-err}

Given dataset $\{\rho_i, O_i\}_{i=1}^N$, a step size $s$, and the Riemannian gradient $\rg=-\frac{1}{N}\sum_ic_i[\rho_i,O_i]$. Let $W$ be the compiled circuit of Eq.~\eqref{eq:appx-W-def} with the Lie-Trotter step $r$ and an order-$m$ group commutator product formula. Assume $|c_i|\leq1$ for all $i$. Then the error from compiling the update in Eq.~\eqref{eq:update_rule} is bounded as 
    \begin{align}
    \|\exp(-s\rg)-W\|
    &\leq \frac{2s^2\xi^2}{r}+\frac{C_{\rm max}^{(m)}s^{\frac{m+1}{2}}}{(rN)^{\frac{m-1}{2}}} \;\;.
\end{align}
where $\xi$ is defined in Eq.~\eqref{eq:data-const} and
$C^{(m)}_{\rm max}:=\max_i|c_i|^{\frac{m+1}{2}}C_i^{(m)}$ with the constant $C_i^{(m)}$ of Lemma~\ref{lemma:GC_err}. 
\end{theorem}
\begin{proof}
We start by considering
\begin{align}
    \norm{\exp(-s\rg)-W} &=\norm{\exp(\frac s N\sum_ic_i[\rho_i,O_i])-W}\;\;, \nn 
    &\leq \norm{e^{\sum_{i=1}^N\t_iA_i}-\left(\prod_{i=1}^Ne^{\th_iA_i}\right)^r}+\norm{\left(\prod_{i=1}^Ne^{\theta_iA_i}\right)^r-\left(\prod_{i=1}^NS_i^{(m)}(\th_i)\right)^r}\;\;,\nn
    &\leq \frac{4s^2}{rN^2}\sum_{i< j}|c_i||c_j|\|O_i\|\|O_j\|+\frac{(s/N)^{(m+1)/2}}{r^{(m-1)/2}}\sum_{i=1}^N|c_i|^{(m+1)/2}C_i^{(m)}\;\;.
\end{align}
Here we use triangle inequality to get the second line with $W=\left(\prod_{i=1}^N S_i^{(m)}(\th_i)\right)^r$ and $A_i={\rm sign}(c_i)[\rho_i,O_i]$. Then, the third line follows from Lemma~\ref{lemma:trott_err} and Lemma~\ref{lemma:GC_err} with $\|\tilde{O}_i\|=\|O_i\|$. 

Furthermore, we bound the first term as follows
\begin{equation}
    \sum_{i< j}|c_i||c_j|\|O_i\|\|O_j\|\leq \frac{1}{2}\left(\sum_{i=1}^N|c_i|\|O_i\|\right)^2=\frac{N^2\xi^2}{2}\;\;,
\end{equation}
where we use $(\sum_{i}a_i)^2=\sum_ia_i^2+2\sum_{i<j}a_ia_j$ for $a_i\in\Rbb$. We bound the second term via 
\begin{equation}
    \sum_{i=1}^N|c_i|^{\frac{m+1}{2}}C_i^{(m)}\leq N\max_{i}\left(|c_i|^{\frac{m+1}{2}}C_i^{(m)}\right)=NC_{\rm max}^{(m)}\;\;,
\end{equation}
using $|c_i|\leq 1$ and definition of $C_{\rm max}^{(m)}$. This completes the proof.
\end{proof}

\subsection{Accumulated Error $\Delta_k$}
\label{appx:acc-err}

Given the trained unitary at the $k$-th iteration $U_k$ and its compiled circuit $\tilde{U}_k$, we define the accumulated error to implement $U_k$
\begin{equation}
\label{eq:accum-err}
    \toterr_k := \|U_k-\tilde{U}_k\|\;\;.
\end{equation} 
According to the protocol in Section~\ref{sec:algo}, this is the accumulated error from approximating $e^{-s_{k-l}\tilde{\rg}_{k-l}}$ with $W_{k-l}$ (see analysis in Appendix~\ref{appx:per-step-err}) for $l=1,\dots,k$, which generate $\tilde{U}_1,\dots,\tilde{U}_k$ from $U_0$. Crucially, each $\tilde{\rg}_j$ for $j\geq1$ is itself an approximation as it is constructed because it is evaluated at the compiled unitary $\tilde{U}_{j}$ rather than at the exact $U_{j}$. Only the first step is exact, $\tilde{G}_0=G_0$, since $\tilde{U}_0=U_0$. In other words, we apply
\begin{equation}
    \tilde{U}_{k} = W_{k-1}\tilde{U}_{k-1}\;\;,
\end{equation}
starting from $\tilde{U}_1=W_0U_0$, to approximate the ideal update rule in Eq.~\eqref{eq:appx-update-rule}. This recurrence relation leads to the following theorem:

\begin{theorem}[Accumulated Error Bound]
\label{thm:appx-accum-err}
    Let $\{\rho_i,O_i\}_{i=1}^N$ be a training dataset, $U_0$ be the initial unitary, and $s_k$ be the step size at iteration $k$. Let $W_k$ be the compiled circuit approximating $e^{-s_k\tilde{G}_k}$, where $\tilde{G}_k=-\frac{1}{N}\sum_{i=1}^Nc_i[\tilde{\rho}_i\k,O_i]$ is the Riemannian gradient from the compiled state $\tilde{\rho}_i\k=\tilde{U}_k\rho_i\tilde{U}_k\ad$. Then $\toterr_k$ of Eq.~\eqref{eq:accum-err} satisfies
    \begin{equation}
    \toterr_k \leq \sum_{j=0}^{k-1}\prod_{a=j+1}^{k-1}(4\xi s_{a}+1)\vep_{j}\;\;,
    \end{equation}
    where $\xi$ is defined in Eq.~\eqref{eq:data-const} and $\vep_j$ is the per-step error at the iteration $j$ defined as 
    \begin{equation}
        \vep_j := \| e^{-s_j\tilde{\rg}_j}-W_{j}\|\;\;.
    \end{equation}
    as in Appendix~\ref{appx:per-step-err} with the quantum state $\rho_i$ replaced by the compiled state at the $j$-th iteration $\tilde{\rho}_i^{(j)}$ and the step size $s$ replaced by the iteration-dependent step size $s_j$. 
\end{theorem}
\begin{proof}
Using a telescoping sum following Ref.~\cite{wright2026double}, we have
\begin{equation}
    U_k-\tilde{U}_k = e^{-s_{k-1}\rg_{k-1}}(U_{k-1}-\tilde{U}_{k-1}) +(e^{-s_{k-1}\rg_{k-1}}-e^{-s_{k-1}\tilde{\rg}_{k-1}})\tilde{U}_{k-1}+(e^{-s_{k-1}\tilde{\rg}_{k-1}}-W_{k-1})\tilde{U}_{k-1}\;\;,
\end{equation}
where $W_{k}$ is a compiled circuit for updating the unitary from iteration $k$ to $k+1$ and $\tilde{\rg}_k=-\frac1 N\sum_ic_i[\tilde{U}_k\rho_i\tilde{U}\ad_k,O_i]$. Here, the sources of error come from the difference between
\begin{enumerate}
    \item the real $(k-1)^{th}$-iterating unitary and the synthesized unitary,
    \item the ${(k-1)}^{th}$ discretized update along the Riemannian gradient direction for $U$ and $\tilde{U}$,
    \item the discretized update along the Riemannian direction and its synthesized circuit. 
\end{enumerate}

Next, we apply the triangle inequality to arrive at 
\begin{align}
    \norm{U_k-\tilde{U}_k}&\leq \norm{U_{k-1}-\tilde{U}_{k-1}} +\norm{e^{-s_{k-1}\rg_{k-1}}-e^{-s_{k-1}\tilde{\rg}_{k-1}}} + \norm{e^{-s_{k-1}\tilde{\rg}_{k-1}}-W_{k-1}} \;\;,\nn 
    &\leq \toterr_{k-1}+s_{k-1}\|\rg_{k-1}-\tilde{\rg}_{k-1}\|+\vep_{k-1} \;\;,\nn 
    &\leq (4s_{k-1}\xi+1)\toterr_{k-1} + \vep_{k-1}\;\;,
\end{align}
where we use the unitary invariance of the operator norm to obtain the first line, i.e., $\|AU\|=\|UA\|=\|A\|$ for a unitary $U$. Then, we apply $\|e^{sA}-e^{sB}\|\leq s\|A-B\|$ to get the second line (see Eq.~(116) in Lemma~9 in Ref.~\cite{gluza2024doublebracket}). We use the notation $\vep_{k-1}$ as the local error in Appendix~\ref{appx:per-step-err} given $\rho_i=\tilde{\rho_i}^{(k-1)}$ and $s=s_{k-1}$. Finally, the last line is obtained from 
\begin{align}
    \|\rg_{k}-\tilde{\rg}_k\|&=\norm{\left(-\frac{1}{N}\right)\sum_{i=1}^Nc_i[U_k\rho_i U\ad_k-\tilde{U}_k\rho_i \tilde{U}\ad_k,O_i]}\;\;,\nn &\leq \frac{2}{N}\sum_{i=1}^N|c_i|\| U_k\rho_i (U\ad_k -\tilde{U}\ad_k)+(U_k-\tilde{U}_k)\rho_i \tilde{U}_k\ad\|\| O_i\| \;\;, \nn 
    &\leq \frac{4}{N}\|U_k-\tilde{U}_k\|\sum_{i=1}^N|c_i|\|O_i\|=4\xi\D_k \;\;,
\end{align}
using the bound of the commutator $\|[A,B]\|\leq2\|A\|\|B\|$, then submultiplicative and unitary invariant of the operator norm, $\|U\rho A\|\leq \|U\rho\|\|A\|\leq\|A\|$), together with $\|A\ad\|=\|A\|$, and $\|\rho\|\leq 1$ for any density matrix $\rho$.

Recursively, we have
\begin{equation}
    \toterr_k \leq \prod_{j=0}^{k-1}(4\xi s_j+1)\toterr_0 + \sum_{j=0}^{k-1}\prod_{a=j+1}^{k-1}(4\xi s_{a}+1)\vep_{j} =\sum_{j=0}^{k-1}\prod_{a=j+1}^{k-1}(4\xi s_{a}+1)\vep_{j}\;\;,
\end{equation}
where $\toterr_0=0$ because the compiled circuit is initialized with the exact initial unitary.
\end{proof}

\subsection{Number of Oracles (Proof of Theorem~\ref{thm:resource})}
\label{appx:oracle-count}

To count the number of oracle queries ($e^{-it\rho_i}$, $e^{-it O_i}$, $U_0$, and $U_0\ad$), we proceed in three steps. First, we count the number of oracles used to compile a single Riemannian gradient update and unroll the recursion over the desired $k$ iterations by leaving the Lie-Trotter steps $r$ and the order $m$ of the group commutator product formula as free parameters. Second, we determine the number of Lie-Trotter step $r_{\rm min}$ required to achieve a uniform per-step error $\vep>\vep_j$ for all $j$ for a given $m$. Third, we tie the per-step error $\vep$ to the target precision $\ep$ for compiling $U_k$, through Theorem~\ref{thm:appx-accum-err}. Substituting $r_{\min}$ back to the oracle counts in the first step yields the oracle scaling for the protocol described in Section~\ref{sec:algo}.

Suppose $e^{-it\rho_i}$ and $e^{-it O_i}$ are oracles which can be exactly implemented for any sample index $i$ and evolution time $t$ and the number of Lie-Trotter step $r$ is used at every iteration. Then, the number of oracles calls required for implementing the Riemannian gradient update can be counted from the unrolled form of Eq.~\eqref{eq:appx-W-def} (for readability we write $U_k$ for $\tilde{U}_k$) 
\begin{equation}
    U_{k+1} = \left(\prod_{i=1}^N e^{i\a_1\sqrt{\th_i}\tilde{O}_i}U_k e^{i\a_2\sqrt{\th_i}\rho_i}U_k\ad \cdots e^{i\a_{N_m-1}\sqrt{\th_i}\tilde{O}_i}U_k e^{i\a_{N_m}\sqrt{\th_i}\rho_i}U_k\ad \right)^rU_k\;\;,
\end{equation}
where $\tilde{O}_i={\rm sign}(c_i)O_i$ and $\th_i=|c_i|s_k/Nr$, and we use $U_ke^{it\rho_i}U_k\ad=e^{it\rho_i\k}$. This leads to $(rNN_m+1)$ queries to $U_k$ or $U_k\ad$ and $N_mNr/2$ queries to each of $e^{-it\rho_i}$ and $e^{-it O_i}$. Given $U_0$, the total number of oracles calls, counting both $e^{-it\rho_i}$ and $e^{-itO_i}$, to compile $U_k$ is given by the following recursion relation:
\begin{align}
\label{eq:n_ora}
    \Nora\k &= (rNN_m+1)\Nora^{(k-1)} + N_mNr\;\;,\nn 
    &= (rNN_m+1)((rNN_m+1)\Nora^{(k-2)}+rNN_m)+rNN_m\;\;, \nn 
    &= \sum_{j=0}^{k-1}(rNN_m+1)^j(rNN_m)\;\;,\nn 
    &= rNN_m\left(\frac{(rNN_m+1)^k-1}{rNN_m}\right) \;\;,\nn
    &= (rNN_m+1)^k-1\;\;.
\end{align}
Here, we use $\Nora^{(0)}=0$ to get the third line since $U_0$ is given and consumes no $e^{it\rho_i}$ and $e^{itO_i}$. Then, the fourth line comes from the geometric summation. Eq.~\eqref{eq:n_ora} expresses the costs for fixed $r$ and $m$. 

Next, the number of $U_0$ and $U_0\ad$ queries, counted together in $\NU$, is 
\begin{align}
\label{eq:n_U0}
    \NU\k&=(rNN_m+1)\NU^{(k-1)}  \nn  
    &= (rNN_m+1)^k
\end{align}
using $\NU^{(0)}=1$. Combining this analysis with the results in previous sections, we arrive at the following theorem.

\begin{theorem}
    Consider the setting of Theorem~\ref{thm:sufficient-cond-autonomous-qml} with the training data ${\rho_i,O_i}$ and $|c_i|\leq1$. Let $s$ be a uniform upper bound on the step size $s_k<s$ for all $k$ and let $\xi$ be as in Eq.~\eqref{eq:data-const}. Suppose the update in Eq.~\eqref{eq:update_rule} is compiled by the Lie-Trotter product formula with $r$ steps and an order-$m$ group commutator formula for $m\geq2$ with $N_m$ exponential terms, as in Eq.~\eqref{eq:appx-W-def}. Then, the number of oracle calls $\Nora\k$ for  $e^{-it\rho_i}$ and $e^{-itO_i}$ and the number of oracle calls $\NU$ for $U_0$ and $U_0\ad$ that suffice to guarantee the total error $\|U_k-\tilde{U}_k\|\leq \ep$ satisfy 
    \begin{equation}
        \Nora\k,\NU\k \leq [s\cdot B_m(\ep)]^k\;\;,
    \end{equation}
    where 
    \begin{equation}
        B_m(\ep)\in\OC\left(\max\left(\frac{N(4\xi s+1)^k}{\eptt},\left[\frac{(4\xi s+1)^k}{\epgc}\right]^\frac{2}{m-1}\right)\right)\;\;,
    \end{equation}
   and $\ep=\epgc+\eptt$ is any split of error budget between the Lie-Trotter and the group commutator product formulas. Choosing $\epgc=\eptt=\ep/2$ gives the statement in Theorem~\ref{thm:resource}.
\end{theorem}
\begin{proof}
We begin by considering the number of oracle calls required to achieve the precision $\ep$. We split the errors from two approximating mechanisms as
\begin{equation}
    \ep=\eptt+\epgc\quad{\rm and}\quad\vep=\ett+\egc\;\;,
\end{equation}
where $\ett$ and $\egc$ are uniform per-step error tolerances needed to achieve the accumulated errors $\eptt$ and $\epgc$, respectively. According to Theorem~\ref{thm:appx-accum-err}, the accumulated error is bounded by
\begin{equation}
    \D_k \leq \sum_{j=0}^{k-1}\prod_{a=j+1}^{k-1}(4\xi s_a+1)\vep_j\leq \left[\frac{(4\xi s+1)^k-1}{4\xi s}\right]\vep\;\;,
\end{equation}
where we apply the uniform bounds $s_a\leq s $ and $\vep_j\leq\vep$ to evaluate the geometric sum. To ensure that $\Delta_k\leq \ep$ for all $k$, it therefore suffices to take
\begin{equation}
\label{eq:appx-resource-vep-to-ep}
    \vep \leq \Lambda_k\ep \;,\quad\Lambda_k:=\frac{4\xi s}{(4\xi s+1)^k-1}\;\;.
\end{equation}

Both terms of the per-step errors bound in Theorem~\eqref{thm:appx-per-step-err} decrease with the number of Lie-Trotter steps $r$. Therefore, we choose $r$ such that each term is upper bounded by its respective tolerance, i.e., $2s^2\xi^2/r\leq \ett$ and $(C_{\rm max}^{(m)}s^{\frac{m+1}{2}}){(rN)^{-\frac{m-1}{2}}} \leq \egc$. This gives lower bounds for the required Lie-Trotter steps, 
\begin{equation}
\label{eq:appx-resource-required}
    \ttt = \frac{2\xi^2s^2}{\ett} \quad{\rm and}\quad \tgc = \frac{1}{N}\left(\frac{C_{\rm max}^{(m)}s^{\frac{m+1}{2}}}{\egc}\right)^{\frac{2}{m-1}}=\frac{s}{N}\left(\frac{C_{\rm max}^{(m)}s}{\egc}\right)^{\frac{2}{m-1}}\;\;.
\end{equation}
Consequently, it is enough to choose
\begin{equation}
    r_{\rm min} = \lceil\max (\ttt, \tgc)\rceil\;\;.
\end{equation}

Next, we bound the total number of oracle queries required to achieve the desired precision via the quantity $R(m):=r_{\rm min}N N_m$ as we find
\begin{equation}
\label{eq:appx-resource-bound-number}
    \Nora\k = (R(m)+1)^k-1\leq (2R(m))^k\quad{\rm and}\quad \NU\k= (R(m)+1)^k\leq (2R(m))^k\;\;,
\end{equation}
since $R(m)\geq 1$ because $r_{\min}\geq1$. Choosing $\ett=\Lambda_k\eptt$ and $\egc=\Lambda_k\epgc$, which satisfies Eq.~\eqref{eq:appx-resource-vep-to-ep}, and inserting into Eq.~\eqref{eq:appx-resource-required}, we obtain

\begin{align}
\label{eq:appx-resource-scaling-require}
    R(m) &\leq NN_m\left(1+\max\left(\frac{2\xi^2s^2}{\Lambda_k\eptt},\frac{s}{N}\left(\frac{C_{\rm max}^{(m)}s}{\Lambda_k\epgc}\right)^{\frac{2}{m-1}} \right)\right)\;\;,\nn 
    &\leq NN_m\max\left(\frac{((4\xi s+1)^k-1)\xi s}{2\eptt},\frac{s}{N}\left(\frac{C_{\rm max}^{(m)}((4\xi s+1)^k-1)}{4\xi \epgc}\right)^{\frac{2}{m-1}} \right)+NN_m\;\;.
\end{align}
Here, we use $s/\Lambda_k =((4\xi s+1)^k-1)/4\xi$ and $\lceil a + b\rceil\leq a+b+1$. The quantities $\xi$, $N_m$ and $C_{\rm max}^{(m)}$ are bounded independently of $N$, $k$, and $\ep$; $N_m$ and $C_{\rm max}^{(m)}$ depend on the order-$m$ and on the chosen product formula. We conclude that 
\begin{equation}
        \Nora\k,\NU\k \leq [s\cdot B_m(\ep)]^k\;\;,
    \end{equation}
    where 
    \begin{equation}
        B_m(\ep)\in\OC\left(\max\left(\frac{N(4\xi s+1)^k}{\eptt},\left[\frac{(4\xi s+1)^k}{\epgc}\right]^\frac{2}{m-1}\right)\right)\;\;.
    \end{equation}
\end{proof}

\paragraph*{\underline{Efficiency regimes.}}
We now derive the scalings stated after Theorem~\ref{thm:resource} in the main text. We split the error budget equally $\eptt=\epgc=\ep/2$ and define $X=(4\xi s+1)^k$. For $m\geq 3$, we have $0<2/(m-1)\leq 1$. In the regime of small target precision $\ep$, $C_{\max}^{(m)}(X-1)/2\xi\ep\geq1$, the second entry of the maximum in Eq.~\eqref{eq:appx-resource-scaling-require} is at most $sC_{\rm max}^{(m)}X/2N\xi \ep$. 

We bound Eq.~\eqref{eq:appx-resource-scaling-require} by the sum of each two entries of the maximum which gives
\begin{equation}
\label{eq:appx-bound-Rm}
    R(m)\leq \frac{sNN_mX(2N(\ep\xi+\xi^2)+C_{\rm max}^{(m)})}{2\ep\xi}\;\;.
\end{equation}
Substituting Eq.~\eqref{eq:appx-bound-Rm} into Eq.~\eqref{eq:appx-resource-bound-number}, we obtain
\begin{equation}
\label{eq:appx-bound-log-Nora}
    \log(\Nora\k)\leq  k \log(\frac{sNN_m(2N(\ep\xi+\xi^2)+C_{\rm max}^{(m)})}{\ep \xi})+k^2\log(4\xi s+1)\;\;,
\end{equation}
and the same bound holds for $\NU\k$. 

We call the protocol efficient if $\Nora\k\in\OC(\poly(n))$, or equivalently $\log \Nora\k\in\OC(\log n)$. Assume that the step size bound $s$, the data constant $\xi$, the constant $C_{\rm max}^{(m)}$, and the target precision $\ep$ are all independent of number of qubits $n$. Since $\xi>0$ and $s>0$, the term $k^2\log(4\xi s+1)$ is $\OC(\log n)$ if and only if $k\in\OC(\sqrt{\log n}$. For $N\in \OC(1)$, the first term in Eq.~\eqref{eq:appx-bound-log-Nora} is $\OC(k)\subseteq\OC(\log(\sqrt{n})$. Hence, for $N\in\OC(1)$, it is suffices that $k\in \OC(\sqrt{\log n})$. If instead $N\in\OC(\poly(n))$, the first term is $k\log N\in \Th (k\log n)$, and requiring $\log(\Nora)\in \OC(\log n)$ forces $k\in\OC(1)$. 

Note that the right-hand side of Eq.~\eqref{eq:appx-bound-log-Nora} grows faster than any multiple of $\log n$ if $k/\sqrt{\log n}\to\infty$, because the $k^2$ term is present in both entries of the maximum, so $k\in\OC(\sqrt{\log n})$ is also the largest growth of $k$ for which the bound of Theorem~\ref{thm:resource} guarantees efficiency. Finally, for $m=2$, the upper bound of the second entry of the maximum in Eq.~\eqref{eq:appx-resource-scaling-require} change becomes $(s/{N})(C_{\rm max}^{(2)}X/2\xi\ep)^2$. Therefore Eq.~\eqref{eq:appx-bound-log-Nora} change only the prefactors, while the power of $k$ and $N$ remain unchanged. Hence the conclusion also hold for $m=2$. 

\section{Symmetry aspects of the autonomous Riemannian gradient based optimization}
\label{appx:autonomous_sym_inherence}

This appendix proves the symmetry results presented in Section~\ref{sec:symmetry-main} and~\ref{sec:unitary-learning}. Appendix~\ref{appx:symmetry-inherit} establishes \textit{symmetry inheritance}: when a symmetry of the data is shared by the observables and the initialized unitary, the training trajectory is confined within subspace of a unitary group, which is the commutant of an operator $S$, without any prior knowledge of such $S$. Appendix~\ref{appx:sym-generalization} shows that, for unitary learning with training states drawn from a single eigenspace of $S$ and a target unitary commuting with $S$, this confinement improves the generalization performance as a generalization bound depends on the dimension of that eigenspace rather than on the full Hilbert-space dimension. 

\subsection{Symmetry 
inheritance}
\label{appx:symmetry-inherit}

Consider an operator $S$ that commutes with every training state $\rho_i$, every observables $O_i$, and the initial unitary $U_0$,
\begin{equation}
\label{eq:appx-sym-inherit-cond}
    [S,\rho_i]=[S,O_i]=[S,U_0]=0
\end{equation}
for all $i$. The theorem below shows that every iterate $U_k$ of the Riemannian gradient descent then commutes with $S$.

\begin{theorem}[Restated Theorem~\ref{thm:sym-inherence}, Symmetry Inheritance]
    Under the conditions in Eq.~\eqref{eq:appx-sym-inherit-cond}, the unitary updated from Eq.~\eqref{eq:update_rule} satisfies
    \begin{equation}
    [S,U_k]=0\;\;,
    \end{equation} for every iteration $k$.
\end{theorem}
\begin{proof}

    We prove this by induction on $k$. For the case $k=0$, we have $[S,U_0]=0$ by the hypothesis. Assume $[S,U_{k'}]=0$ at some iteration $k'$. Multiply $SU_{k'}= U_{k'}S$ by $U_{k'}^\dagger$ on both side and using unitarity gives $[S,U_{k'}^\dagger]=0$. 
    Together with $[S,\rho_i]=0$ and $\rho_i^{(k')}:=U_{k'}\rho_iU_{k'}\ad$, we obtain $[S,\rho_i^{(k')}]=0$. 
    The Jacobi identity then gives 
    \begin{align}
        [S,[\rho_i^{(k')},O_i]]+[\rho_i^{(k')},[O_i,S]]+[O_i,[S,\rho_i^{(k')}]] &= 0\;\;,\nonumber \\ 
        [S,[\rho_i^{(k')},O_i]] &= 0\;\;.
    \end{align}
    The second line follows from the first because $[S,O_i]=0$ by the hypothesis and $[S,\rho_i^{(k')}]=0$ by the induction step. Since $(\partial f(\ell_i)/\partial \ell_i)\bigr|_{\ell_i=\ell_i(U_{k'})}$ in Eq.~\eqref{eq:generic_RG_grad} are scalar, $[S,G_{k'}]=0$. Because $S$ commutes with $G_{k'}$, it commutes with every power of $G_{k'}$. From $U_{k'+1}=e^{-s\rg_{k'}}U_{k'}$, we obtain $[U_{k'+1},S]=0$. By induction,
    \begin{equation}
        [S,U_k] =0\;\;,
    \end{equation}
    for all $k$.
\end{proof}

We now turn to the consequence of the symmetry inheritance for the unitary learning. Let a Hermitian operator $S$ have the spectral decomposition 
\begin{equation}
    S = \sum_{\mu}\mu\Pi_\mu\;\;,
\end{equation} 
where $\Pi_\mu$ is the projector onto the eigenspace of $S$ with eigenvalue $\mu$. 
For a $\deff$-dimensional eigenspace of $S$, we denoted this subspace by $\HC_{\rm eff}=\Pi_{\deff}\HC$, where $\Pi_{\deff}$ is the projector onto it. Then, we make the following assumption.
\begin{assumption}
\label{assump:sym-gen}
    Consider the unitary learning problem with the loss in Eq.~\eqref{eq:unitary-loss-fn}, training states $\rho_i=\ketbraq{\psi_i}$, and target states $\sigma_i=V\rho_iV^\dagger$.
    \begin{enumerate}
    \item Every training state lies in $\HC_{\rm eff}$: $\Pi_{\deff}\ket{\psi_i}=\ket{\psi_i}$ for all $i$. 
    \item The initial training unitary $U_0$ and the target unitary $V$ commute with $S$: $[S,U_0]=[S,V]=0$.
    \end{enumerate}
\end{assumption}
Point~$1$ in the assumption implies that the loss in Eq.~\eqref{eq:unitary-loss-fn} depends on $U$ only through its restriction to $\HC_{\rm eff}$: for every unitary $U$,
\begin{equation}
    \LC(U)=1-\frac{1}{N}\sum_{i=1}^N\lvert\bra{\psi_i}U^\dagger V\ket{\psi_i}\rvert^2 = 1-\frac{1}{N}\sum_{i=1}^N\lvert\bra{\psi_i}\Pi_{\deff} U^\dagger V\Pi_{\deff}\ket{\psi_i}\rvert^2\;\;.
    \label{eq:appx-loss-restricted}
\end{equation}
Beyond this restriction, point~$1$ and ~$2$ together imply the following corollary. 

\begin{corollary}[Sector-restricted mismatched unitary]
\label{coro:W-is-unitary}
    Under the point~$2$ in Assumption~\ref{assump:sym-gen}, for every unitary $U$ with $[S,U]=0$, the restricted mismatch operator $W:=\Pi_{\deff} {U}\ad V\Pi_{\deff}$ is a unitary operator on $\HC_{\rm eff}$,
    \begin{equation}
        W(U)^\dagger W(U)=\Pi_{\deff}\;\;,
    \end{equation}
\end{corollary}
\begin{proof}
    Since $\Pi_{\deff}\ket{\psi_i}=\ket{\psi_i}$, we have $S\rho_i=\l\rho_i=\rho_iS$. So $[S,\rho_i]=0$. Consequently, $[S,V\rho_iV\ad] =0$ because $[S,V]=0$. Together with $[S,U_0]=0$, Eq.~\eqref{eq:appx-sym-inherit-cond} holds with $O_i=V\rho_iV\ad$ and Theorem~\ref{thm:sym-inherence} gives $[S,U_k]=0$ for every $k$. Because $S=S\ad$, the spectral projector $\Pi_{\deff}$ is a polynomial in $S$, so $[\Pi_{\deff},V]=[\Pi_{\deff},U]=0$ for every $U$ commuting with $S$. Hence 
    \begin{equation}
    W\ad(U) W(U)=\Pi_{\deff} V^\dagger U\Pi_{\deff} U_k^\dagger V\Pi_{\deff}=\Pi_{\deff} V^\dagger UU^\dagger V\Pi_{\deff}=\Pi_{\deff}
    \end{equation}
    where the second equality uses $[\Pi_{\deff},U^\dagger V]=0$ and $\Pi_{\deff}^2=\Pi_{\deff}$, and the third uses the unitarity of $U$ and $V$. 
\end{proof}

\subsection{Symmetric generalization error bound for learning unitary}
\label{appx:sym-generalization}

Building upon Corollary~\ref{coro:W-is-unitary}, we now prove Theorem~\ref{thm:sample-complexity-u-learning}. We need three tools: Hoeffding's inequality, which provides the concentration of the loss value at a fixed unitary, an $\ep$-net of $\UC(\deff)$, which makes the concentration uniform over all unitaries, and a bound on the size of that net.

\begin{proposition}[Hoeffding's Inequality]
\label{prop:hoeffding}
    Let $Z_1,\dots,Z_N$ be independent real random variables with $Z_i\in [a_i,b_i]$, and let  
    \begin{equation}
        \Gamma_N := \sum_{i=1}^NZ_i\;\;.
    \end{equation}
    Then for every $t>0$, 
    \begin{equation}
        \Pr(|\Gamma_N-\Ebb[\Gamma_N]|\geq t)\leq 2\exp({-\frac{2t^2}{\sum_{i=1}^N(b_i-a_i)^2}})\;\;.
    \end{equation}
\end{proposition}

Next, we define the $\ep$-net, the covering number, and the ball that we use to discretize the unitary group.

\begin{definition}[$\ep$-Net and Covering Number]
    Given a set $T$ with a distance $d(\cdot,\cdot)$, consider a subset $\{t_1,\dots,t_M\}\subseteq T$ such that for all $t\in T$, there exists an index $j$ such that $d(t,t_j)\leq \ep$. We call such a subset an $
    \vep$-net and the smallest possible $M$ to cover $T$ the covering number. 
\end{definition}

\begin{definition}[Ball]
    A ball in a metric space $(T,d)$ centered at $x_0\in T$ with radius $R$ is 
    \begin{equation}
        B(x_0,R)=\{x \in T\bigr| d(x_0,x)\leq R\}\;\;.
    \end{equation}
\end{definition}

For $\deff=1$, the loss vanishes for every unitary commuting with $S$. Therefore, we assume $\deff\geq2$, which excludes this trivial case and ensures that $\log(\deff N)>0$. We are now ready to prove the generalization bound in Theorem~\ref{thm:sample-complexity-u-learning}.

\begin{theorem}
[Generalization bound of Theorem~\ref{thm:sample-complexity-u-learning}] 
\label{thm:appx-sample-complexity}
    Under Assumption~\ref{assump:sym-gen} with $\deff\geq 2$. Let training states $\ket{\psi_1},\dots,\ket{\psi_N}$ be drawn i.i.d. from a distribution $\DC$ on the unit vector of $\HC_{\rm eff}$. Let $\LC$ be training loss in Eq.~\eqref{eq:unitary-loss-fn}, $\LC_D(U):=1-\Ebb_{\psi\sim\DC}|\bramatket{\psi}{U\ad V}{\psi}|^2$ be the true risk, and ${\rm gen}(U)=\LC(U)-\LC_\DC(U)$ be the generalization error. Then, under the unitary update rule in Eq.~\eqref{eq:update_rule} with the probability at least $1-\d$ over the training sampling for every iteration $k$,
    \begin{align}
    \label{eq:appx-gen-bound}
        |{\rm gen}(U_k)|&\leq \frac{4}{\sqrt N}+\sqrt{\frac{\deff^2\ln(1+2\pi\sqrt{\deff N})+\ln(2/\d)}{2N}} 
        \nn &\in \OC\left(\sqrt{\frac{\deff^2\log(\deff N)+\log(1/\d)}{N}}\right)
    \end{align}
\end{theorem}
\begin{proof}
    Let $W=W(U)=\Pi_{\deff} U\ad V \Pi_{\deff} \in \UC(\deff)$ and set $f_i(W):=|\bramatketq{\psi_i}{W}|^2\in[0,1]$. By corollary~\ref{coro:W-is-unitary}, we write the training loss and the true risk as functions of $W$,
    \begin{align}
        \LC(U)=\tilde{\LC}(W):=1-\frac{1}{N}\sum_{i=1}^Nf_i(W),\quad\LC_\DC(U)=\tilde{\LC}_{\DC}(W):=1-\Ebb_{\psi\sim\DC}|\bramatketq{\psi}{W}|^2\;\;.
    \end{align}
    Since $\ket{\psi_i}$ are i.i.d., the variable $Z_i:=f_i(W)/N\in[0,1/N]$ are also i.i.d. with $\Ebb[\sum_{i=1}^NZ_i]=1-\tilde{\LC}_{\DC}(W)$. Therefore, Proposition~\ref{prop:hoeffding} with $\sum_{i=1}^N(b_i-a_i)^2=\frac 1 N$ gives
    \begin{equation}
    \label{eq:hoeffding-in-gen}
        \Pr(|\tilde{\LC}(W)-\tilde{\LC}_D(W)|\geq t)\leq 2e^{-2Nt^2}\;\;.
    \end{equation}

    Let $\{W_1,\dots,W_M\}$ be $\ep$-net of $\UC(\deff)$ in the Schatten 2-norm $\|\cdot\|_2$ and let 
    \begin{equation}
        E := \bigcup_{m=1}^M\{|\tilde{\LC}(W_m)-\tilde{\LC}_{\DC}(W_m)|>t\}
    \end{equation}
    be the event that some net point deviates by $t$. 
    By the union bound and Eq.~\eqref{eq:hoeffding-in-gen}, 
    \begin{equation}
        \Pr(E)\leq \sum_{m=1}^M\Pr(|\tilde{\LC}(W_m)-\tilde{\LC}_D(W_m)|\geq t)\leq 2Me^{-2Nt^2}\;\;.
    \end{equation}
    Equating the right-hand side to $\d$, that is, choosing
    \begin{equation}
    t=\sqrt{\frac{\ln M+\ln\frac{2}{\d}}{2N}}\;\;,
    \end{equation}
    guarantees that every net point obeys $|\tilde{\LC}(W_m)-\tilde{\LC}_\DC(W_m)|\leq t$ simultaneously, with the probability at least $1-\d$. 

    Next we show that both losses are $2$-Lipschitz in $W$. For $a=|\bramatketq{\psi}{W}|$ and $b=|\bramatketq{\psi}{W'}|$, which lie in $[0,1]$,
    \begin{equation}
        |a^2-b^2|=(a+b)|a-b|\leq 2|a-b|\leq 2|\bramatketq{\psi}{(W-W')}|\leq 2\|W-W'\|_{\infty}\leq 2\|W-W'\|_2
    \end{equation}
    where the second inequality is the reverse triangle inequality. Averaging over the sample gives 
    \begin{equation}
        |\tilde{\LC}(W)-\tilde{\LC}(W')|\leq \frac{1}{N}\sum_{i=1}^N|f_i(W)-f_i(W')|\leq 2\|W-W'\|_2\;\;,
    \end{equation}
    while taking the expectation over $\psi\sim\DC$ instead of the sample average gives the same bound for $\tilde{\LC}_{\DC}$. Let us now take any $W\in \UC(\deff)$ and a new point $W_j$ with $\|W-W_j\|\leq{\ep}$. On the complementary of $E$, the triangle inequality gives 
    \begin{align}
    \label{eq:appx-gen-net-bound}
        |\tilde{\LC}(W)-\tilde{\LC}_\DC(W)|
        &\leq |\tilde{\LC}(W)-\tilde{\LC}(W_m)|+|\tilde{\LC}(W_m)-\tilde{\LC}_\DC(W_m)|+|\tilde{\LC}_\DC(W_m)-\tilde{\LC}_\DC(W)|\;\;,\\
        &\leq2\ep+t+2\ep=4\ep+\sqrt{\frac{\ln M+\ln\frac{2}{\d}}{2N}}\;\;.
    \end{align}
    with the probability at least $1-\d$. 

     It remains to bound the covering number $M$. 
     We use a volumetric packing argument in Euclidean space $\Rbb^L$. Let $\{x_1,\cdots,x_M\}$ be a maximal subset of the ball $B(x_c,R)$ with pairwise distances greater than $\ep$. By maximality, this subset is an $\ep$-net of the ball, so the covering number of the ball is at most $M$. Therefore, the balls $B(x_i,\ep/2)$ are pairwise disjoint and contains in $B(x_c,R+\ep/2)$. Comparing their volumes gives 
    \begin{equation}
        M(\ep/2)^L\leq (R+\ep/2)^L\;\;,
    \end{equation}
    that is,
    \begin{equation}
        M\leq \left(\frac{R+\ep/2}{\ep/2}\right)^L=\left(1+\frac{2R}{\ep}\right)^L\;\;.
    \end{equation}
    To transfer this bound to the unitary group, consider the ball
    \begin{equation}
        B(W_0,R):=\{e^AW_0: A\ad=-A,\;\|A\|_2\leq R\}
    \end{equation}
    where $W_0\in\UC(\deff)$ is fixed. For any $W\in\UC(\deff)$, write the unitary $WW_0\ad=\sum_{m=1}^{\deff}e^{i\th_m}\ketbraq{v_m}$ in its eigenbasis and set $A:=i\sum_{m=1}^{\deff}\th_m\ketbraq{v_m}$, so that $A\ad=-A$ and $W=e^{A}W_0$. We choose the eigenphases in $(-\pi,\pi]$, this choice minimizes $\|A\|_2$  among all logarithms of $WW_0\ad$ and gives 
    \begin{equation}
        \|A\|^2_2=\sum_{m=1}^{\deff}\th_m^2\leq \deff \pi^2=R^2\;\;.
    \end{equation}
    Hence, every $W\in\UC(\deff)$ lies in $B(W_0,R)$ with the  radius $R=\pi\sqrt{\deff}$. Since the map $A\to e^{A}W_0$, both equipped with the Schatten 2-norm, is 1-Lipschitz, the image under $A\to e^{A}W_0$ of an $\ep$-net of the Lie-algebra ball $\{A:A\ad =-A\;, \|A\|_2\leq R\}$ is an $\ep$-net of $B(W_0,R)$. So we take $M$ to be the covering number of the Lie-algebra ball. To bound this covering number, we note that the space of anti-Hermitian matrices is a real vector space of dimension $\deff^2$, since an anti-Hermitian matrix is determined by $\deff$ imaginary diagonal entries and by the real and imaginary parts of its $\deff(\deff-1)/2$ upper-triangular entries. Thus, the volumetric bound applies with $L=\deff^2$ and $R=\pi\sqrt{\deff}$:
    \begin{equation}
        \ln(M)\leq \deff^2\ln(1+\frac{2\pi\sqrt{\deff}}{\ep})\;\;,
    \end{equation}
    and Eq.~\eqref{eq:appx-gen-net-bound} becomes 
    \begin{equation}
     |\tilde{\LC}(W)-\tilde{\LC}_\DC(W)|\leq 4\ep +\sqrt{\frac{\deff^2\ln(1+2\pi\sqrt{\deff}/\ep)+\ln(2/\d)}{2N}}\;\;,
    \end{equation}
    Since $\vep>0$ is arbitrary, choosing $\ep=1/\sqrt{N}$. For  $\deff\geq 2$ and $N\geq1$, we express all the terms by the multiple of the square-root term 
    \begin{align}
    \label{eq:appx-uniform-bound}
        |\tilde{\LC}(W)-\tilde{\LC}_\DC(W)| \leq \frac 4 {\sqrt{N}}+\sqrt{\frac{\deff^2\ln(1+2\pi\sqrt{\deff N})+\ln(2/\d)}{2N}}\;\;.
    \end{align}

    Finally, by Corollary~\ref{coro:W-is-unitary}, $W_k=\Pi_{\deff} U_k^\dagger V\Pi_{\deff}\in\UC(\deff)$ for every $k$, with $\LC(U_k)=\tilde{\LC}(W_k)$ and $\LC_\DC(U_k)=\tilde{\LC}_\DC(W_k)$. The bound in Eq.~\eqref{eq:appx-uniform-bound} holds for all $W\in\UC(\deff)$ on a single event of probability at least $1-\delta$ that depends only on the training sample. It therefore applies to every data-dependent iterate $W_k$ simultaneously, with no assumption on the value of the training loss, and gives
    \begin{align}
        \lvert{\rm gen}(U_k)\rvert=\lvert\LC(U_k)-\LC_\DC(U_k)\rvert&\leq\frac{4}{\sqrt N}+\sqrt{\frac{\deff^2\ln\!\big(1+2\pi\sqrt{\deff N}\big)+\ln(2/\delta)}{2N}}\;\;,\nn
        &\in \OC\left(\sqrt{\frac{\deff^2\log(\deff N)+\log(1/\d)}{N}}\right)\;\;,
    \end{align}
    for every iteration $k$, where we use $\ln(\deff N)\geq \ln 2 $ and $1+2\pi \sqrt{\deff N}\leq (1+2\pi)\sqrt{\deff N}$ to get $\ln(1+2\pi \sqrt{\deff N})+\ln(2)<c \ln (\deff N)$, and  $\ln(16)\leq \deff^2 \ln (\deff N)$ for $\deff\geq 2$ and $N\geq 1$. This proves Eq.~\eqref{eq:appx-gen-bound}.
\end{proof}

Theorem~\ref{thm:appx-sample-complexity} bounds the gap between the training loss and the true risk, but does not determine the value of the true risk itself. Using the second-moment formula for a 2-design~\cite{mele2023introduction}, we now derive a closed form for $\LC_{\DC}(U)$ when $\DC$ is a state-2 design on $\HC_{\rm eff}$, which is the second part of Theorem~\ref{thm:sample-complexity-u-learning}.

\begin{proposition}[Second moment of a state 2-design]
\label{prop:second-moment-2-design}
    Let $\DC$ be a state 2-design on 
    $\HC_{\rm eff}$ and let $O$ be a linear operator in that space. Then
    \begin{equation}
        \Ebb_{\psi\sim \DC}|\bramatketq{\psi}{O}|^2=\frac{|\Tr(O)|^2+\Tr(O\ad O)}{\deff(\deff+1)}\;\;.
    \end{equation}
\end{proposition}

\begin{theorem}[True risk under a 2-design in Theorem~\ref{thm:sample-complexity-u-learning} ]
    Let Assumption~\ref{assump:sym-gen} holds and let $\DC$ be a state $2$-design on the $\deff$-dimensional Hilbert space associated on $\HC_{\rm eff}$. Then, for every $U$ with $[S,U]=0$,  
    \begin{equation}
        \LC_\DC(U)=1-\frac{|\Tr(\Pi_{d_{\rm eff}} U\ad V)|^2+\deff}{\deff(\deff+1)}\;\;.
    \end{equation}
\end{theorem}

\begin{proof}
From Eq.~\eqref{eq:appx-loss-restricted}, replacing the sample average by the expectation gives $\LC_{\DC}(U)=1-\Ebb_{\psi\sim \DC}[|\bramatketq{\psi}{W}|^2]$ with $W=\Pi_{d_{\rm eff}}U\ad V\Pi_{d_{\rm eff}}$. Proposition~\ref{prop:second-moment-2-design} with $O=W$ then gives 
\begin{align}
    \LC_{\DC}(U)&=1-\frac{|\Tr(W)|^2+\Tr(W\ad W)}{\deff(\deff+1)}\;\;,\nn &=1-\frac{|\Tr(W)|^2+\Tr(\Pi_{d_{\rm eff}})}{\deff(\deff+1)}\;\;,\nn 
    &=1-\frac{|\Tr(W)|^2+\deff}{\deff(\deff+1)}\;\;,\nn
    &=1-\frac{|\Tr(\Pi_{d_{\rm eff}} U\ad V)|^2+\deff}{\deff(\deff+1)}\;\;.
\end{align}
The second line use $W\ad W =\Pi_{d_{\rm eff}}$ from Corollary~\ref{coro:W-is-unitary} which is hold because $[S,U]=0$. Then, the last line uses the cyclicity of the trace together with $\Pi_{d_{\rm eff}}^2=\Pi_{d_{\rm eff}}$.
\end{proof}


\section{Extended Numerical Results} 
\label{appx:extended_numerics}

\begin{figure}[!h]
    \centering
    \includegraphics[width=\linewidth]{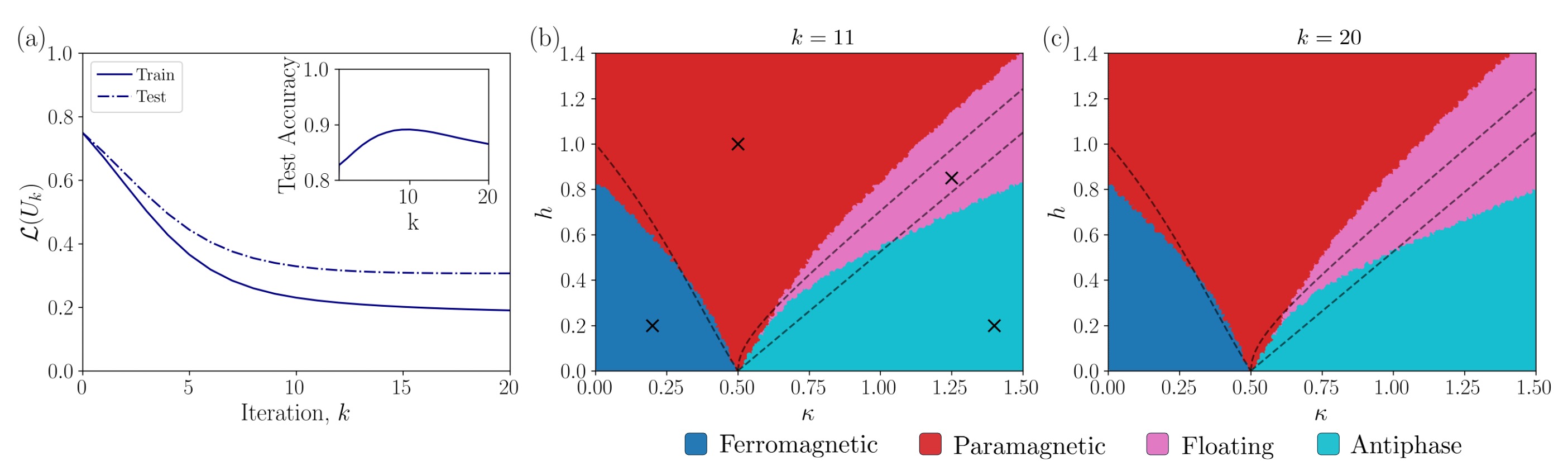}
    \vspace{-0.5 cm}
    \caption{\textbf{Performance of phase classification for the ANNNI model} (a) The training and testing loss functions are plotted against the number of training iterations $k$. In the inset of (a), the test accuracy is plotted against $k$ from $k=1$ to $k=20$. (b-c) The predicted phases in $\kappa$-$h$ plane after $k=11$ and $k=20$ iterations are shown for all testing points, respectively. Each color represents a different predicted phase (blue, red, pink, and cyan are ferromagnetic,  paramagnetic, floating, and antiphase phases, respectively), and black crosses ($\times$) indicate the four training samples. The dashed lines are the thermodynamic-limit phase boundaries of Ref.~\cite{cea2024exploring}. We show $k=11$ because it is the iteration that achieve the maximal test accuracy.}
    \label{fig:p3_classif_predicted_phase_diagram}
\end{figure}

Here we show that the phase-classification results in Section~\ref{sec:numerics_classif} are not specific to the 1D generalized cluster model by extending the numeric to the axial next-nearest-neighbour Ising (ANNNI) model. The Hamiltonian for the system size $n'$ takes the form of 
\begin{equation}
    H_{\rm ANNNI} = -J_1 \sum_{i=1}^{n'-1} X_i X_{i+1} - J_2 \sum_{i=1}^{n'-2} X_i X_{i+2} - B \sum_{i=1}^{n'} Z_i\;\;,
\end{equation}
parameterized by $J_1$, $J_2$, and $B$. This model also exhibits four distinct phases~\cite{cea2024exploring}, characterized in the coordinates of $h=B/J_1$ and $\kappa=-J_2/J_1$. Throughout our analysis of this model, we set $J_1=1$. As in Section~\ref{sec:numerics_classif}, each input is the ground state of $H_{\rm ANNNI}$ at a point $(h,\kappa)$ encoded in a two-qubit ancilla in the uniform superposition over the $Q=4$ labels. We simulate an $8$ qubit system (with the total of $n=10$ qubits to run the algorithm) using a step size of $s=1$ and $U_0=\mathbb{1}$. We train the model using only $4$ data points, each point representing one of the four phases, indicated by the black crosses in Fig.~\ref{fig:p3_classif_predicted_phase_diagram}(b). The test set is a grid of $20,974$ points over $h\in[0.01,1.40]$ and $\kappa\in[0.01,1.50]$, with the label taken from the phase boundaries of Ref.~\cite{cea2024exploring}. As a result, we obtain the loss values of $0.19$ and $0.31$ for training and testing dataset, respectively, after $k=20$ training iterations as shown in Fig.~\ref{fig:p3_classif_predicted_phase_diagram}(a). For this model, the test accuracy increases with $k$ until it reaches its maximum of $0.89$ at $k=11$ with the predicted phase diagram in Fig.~\ref{fig:p3_classif_predicted_phase_diagram}(b). After that, the test accuracy reduces to $0.87$ at $k=20$ with the predicted phase diagram in Fig.~\ref{fig:p3_classif_predicted_phase_diagram}(c). This slight decrease is a mild overfitting effect of the four-sample training set with the misclassified points lie mostly near the phase boundaries, similar to the generalized cluster model in Section~\ref{sec:numerics_classif}.

\end{document}